\documentclass[useAMS,usenatbib,twocolumn]{mn2e}

\usepackage{textcomp,graphicx,amsmath,pdflscape,amssymb,setspace}

\def\gtsim{\raisebox{-.5ex}{$\;\stackrel{>}{\sim}\;$}}
\def\B{\textit{B}}
\def\V{\textit{V}}
\def\R{\textit{R}}
\def\I{\textit{i$^\prime$}}
\def\z{\textit{z$^\prime$}}
\def\Ne{\textit{NB}816}
\def\Nn{\textit{NB}921}
\def\J{\textit{J}}
\def\K{\textit{K}}
\def\sex{\normalsize SE\scriptsize XTRACTOR\normalsize} 

\newcommand\aj{{AJ}}%
\newcommand\araa{{ARA\&A}}%
\newcommand\apj{{ApJ}}%
\newcommand\apjl{{ApJ}}%
\newcommand\apjs{{ApJS}}%
\newcommand\aap{{A\&A}}%
\newcommand\aaps{{A\&AS}}%
\newcommand\mnras{{MNRAS}}%
\newcommand\pasp{{PASP}}%
\newcommand\pasj{{PASJ}}%
\newcommand\ssr{{Space~Sci.~Rev.}}%
\newcommand\nat{{Nature}}%
\newcommand\na{{NewA}}%
\newcommand\raa{{RAA}}%

\title[Supernovae in the Subaru Deep Field]{Supernovae in the Subaru Deep Field: the rate and delay-time distribution of type~Ia supernovae out to redshift 2}

\author[Graur et al.]
{O.~Graur,$^1$\thanks{E-mail: orgraur@wise.tau.ac.il}
D.~Poznanski,$^{2,3,4}$
D.~Maoz,$^1$
N.~Yasuda,$^5$
T.~Totani,$^6$ 
M.~Fukugita,$^5$ \newauthor 
A.~V.~Filippenko,$^3$
R.~J.~Foley,$^{3,7}$
J.~M.~Silverman,$^3$ 
A.~Gal-Yam,$^8$ 
A.~Horesh,$^{9}$ \newauthor
and B.~T.~Jannuzi$^{10}$
\\
$^1$School of Physics and Astronomy, Tel-Aviv University, Tel-Aviv 69978, Israel \\
$^2$Lawrence Berkeley National Laboratory, 1 Cyclotron Rd., Berkeley, CA 94720, USA \\
$^3$Department of Astronomy, University of California, Berkeley, CA 94720-3411, USA \\
$^4$Einstein Fellow \\
$^5$Institute for the Physics and Mathematics of the Universe, University of Tokyo, Kashiwa 2778583, Japan \\
$^6$Department of Astronomy, School of Science, Kyoto University, Sakyo-ku, Kyoto 606-8502, Japan \\
$^7$Harvard/Smithsonian Center for Astrophysics, 60 Garden St., Cambridge, MA 02138, USA \\
$^8$Department of Particle Physics and Astrophysics, Weizmann Institute of Science, 76100 Rehovot, Israel \\
$^9$Cahill Center for Astrophysics, California Institute of Technology, Pasadena, CA 91125, USA \\
$^{10}$National Optical Astronomy Observatory, Tucson, AZ 85726-6732, USA
}
\date{2011 June 20}

\begin{document}

\maketitle


\setstretch{1}

\begin{abstract}
\noindent
The type~Ia supernova (SN~Ia) rate, when compared to the cosmic star formation history (SFH), can be used to derive the delay-time distribution (DTD, the hypothetical SN~Ia rate vs. time following a brief burst of star formation) of SNe~Ia, which can distinguish among progenitor models.
We present the results of a SN survey in the Subaru Deep Field (SDF).
Over a period of three years, we have observed the SDF on four independent epochs with Suprime-Cam on the Subaru 8.2-m telescope, with two nights of exposure per epoch, in the \R, \I, and \z\ bands.
We have discovered 150 SNe out to redshift $z \approx 2$.
Using 11 photometric bands from the observer-frame far-ultraviolet to the near-infrared, we derive photometric redshifts for the SN host galaxies (for 24 we also have spectroscopic redshifts). 
This information is combined with the SN photometry to determine the type and redshift distribution of the SN sample.
Our final sample includes 28 SNe~Ia in the range $1.0<z<1.5$ and 10 in the range $1.5<z<2.0$.
As our survey is largely insensitive to core-collapse SNe (CC~SNe) at $z>1$, most of the events found in this range are likely SNe~Ia.
Our SN~Ia rate measurements are consistent with those derived from the {\it Hubble Space Telescope} ({\it HST}) GOODS sample, but the overall uncertainty of our $1.5<z<2.0$ measurement is a factor of 2 smaller, of 35--50 per cent.
Based on this sample, we find that the SN~Ia rate evolution levels off at $1.0<z<2.0$, but shows no sign of declining. 
Combining our SN~Ia rate measurements and those from the literature, and comparing to a wide range of possible SFHs, the best-fitting DTD (with a reduced $\chi^2 = 0.7$) is a power law of the form $\Psi(t) \propto t^\beta$, with index $\beta = -1.1 \pm 0.1$ (statistical) $\pm 0.17$ (systematic).
This result is consistent with other recent DTD measurements at various redshifts and environments, and is in agreement with a generic prediction of the double-degenerate progenitor scenario for SNe~Ia.
Most single-degenerate models predict different DTDs.
By combining the contribution from CC~SNe, based on the wide range of SFHs, with that from SNe~Ia, calculated with the best-fitting DTD, we predict that the mean present-day cosmic iron abundance is in the range $Z_{\rm{Fe}}=(0.09$--$0.37)~Z_{\rm{Fe,\odot}}$.
We further predict that the high-$z$ SN searches now beginning with {\it HST} will discover 2--11 SNe~Ia at $z > 2$.

\end{abstract}



\section{INTRODUCTION}
\label{sec:intro}

Supernovae (SNe) play important roles in a variety of astrophysical settings, from galaxy evolution to the metal enrichment of the interstellar medium, as catalysts of star formation, and as distance indicators.
SNe are separated into two main physical classes: core-collapse SNe (CC~SNe), which include all type~II SNe (i.e., those objects which exhibit obvious H lines in their spectra) and Type~Ib/c SNe (i.e., spectra lacking H and with weak Si and S lines); and type~Ia SNe (SNe~Ia), which show strong Si and S, but no H, lines in their spectra (see \citealt{1997ARA&A..35..309F} for a review; see \citealt{2010Natur.465..322P} for a possible exception). 
CC~SNe occur in massive stars that have reached the end of their fuel cycles. 
Pre-explosion images have revealed directly the progenitors of some CC~SNe, confirming that the progenitors of SNe~II-P and SNe~IIn are red and blue supergiants (or luminous blue variables), respectively; that most SNe~Ib/c are the result of moderate-mass interacting binaries; and that broad-lined SNe~Ic are the explosions of massive Wolf-Rayet stars (see \citealt{2009ARA&A..47...63S} for a review).

In contrast, SNe~Ia are thought to be the result of the thermonuclear combustion of carbon-oxygen white dwarfs (WDs) that approach the Chandrasekhar limit through mass accretion in close binary systems (see \citealt{2000ARA&A..38..191H} and \citealt{2010arXiv1011.0441H} for reviews).
Two basic routes have been suggested for the WD to grow in mass.
The single-degenerate model postulates mass accretion from a main-sequence or giant companion star \citep{1973ApJ...186.1007W}, whereas the double-degenerate (DD) model invokes the merger of two WDs \citep{1984ApJS...54..335I, 1984ApJ...277..355W}.
However, there have been no unambiguous identifications of SN~Ia progenitors in pre-explosion images, or of remaining companions in historical SN~Ia remnants (e.g., \citealt{2008Natur.451..802V, 2008MNRAS.391..290R, 2009ApJ...691....1G, 2009ApJ...701.1665K}).
Programmes that seek to determine the DD merger rate by surveying for WD binaries \citep{2004ASPC..318..402N, 2007A&A...464..299G, 2009ApJ...707..971B} have yet to conclude whether this channel can account for some or all of the SN~Ia rate.
\citet{2010arXiv1011.4322T} has recently proposed that at least some of the SN~Ia progenitors may be triple systems, comprised of a WD--WD inner binary and a tertiary that induces \citet{Kozai1962} oscillations in the inner binary, driving it to higher eccentricity and shortening the time until a gravitational-wave-driven merger between the two WDs.
The possibility of detecting such triple systems through their gravitational-wave signals is explored by \citet{2011ApJ...729L..23G}.
 
One way to constrain indirectly the different SN~Ia progenitor models is through their delay-time distribution (DTD) --- the distribution of times between a hypothetical $\delta$-function-like burst of star formation, and the subsequent SN~Ia explosions.
Different progenitor and explosion models predict different forms of the DTD (e.g., \citealt{2000ApJ...528..108Y, 2004MNRAS.350.1301H}; \citealt*{2009ApJ...699.2026R}; \citealt{2010A&A...515A..89M}).
Metallicity effects can also affect the DTD in some models (e.g., \citealt{2009ApJ...707.1466K}).
There are various ways to estimate the DTD observationally.
\citet{2005A&A...433..807M} compared the SN~Ia rate per unit mass in different types of galaxies and found that the rate in blue galaxies is a factor of 30 larger than in red galaxies.
This result led to the so-called \textquoteleft $A+B$\textquoteright\ model \citep{2005ApJ...629L..85S}, which reproduces the SN~Ia rate with a term proportional (through $A$) to the total stellar mass of the SN host population, and a second term which is proportional (through $B$) to the star formation rate (SFR) of the host population.
The $A+B$ model is effectively a two-time-bin approximation of the DTD.

\citet{2008PASJ...60.1327T} compared the SN rates in elliptical galaxies in the Subaru-XMM Deep Field (SXDF) to the mean ages of their stellar populations, and deduced a power-law shape of the form $\Psi(t) \propto t^{\beta}$ for the DTD, with $\beta \approx -1$ in the delay-time range of 0.1--4~Gyr.
\citet{maoz2010loss} compared the SN rate and the star formation histories (SFHs) of a subset of the galaxies monitored by the Lick Observatory SN Search \citep{2011MNRAS.412.1419L}.
They reconstructed a falling DTD, with significant detections of both \textquoteleft prompt\textquoteright\ SNe~Ia (with delays of $<420$~Myr) and \textquoteleft delayed\textquoteright\ ones ($>2.4$~Gyr). 
Similar results were obtained by \citet{2010AJ....140..804B}, analysing the SNe~Ia from the Sloan Digital Sky Survey II (SDSS-II; \citealt{2000AJ....120.1579Y}).
\citet{maoz2010magellan} compared between the SN rate in the Magellanic Clouds as inferred from SN remnants and the SFHs of their resolved stellar populations, and detected a prompt component 
in the DTD.
Comparisons of the rates of SNe~Ia and the luminosity-weighted mean ages of their host populations have been undertaken by \citet{2008A&A...492..631A}; \citet{2009ApJ...707...74R}; \citet*{2009ApJ...704..687C}; \citet{Schawinski2009}; and \citet{2010AJ....139...39Y}.
While some of these studies may be susceptible to biases resulting from the choices of \textquoteleft control samples\textquoteright\ (see, e.g., \citealt{2008MNRAS.384..267M}), they have generally also found evidence for a population of SNe~Ia with short delays.

Measurement of SN rates versus redshift in galaxy clusters has provided another powerful probe of the DTD.
Cluster SFHs are relatively simple, and thus the form of the DTD is obtainable almost directly from the SN rate as a function of cosmic time. 
Furthermore, the deep gravitational potentials mean that the total metal content of clusters, as quantified by optical and X-ray measurements, provide a record of the time-integrated contributions, and hence numbers, of SNe over the cluster histories. 
This sets the integral of the DTD.
\citet*{maoz2010clusters} have recently compiled and analysed cluster SN rates from a number of surveys in the redshift range $0<z<1.5$ (\citealt*{2002MNRAS.332...37G}; \citealt{2007ApJ...660.1165S, 2010ApJ...718..876S, 2008AJ....135.1343G, 2008MNRAS.383.1121M, 2010ApJ...715.1021D, 2010arXiv1010.5786B}). 
They find that the best-fitting DTD is a power law with an index of $\beta = -1.1 \pm 0.2$ or $\beta = -1.3 \pm 0.2$, depending on the assumed value of the present-day stellar-to-iron mass ratio in clusters. 
Thus, a variety of recent attempts to recover the DTD, involving a range of techniques, redshifts, and environments, consistently indicate a power-law DTD with index $\beta \approx -1$
(see \citealt{maoz2010clusters} for an intercomparison of these results).  

There is, however, one approach to recover the DTD that has produced some conflicting results. 
The SN rate in field galaxies at cosmic time $t$, $R_\textrm{Ia}(t)$, is the convolution of the SFH, $S(t)$, with the DTD, $\Psi(t)$:
\begin{equation}
\label{eq:dtd_intro}
R_{\textrm{Ia}}(t) = \int_0^t S(t-\tau) \Psi(\tau)d\tau.
\end{equation}
\noindent
The DTD can therefore be recovered, in principle, by comparing the field SN~Ia rate vs. redshift to the cosmic SFH. 
The cosmic SFH has been measured out to $z \approx 6$ (see, e.g., the compilation of \citealt{2006ApJ...651..142H}, hereafter HB06), and several surveys have attempted to extend these measurements out to $z \approx 8$ (\citealt{2007MNRAS.377.1024V, 2008ApJ...683L...5Y}, hereafter Y08; \citealt{2008ApJ...686..230B, 2009ApJ...692..778R, 2009ApJ...705L.104K, 2010RAA....10..867Y}).
While all surveys observe a rise in the SFH towards $z = 1$--2.5, to date estimates of the SFH based on the ultraviolet (UV) emission of field galaxies (e.g., \citealt{2010arXiv1006.4360B}) have produced shallower evolutions than those based on the far-infrared (IR) continuum of galaxies, (e.g., \citealt{2005ApJ...632..169L, 2010ApJ...718.1171R}).
This is due to the systematic uncertainty introduced by the need to correct the observed UV luminosity for extinction by dust.
A recent attempt by \citet[hereafter O08]{2008PASJ...60..169O} to derive the cosmic SFH using CC~SN and SN~Ia rate measurements found constraints which are consistent with the latest IR-based SFH measurements, and slightly higher than the latest UV-based measurements. 

\citet{2004MNRAS.347..942G} were the first to set constraints on the DTD with this approach, based on a small sample of SNe~Ia out to $z=0.8$.
A number of surveys over the past decade have measured the SN~Ia rate out to $z \approx 0.2$ (\citealt*{cappellaro1999}; \citealt{hardin2000, 2002ApJ...577..120P, 2003ApJ...594....1T, blanc2004, botticella2008}, hereafter B08; \citealt{horesh2008, li2011rates}).
Additional surveys, such as the SDSS \citep{2003ApJ...599L..33M, dilday2008, dilday2010a} and the Supernova Legacy Survey (SNLS; \citealt{neill2006}, hereafter N06 \citealt{neill2007}) have added measurements out to $z \approx 0.8$.
The previously discordant measurements of the Institute for Astronomy (IfA) Deep Survey \citep{2006ApJ...637..427B} have recently been corrected and extended to redshift $z=1.05$ by \citet{2010ApJ...723...47R}.

Measurements of the SN rate at $z>1$ were first realized by \citet[hereafter D04]{dahlen2004}, using the {\it Hubble Space Telescope} ({\it HST}) Advanced Camera for Surveys (ACS) observations of the GOODS fields.
Additional data were analysed by \citet*[hereafter D08]{dahlen2008}.
D04 and D08 argued that their data indicate a peak in the SN rate at $z \approx 0.8$, with a steep decline at higher redshifts.
Based on this rate evolution, \citet{2004ApJ...613..200S}, D04, and D08 deduced a best-fitting narrow Gaussian-shaped DTD, centred at a delay time of 3.4~Gyr.
Similarly, \citet*{2010ApJ...713...32S} adopted a unimodal, skew-normal function (see their equation~6) for the DTD, from which they inferred that the DTD should be confined to a delay-time range of 3--4~Gyr.
However, analysing much of the same data, \citet{kuznetsova2008goods} found that they could not distinguish between a flat SN rate at $z>0.5$ and a decline at $z>1$, due to the large statistical and systematic uncertainties in the {\it HST}/GOODS dataset.

\citet{horiuchi2010} recently found that when coupled with the Y08 SFH, the Gaussian DTD proposed by D08, along with the bimodal DTD from \citet*{2006MNRAS.370..773M}, underpredicted precise SN~Ia rate measurements at $z<0.3$.
A power-law DTD with index $\beta=-1.0 \pm 0.3$, however, fit the data well.
A similar attempt by \citet{2008NewA...13..606B} to couple between the cosmic SFH and the SN~Ia rates from the above data also led to the conclusion that a broad range of DTD models could be accomodated by the data, including power-law DTDs, due to small-number statistics. 
In three {\it HST} cycles, GOODS found 53 SNe~Ia, of which only 3 were in the $1.4<z<1.8$ range.
Larger SN~Ia samples are clearly needed in order to determine precise rates at these redshifts, to recover the DTD, and to compare it to other measurements.

To address this problem, in 2005 we initiated a ground-based high-$z$ SN survey with the objective of determining the SN~Ia rate at $z>1$. 
Our survey is based on single-epoch discovery and classification of SNe in the Subaru Deep Field (SDF; \citealt{2004PASJ...56.1011K}, hereafter K04).
In 2007 we presented initial results from our survey for SNe~Ia out to $z=1.6$, based on the first epoch of observations (\citealt{poznanski2007sdf}, hereafter P07b).
This first epoch produced a number of SNe~Ia that was similar to that found by D04 in GOODS. 
The high-$z$ rates we found were also consistent with those of D04 and D08, with similar uncertainties, but our results suggested a flat rather than a declining SN~Ia rate at high redshifts.

In this paper, we present our final sample of 150~SNe, based on four SDF epochs, and derive the most precise SN~Ia rates to date at $1<z<2$.
In Section \ref{sec:obs} we describe our observations of the SDF and spectroscopy of several of our SN host galaxies. 
Sections \ref{sec:cands} and \ref{sec:gal} detail our methods for discovering the SNe and their host galaxies. 
In Section \ref{sec:classify} we classify the SN candidates into SNe~Ia and CC~SNe with the SN Automatic Bayesian Classifier (SNABC) algorithm of \citet*[hereafter P07a]{poznanski2007snabc}. 
The distribution of SNe among types and redshift bins is examined in Section \ref{sec:debias}, and corrected for biases introduced by the SNABC. 
We derive the SN~Ia and CC~SN rates in Section~\ref{sec:rates}. 
The SN~Ia rates, along with rates collected from the literature, are then used to constrain the DTD in Section~\ref{sec:DTD}.
The best-fitting DTD is used to predict the SN~Ia rate at $z>2$ and calculate the accumulation of iron in the Universe, as a function of redshift, in Section \ref{sec:iron}.
We summarise and discuss our results in Section \ref{sec:discuss}.
Throughout this paper we assume a $\Lambda$-cold-dark-matter ($\Lambda$CDM) cosmological model with parameters $\Omega_{\Lambda} = 0.7$, $\Omega_m = 0.3$, and $H_0 = 70$~km~s$^{-1}$~Mpc$^{-1}$.
Unless noted otherwise, all magnitudes are on the AB system \citep{1983ApJ...266..713O}.


\section{OBSERVATIONS AND REDUCTIONS}
\label{sec:obs}

\subsection{Imaging}
\label{subsec:imaging}

\begin{table*}
 \begin{minipage}{\textwidth}
\center
\hspace{1.5in}\parbox{5.8in}{\caption{Summary of optical imaging data for epochs 2 through 5}\label{table:exp_time}}
 \begin{tabular}{l|c|c|c|c|c|c|l}
  \hline
  \hline
  {Epoch} & {Band} & {Exp.} & {Seeing} & {3$\sigma~m_{\rm{lim}}~^a$} & {5$\sigma~m_{\rm{lim}}~^b$} & {$m_0~^c$} & {UT Date} \\
  {} & {} & {[s]} & {[arcsec]} & {[mag]} & {[mag]} & {[mag/count]} & {} \\
  \hline
  2 & \R & 7,920   & 1.06 & 27.18 & 26.63 & 33.93 & 2005 Mar. 5/6 \\
  {}  & \I & 10,800  & 0.99 & 27.00 & 26.45 & 33.99 & 2005 Mar. 5/6 \\
  {} & \z & 18,240  & 1.03 & 26.33 & 25.77 & 32.92 & 2005 Mar. 5/6 \\
  3 & \R & 11,460  & 0.79 & 27.98 & 27.43 & 34.08 & 2007 Feb. 12/13/14/15 \\
  {}  & \I & 15,000  & 0.80 & 27.79 & 27.24 & 34.11 & 2007 Feb. 12/13/14/15 \\
  {} & \z & 27,240  & 0.85 & 26.90 & 26.35 & 33.01 & 2007 Feb. 12/13/14/15 \\
  4 & \R & 8,220 & 0.90 & 27.36 & 26.80 & 33.14 & 2007 May 15/16 \\
  {}  & \I & 7,960   & 0.84 & 27.17 & 26.62 & 33.16 & 2007 May 15/16 \\
  {} & \z & 17,150  & 0.73 & 26.86 & 26.30 & 31.87 & 2007 May 15/16  \\
  5 & \R & 10,550  & 0.83 & 27.70 & 27.14 & 34.00 & 2008 Jun. 1/2/3/4 \\
  {}  & \I & 12,960  & 0.81 & 27.50 & 26.94 & 34.06 & 2008 Jun. 1/2/3/4 \\
  {} & \z & 23,500  & 0.73 & 27.21 & 26.66 & 32.99 & 2008 Jun. 1/2/3/4 \\
  \hline 
  \multicolumn{8}{l}{$^a 3\sigma$ limiting magnitude, within a circular aperture having a radius the size of the image's seeing FWHM.} \\
  \multicolumn{8}{l}{$^b 5\sigma$ limiting magnitude.} \\
  \multicolumn{8}{l}{$^c$Magnitude zero point, i.e., the magnitude of a source in the image with 1 count (2.6 $e^-$).} \\
 \end{tabular}
\end{minipage}
\end{table*}

The SDF ($\alpha=13^{\rm h}24^{\rm m}39^{\rm s}, \delta=+27^{\circ}29'26''$; J2000) was first imaged by K04 with the Suprime-Cam camera on the Subaru 8.2-m telescope on Mauna Kea, Hawaii. 
Suprime-Cam is a $5 \times 2$ mosaic of $2\rm{k} \times 4\rm{k}$ pixel CCDs at the prime focus of the telescope, with a field of view of $34 \times 27$ arcmin$^2$, and a scale of 0.202 arcsec pixel$^{-1}$ \citep{2002PASJ...54..833M}.
K04 imaged the SDF in five broad-band filters (\B, \V, \R, \I, and \z) and two narrow-band filters (\Ne\ and \Nn), over an area of $30 \times 37$ arcmin$^2$, down to $3\sigma$ limiting magnitudes of $\B=28.45$, $\V=27.74$, $\R=27.80$, $\I=27.43$, $\z=26.62$, $\Ne=26.63$, and $\Nn=26.54$ (5$\sigma$ limits of $\B=27.87$, $\V=27.15$, $\R=27.24$, $\I=27.01$, $\z=26.06$, $\Ne=26.24$, and $\Nn=26.07$), as measured in circular apertures having radii of 1 arcsec.
See K04 for details of those observations.
This initial epoch of optical observations is denoted here as \textquoteleft epoch 1.\textquoteright\

In our analysis, we also make use of additional existing data on the SDF, particularly for estimating the properties of the galaxies hosting the SNe we find. 
Near-infrared (NIR) photometry, in \J\ and \K, was obtained with the Wide-Field Camera on the United Kingdom Infrared Telescope (UKIRT; \citealt{2009ApJ...691..140H}; Motohara et al., in preparation) down to 3$\sigma$ limiting magnitudes of $\J=24.67$ and $\K=25.07$ in apertures with radii of 1 arcsec (5$\sigma$ limits of $\J=24.33$ and $\K=24.52$ mag). 
While the \K-band data cover the same area of the SDF as the optical observations, the \J-band data cover only $\sim 40$ per cent of the field.
UV observations of the SDF were obtained by the {\it Galaxy Evolution Explorer} ({\it GALEX}; \citealt{Ly2009}), with total exposures of 81~ks in the far-UV ({\it FUV}) band ($\lambda \approx 1530$~\AA) and 161~ks in the near-UV ({\it NUV}) band ($\lambda \approx 2270$~\AA). 
These integration times result in 3$\sigma$ limiting magnitudes of 26.42 and 26.49 in the ({\it FUV}) and ({\it NUV}) bands, respectively, in apertures with radii of 7.5 arcsec (or 5$\sigma$ limits of 25.86 and 25.93 mag).

We reimaged the field on four separate epochs (UT dates are used throughout this paper): 2005 March 5 and 6 (epoch 2, analysed by P07b); 2007 February 12--15 (epoch 3); 2007 May 15 and 16 (epoch 4); and 2008 June 1-4 (epoch 5).
During epochs consisting of two nights, the SDF was observed during most of the night. 
On the epochs that were spread over four nights, either the first or the second half of each night was 
dedicated to the SDF programme. 
In either case, we consider the consecutive nights to be a single epoch, whose nightly images can be coadded, given the longer time scales on which SNe evolve at the redshifts we probe. 
On all occasions, we imaged the field in the three reddest Suprime-Cam broad bands: \R, \I, and \z.
These filters, which probe the rest-frame blue emission of SNe at $z=1$--2, are the most suitable for discovering and classifying such SNe (e.g., \citealt{2002PASP..114..833P, 2004PASP..116..597G, 2004ApJ...600L.163R}).
We followed a dithering pattern similar to the one described by K04.
Table~\ref{table:exp_time} lists the exposure times, average seeing, and limiting magnitudes in each band, for epochs 2 through 5.
In general, the average seeing for each night ranged between 0.7 and 1 arcsec full width at half-maximum intensity (FWHM).

We reduced the Subaru observations with the Suprime-Cam pipeline {\it SDFRED} \citep{2002AJ....123...66Y, 2004ApJ...611..660O}.
Briefly, the individual frames were overscan subtracted, flat fielded using superflats, distortion corrected, sky subtracted, registered, and combined. 
In contrast to K04 and P07b, we did not apply point-spread function (PSF) degradation on the new images, since it reduces the frame depth.
The combined image was then matched to the \I-band image from K04 by using the {\it astrometrix}\footnote{http://www.na.astro.it/$\sim${radovich}} code to find the astrometric correction, and the IRAF\footnote{IRAF is distributed by the National Optical Astronomy Observatories, which are operated by the Association of Universities for Research in Astronomy, Inc., under cooperative agreement with the National Science Foundation (NSF).} \citep{1986SPIE..627..733T} task {\it wregister} to register the two images.
The photometric calibration of the images from epoch 1 was done by K04, achieving a precision for the zero points of $\sim 0.05$ mag (see section~4.2 of K04).
We calibrated our images relative to epoch 1 by comparing the photometry of all the objects detected with \sex\ \citep{sextractor} in both epochs.
The mean of the differences between the two measurements was taken to be the difference in zero points.

In order to create a reference image to be compared to each epoch, the images of all the other epochs were scaled, weighted according to their depth, and stacked using the IRAF task {\it imcombine}. 
The stacking process included a sigma-clipping procedure that excluded any transient or highly variable objects from the resulting summed image.
Four \textquoteleft master\textquoteright\ images were created in this fashion, for each search epoch, where each such image is composed of all other epochs, except the search epoch in question.
These master images proved deeper and sharper than the original epoch-1 images used by P07b as reference images for the subtraction process.
For example, the epoch-5 master image has a $3\sigma$ limiting magnitude of $\z=27.01$, as measured in an aperture having a radius the size of the image's PSF FWHM of 0.96 arcsec, and is the deepest of the master images.
As discussed in Section~\ref{subsec:cands_select} below, the use of the new master images as reference images resulted in the discovery of SNe in epoch~2 that went undiscovered by P07b.

We performed PSF matching, scaling, and image subtraction between the target and reference images in each Subaru epoch in all bands, using the software HOTPANTS\footnote{http://www.astro.washington.edu/users/becker/hotpants.html}, an implementation of the ISIS algorithm of \citet{1998ApJ...503..325A} for image subtraction (as described by \citealt{Becker2004hotpants}).
Briefly, HOTPANTS divides the images into a predetermined number of regions, and in each region finds the convolution kernel necessary to match the PSF of one image to that of the other. 
HOTPANTS is similar to ISIS, which was used by P07b, but allows more control over the subtraction process.
For example, each region of the image is subdivided into stamps and substamps, where the substamps are centred on astronomical objects.
The kernel is then computed for each substamp, producing a distribution of values used to sigma-clip outliers, thus ensuring a more accurate determination of the kernel in each stamp, and ultimately a better mapping of the spatial variations of the kernel across the image.
We also made use of the software's ability to mask saturated pixels, which vastly reduced the number of residuals in the difference images.

As a consequence of the dithering, the final images have a field of view of 0.31 deg$^2$; however, due to the different effective exposures in the fringes of the field, a substantial region along the edges suffers from a significantly lower signal-to-noise ratio (S/N).
We therefore crop the edges of the difference image, ending with a total subtraction area of 0.25 deg$^2$.

\subsection{Spectroscopy}
\label{subsec:spectro}

As detailed in section 2.2 of P07b, we obtained spectra of 17 of the SN host galaxies from epoch 2, together with several hundred random galaxies in the SDF, using the Low-Resolution Imaging Spectrometer (LRIS; \citealt{1995PASP..107..375O}) on the Keck~I 10-m telescope, and the Deep Imaging Multi-Object Spectrograph (DEIMOS; \citealt{2003SPIE.4841.1657F}) on the Keck~II 10-m telescope.

In addition to the SN host spectra published by P07b, we obtained spectra of 7 additional SN host galaxies. 
These spectra were taken during observations carried out on the night of 2010 February 15 with DEIMOS on the Keck~II telescope. 
The single mask utilised for these observations contained 16 SN host galaxies, as well as the positions of tens of filler galaxies. 
The mask was observed for a total of $3 \times 30$~min.
We used the 600 line mm$^{-1}$ grating, with the GG455 order-blocking filter and a wavelength range of $\sim 4400$--9600~\AA, with the exact limits depending on each individual spectrum. 

The 600 line mm$^{-1}$ grating yields a FWHM intensity resolution of $\sim 3$~\AA, or $\sim 120$ km~s$^{-1}$, at 7500~\AA. 
This resolution is sufficient to resolve many night-sky lines and the [O~II] $\lambda\lambda3726$, 3729 doublet. 
By resolving night-sky lines, one can find emission lines in the reddest part of the spectrum, where sky lines are blended in low-resolution spectra.
Furthermore, by resolving the [O~II] doublet, we can confidently identify an object's redshift, even with only a single line.

The DEIMOS data were reduced using a modified version of the DEEP2 data-reduction pipeline\footnote{http://astro.berkeley.edu/$\sim$cooper/deep/spec2d/}, which bias corrects, flattens, rectifies, and sky subtracts the data before extracting a spectrum \citep{2007ApJ...657L.105F}. 
The wavelength solutions were derived by low-order polynomial fits to the lamp spectral lines, and shifted to match night-sky lines at the positions of the objects. 
Finally, the spectra were flux calibrated by scaling them to the mean fluxes in the \R\ and \I\ bands.
Consequently, the displayed continuum spectral shape is not precisely calibrated.
In any event, the continuum emission of the host galaxies is weak and noisy, and therefore we rely on spectral lines alone for redshift identification.

\begin{figure*}
\begin{minipage}{\textwidth}
\center
\vspace{0.2cm}
\includegraphics[width=0.8\textwidth]{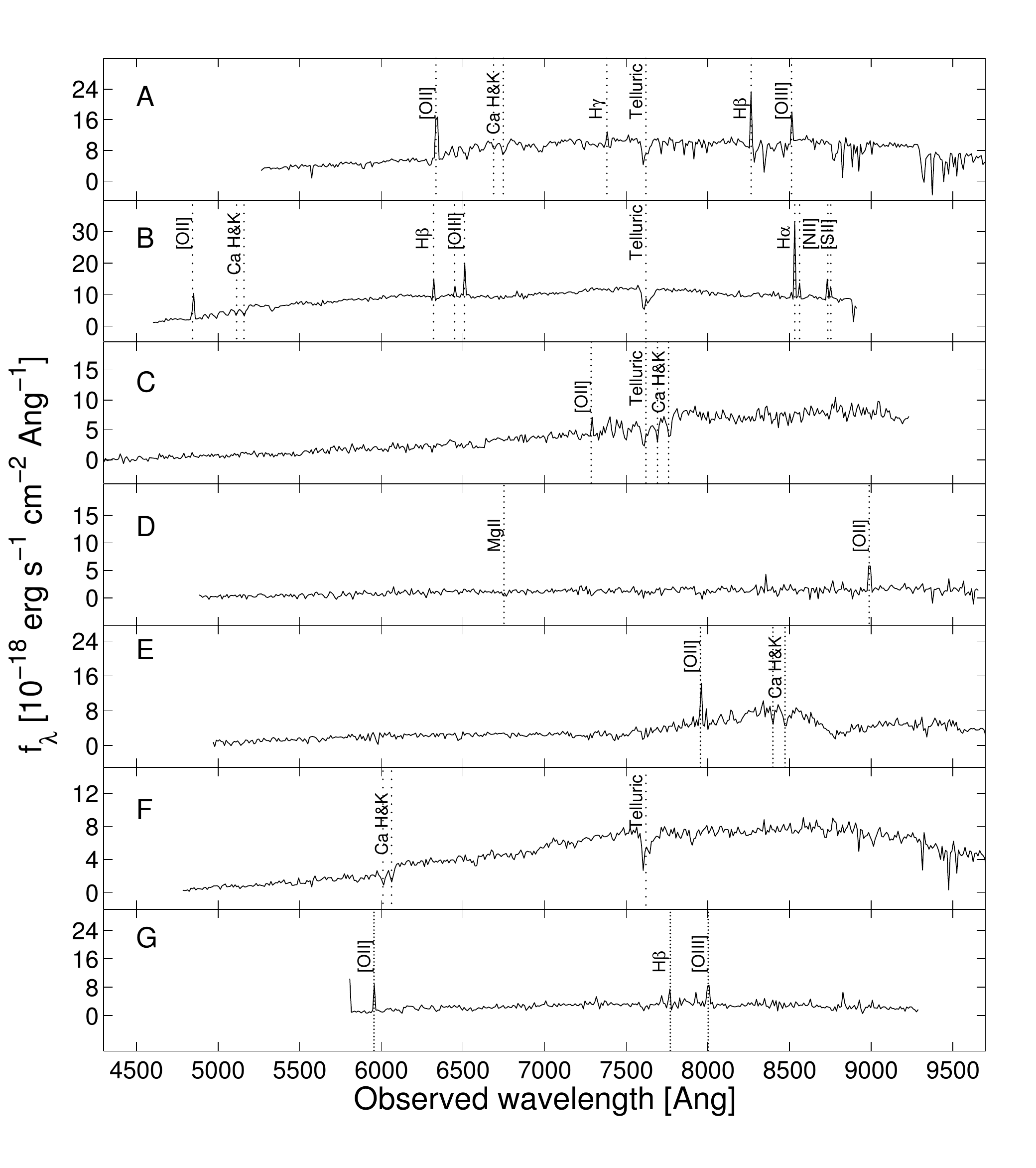}
\caption{SN host-galaxy spectra from the 2010 February 15 Keck DEIMOS observations, with the prominent emission and absorption features marked. The spectra have been rebinned into 10~\AA\ bins. (a) hSDF0702.03, $z=0.70$; (b) hSDF0702.21, $z=0.30$; (c) hSDF0702.23, $z=0.96$; (d) hSDF0705.18, $z=1.41$; (e) hSDF0806.48, $z=1.13$; (f) hSDF0806.54, $z=0.53$; and (g) hSDF0806.55, $z=0.60$.}
\label{fig:spectra}
\end{minipage}
\end{figure*}


\section{SUPERNOVA CANDIDATES}
\label{sec:cands}

In this section we describe the methods by which we have discovered the SN candidates in our sample, derive the detection efficiency of the survey, and measure the photometric and astrometric properties of the candidates and their uncertainties.
We have discovered a total of 163 transient objects, of which 150 are most likely SNe.
The luminosities of the transients, inferred from their measured photometry and the redshifts of their associated host galaxies (as derived in Section~\ref{subsec:zebra}, below), lead us to conclude that these 150 events are SNe.
In Section~\ref{subsec:cands_select} we describe the criteria according to which the transients were chosen, culling random noise peaks, image subtraction artefacts, and previously known active galactic nuclei (AGNs).
We calculate the probability of contamination by flaring Galactic M dwarfs and unknown AGNs in Section~\ref{subsec:gal_phot}.
The probable contamination by AGNs is compared with the number of actual possible AGNs among the candidates in Section~\ref{subsec:pec-sne}.
In Section~\ref{subsec:gal_phot} we also calculate the probability of a chance association between a transient object and its surrounding galaxies.

Since our survey classifies SNe based on single-epoch observations without spectroscopic follow-up observations, the SNe we discover do not satisfy the International Astronomical Union's criteria for a \textquoteleft standard\textquoteright\ designation.
As in P07b, we will continue to use the following naming conventions. 
We denote the SNe from epochs 2 through 5 respectively as \textquoteleft SNSDF0503.XX,\textquoteright\ \textquoteleft SNSDF0702.XX,\textquoteright\ \textquoteleft SNSDF0705.XX,\textquoteright\ and \textquoteleft SNSDF0806.XX,\textquoteright\ with the first two digits denoting the year, the next two digits the month, and XX being a serial number ordered according to the SN \z-band apparent magnitude.
The respective host galaxies are referred to as \textquoteleft hSDF0503.XX,\textquoteright\ \textquoteleft hSDF0702.XX,\textquoteright\ \textquoteleft hSDF0705.XX,\textquoteright\ and \textquoteleft hSDF0806.XX.\textquoteright\

\subsection{Candidate selection}
\label{subsec:cands_select}

The \z-band difference image obtained with HOTPANTS was scanned with \sex\ to search for variable objects.
\sex\ was set to identify and extract all objects which had at least 6 connected pixels with flux $3\sigma$ above the local background level.
T. Morokuma (private communication) provided us with a catalogue of 481 AGNs, which were identified in epoch 1 by their long-term \I-band variability. 
In our survey, these galaxies were therefore ignored, as further discussed in Section~\ref{subsec:gal_phot}.
These galaxies still constitute fewer than 1 per cent of all galaxies in the SDF, and therefore this exclusion has negligible effect on our SN survey.

In order to reject other non-SN events, the remaining variable candidates were examined as follows.
\begin{enumerate}
\item \noindent Of the objects identified by \sex, we rejected all those which showed suspect residual shapes, indicative of a subtraction artefact.
For maximum completeness, the threshold for \sex\ detection was set low, and thousands of candidates were inspected by eye by one of us (OG).
\item \noindent  We compared two \z-band difference images of the same field.
The main difference image was obtained by allowing HOTPANTS to calculate the convolution kernel for the subtraction over the entire image.
A second difference image was obtained by forcing HOTPANTS to break the image into four subregions, and calculate the convolution kernel in each one.
This second difference image was generally less clean than the first, but allowed for the rejection of subtraction artefacts in the main difference image, as not all of those would be reproduced in the second subtraction process.
\item \noindent  We compared the main \z-band difference image in a certain epoch with difference images of the other epochs in order to identify and reject AGNs that were not already rejected based
on the Morokuma AGN catalogue, or other objects that exhibited variability over a large stretch of time.
Roughly 40 transients were identified as AGN candidates due to their variability over several epochs. 
These objects were not included in the Morokuma AGN catalogue, and may have been quiescent at the time it was compiled. 
\item \noindent  In order to further reject subtraction artefacts, we stacked the exposures in each epoch into two subepoch images, where each subepoch was composed of half of the observation nights.
These images were then used to obtain new difference images which we compared with the main \z-band images.
As in the previous steps, objects which appeared in the main difference image, but not in the subepoch difference images, were rejected.
We note that solar-system objects were already eliminated in the nightly averaging, since even as far as 30~AU \citep{1997ApJ...490..879S} a Kuiper Belt object would move due to the Earth's motion by $\sim 40$ arcsec, or 200 pixels, in the course of an 8-hour night. 
\item \noindent For every candidate found in the \z\ band, difference images in the \R\ and \I\ bands were also examined, and objects which showed suspect residual shapes were rejected.
We note that no candidate was rejected because of a nondetection in the \R\ or \I\ bands, since at least some high-$z$ SNe are expected to be very faint or undetected in the observed-frame \R\ and \I\ bands.
\item \noindent  Finally, for each SN candidate, we derived the local S/N by dividing the SN counts in an aperture of 1 arcsec radius (before application of an aperture correction) by the standard deviation of the total counts in tens of identical apertures centred on surrounding blank regions.
SN candidates which had a S/N smaller than 3 were rejected as probable noise peaks. \\
We note that steps (ii) and (iii) are selection criteria additional to those followed by P07b.
\end{enumerate} 

In order to apply our new criteria uniformly to the full SN survey, we resurveyed epoch 2.
Of the 33 SNe found by P07b, 28 were recovered.
The SN candidates listed in P07b as SNSDF0503.27, SNSDF0503.33, and SNSDF0503.40 were not detected by \sex, because the S/N was too low.
While the first two SN candidates listed above appear in the difference images, we do not detect the third one in our renewed analysis.
SNSDF0503.29 was detected by \sex, but whereas in the main difference image it appears as a point source, in the secondary difference image it is extended, and the position of its centre is offset by $\sim 0.35$ arcsec.
SNSDF0503.32 was not detected by \sex, and while it appears in the main difference image, it is absent from the secondary difference image.
Thus, with our improved reference images and image subtraction procedures, these events from P07b do not pass our current selection criteria.

On the other hand, we have discovered 8 new SN candidates in epoch 2, not reported by P07b.
In this work, these SN candidates are listed as SNSDF0503.06, SNSDF0503.16, SNSDF0503.19, SNSDF0503.27, SNSDF0503.31, SNSDF0503.32, SNSDF0503.33, and SNSDF0503.34.
The differences between the P07b sample and the present sample are due to two reasons: (a) the use of HOTPANTS in the current work, which provides cleaner subtractions than ISIS, and (b) the use of deeper \z-band master images with better image quality, instead of the shallower epoch-1 \z-band image, as references.
In any event, the list of epoch-2 SNe that we report in Table~\ref{table:SNe} supersedes the one presented by P07b.

\subsection{Detection efficiency simulation}
\label{subsec:fakes_eff}

\begin{figure*}
\begin{minipage}{\textwidth}
\centering
\vspace{0.2cm}
\includegraphics[width=0.8\textwidth]{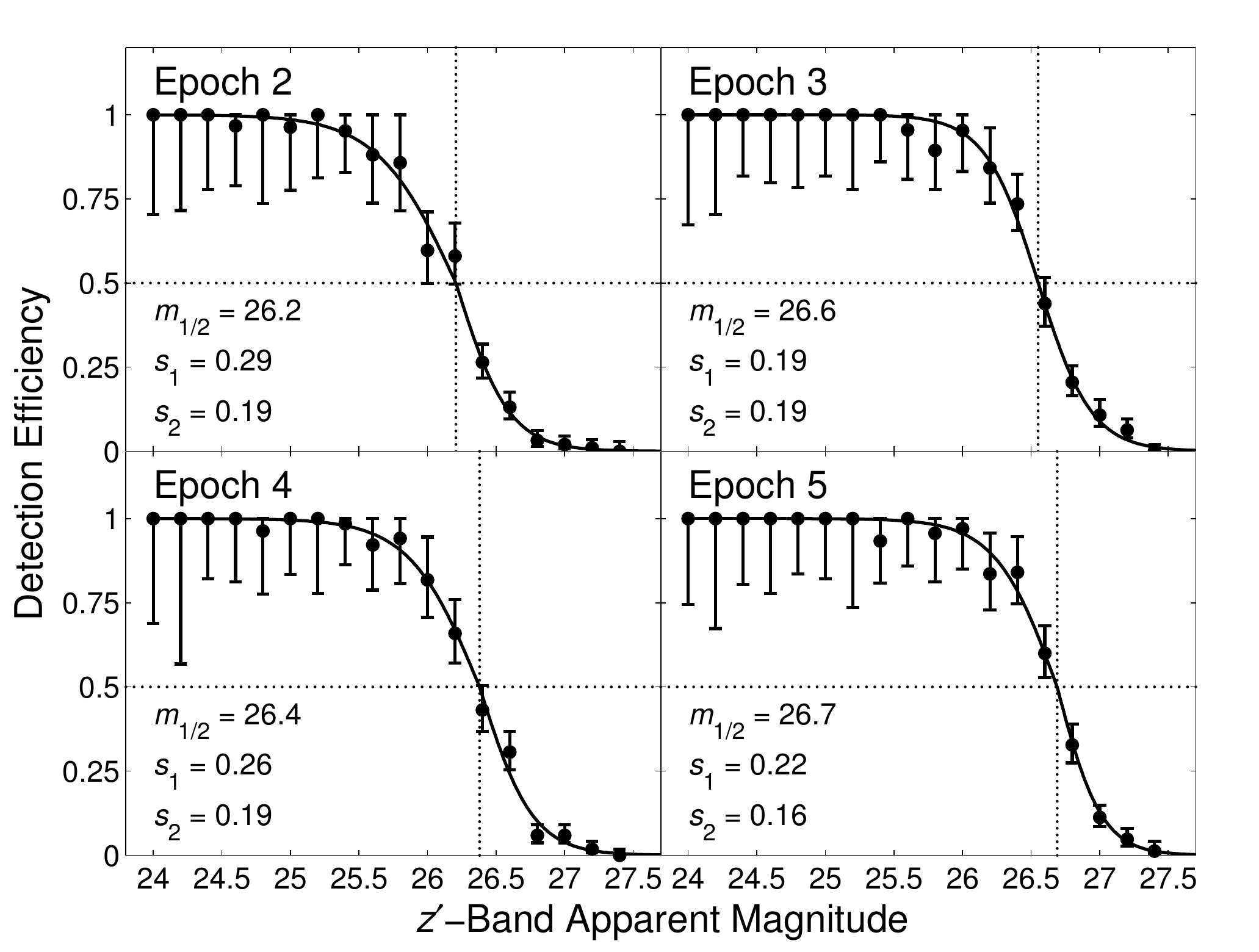}
\caption{Fraction of simulated SNe recovered as a function of \z-band magnitude. Error bars indicate 1$\sigma$ binomial uncertainties. The dotted lines mark the 50 per cent efficiency mark.}
\label{fig:fakes_eff}
\end{minipage}
\end{figure*}

In our survey, SNe may be missed as a result of many effects, including imperfect subtractions, noise fluctuations, and human error.
In order to quantify these systematic effects, we measure our detection efficiency by blindly planting artificial point sources, which match the SN population in our survey as closely as possible, in the presubtraction \z-band images, and then discovering them along with the real SNe. 
The simulated SN sample was constructed as detailed in section 3.2 of P07b.
Our resulting efficiency as a function of magnitude, in each epoch, can be seen in Fig.~\ref{fig:fakes_eff}.
We follow \citet{2007ApJ...660.1165S} and fit the following function to the data:
\begin{equation}\label{eq.efficiency}
\eta(m;m_{1/2},s_1,s_2) = \left\{ \begin{array}{ll}
      \left(1 + e^{\frac{m-m_{1/2}}{s_1}}\right)^{-1}, & \mbox{$m\le m_{1/2}$}\\
      \left(1 + e^{\frac{m-m_{1/2}}{s_2}}\right)^{-1}, & \mbox{$m>m_{1/2}$},\\
      \end{array} \right.
\end{equation}
where $m$ is the \z-band magnitude of the fake SNe, $m_{1/2}$ is the magnitude at which the efficiency drops to 0.5, and $s_1$ and $s_2$ determine the range over which the efficiency drops from 1 to 0.5, and from 0.5 to 0, respectively.

\subsection{Supernova sample}
\label{subsec:sn_sample}

We have found a total of 150 SNe, with magnitudes in the range $\z=22.9$ to $\z=26.7$. Table~\ref{table:SNe} lists the SNe and their properties.
Apart from these 150 SNe, we detect several tens of candidates at fainter magnitudes, as we would expect based on our efficiency simulations, but these are all objects with $\rm{S/N}<3$.
While some of these objects may be SNe, an unknown number of them could be false positives, such as subtraction artefacts or random noise peaks.
We therefore limit our sample to $\z<26.6$, $\z<26.4$, and $\z<26.7$ mag for epochs 3 through 5, respectively.
These are the values of $m_{1/2}$ in each epoch.
In epoch 2 we reach the 50 per cent efficiency mark at 26.2 mag. 
However, in the interest of backward compatibility with P07b, we lowered the efficiency cutoff for epoch 2 to 26.3 mag.

Using \sex, we have performed aperture photometry of the SNe in the \R, \I\, and \z\ difference images within fixed 1-arcsec-radius circular apertures.
To estimate the aperture correction and photometric uncertainty, we measured the magnitudes of $\sim 600$ simulated point sources, ranging in brightness from 23 to 28 mag, planted in a 4k $\times$ 4k pixel subframe of the SDF \R-, \I-, and \z-band images. 
We took the difference between the average of the magnitude in each bin and the true magnitude as the required aperture correction, and the standard deviation in each magnitude bin to be the minimum photometric statistical error for objects of that magnitude.
For example, the mean aperture correction for the epoch-2 \z-band image was 0.2 mag (i.e., due to aperture losses, the measured photometry was 0.2 mag too faint) and the standard deviation ranged from 0.03 to 0.29 mag from the brightest to the faintest artificial sources, respectively.
The adopted uncertainty for each SN was taken to be the larger among the uncertainty computed by \sex\ and the statistical uncertainty for the given magnitude bin from the simulations.

We also measured the offset of each SN from its host galaxy.
To estimate the uncertainty of the offset, $\sim$\,12,000 simulated point sources, divided into magnitude bins of width 0.3 mag, were planted in the \z-band image of each epoch. 
We then measured their locations, in both the original image and the \z-band difference image, using \sex, and took the mean of the location residuals in each bin as an estimate of the uncertainty of the object's location.
This uncertainty was added, in quadrature, to the uncertainty in the location of the SN host galaxy.
The real SN offsets ranged between $0$ and $3.61$ arcsec, and the uncertainties ranged between $0.02$ and $0.16$ arcsec, with the centres of brighter sources being, of course, better localized.


\section{SUPERNOVA HOST GALAXIES}
\label{sec:gal}

In this section, we determine the host galaxy of each SN and then measure its properties.
The SN host galaxies, including their photometry in all available bands, are presented in Table~\ref{table:SNe_hosts}.

\subsection{Identification and photometry}
\label{subsec:gal_phot}

The SN host galaxies were chosen to be the closest galaxies, in units of those galaxies' half-light radii, as measured with \sex\ in the \I\ band. 
A small number of SNe had several possible hosts.
To choose between them we measured the photometric redshift (photo-$z$) of each host.
If the different hosts were found to be at the same redshift, that redshift was adopted for the SN as well.
If, on the other hand, the different hosts were found to lie at different redshifts, we computed the likelihood of a SN of the type, as classified by SNABC, at those different redshifts being observed at the magnitude measured.
In this manner we were able to eliminate unlikely hosts.

Using \sex, we measured the \citet{1976ApJ...209L...1P} magnitude of the host galaxies in the seven optical bands of epoch 1.
We chose Petrosian photometry, since it measures the flux of resolved objects within a given fraction of the object's light profile, thus enabling one to compare between measurements taken in different filters.
The resulting catalogue was cross-matched with the \J\ and \K\ catalogues. 
Additionally, for each galaxy we checked the corresponding location in the {\it GALEX} {\it FUV} and {\it NUV} background-subtracted images. 
Since the {\it GALEX} PSF is much larger than that of Subaru and UKIRT, most of our galaxies appear as point sources, making it impossible to measure Petrosian magnitudes; hence, any measurement within any aperture would not capture the same percentage of light as in the optical and NIR bands.
Furthermore, owing to the density of sources in the SDF and the size of the {\it GALEX} PSF, in many cases it proved impossible to determine which source in the optical image was associated with the UV signal.
In those cases where we could associate nondetections in the UV bands unambiguously with our host galaxies, we added the limiting magnitudes in the relevant UV bands to the catalogue.
In Section~\ref{subsec:zebra} we detail how we combined these limiting magnitudes with the optical and NIR data to compute the redshifts of the SN host galaxies.
As with the SN photometry, for the host photometry we estimated the uncertainty in each magnitude bin using artificial sources with galactic profiles (created with the IRAF routine {\it gallist}) that we planted in the images.

To test whether any of our chosen host galaxies are merely chance associations, we counted the fractions of the total imaged SDF area that are within 0.1-light-radius-wide annuli of all the galaxies detected in the field. 
From this we conclude that, among the 110 SNe within $\leq 0.5$ light radii of their chosen hosts, $<1$ SN is expected to be a chance association.
These 110 SNe include all 12 SNe in the $1.5<z<2.0$ range, and 24 of the 26 SNe in the $1.0<z<1.5$ range.
At larger host-SN separations, 23, 6, and 1 of our SNe are found within 0.5--1.0, 1.0--1.5, and 1.5--2.0 light radii of their host galaxies, respectively. 
Among these events, we expect 6, 3, and $<1$ (respectively) to be chance associations. 
However, 28 of these 30 large-separation events are at $z<1$. 
Thus, while some fraction of our $z<1$ rate may be due to contamination by chance associations, we
estimate that our $1.0<z<1.5$ rate is affected by such contamination at only the few-percent level, and the $1.5<z<2.0$ negligibly so.

In P07b we found that, assuming a \cite{1968adga.book.....S} model for the galaxy radial profile between $n=4$, the \citealt{1948AnAp...11..247D} law  \citep{2002AJ....124..266P} and $n=1$, an exponential disk \citep{1970ApJ...160..811F,2002AJ....124..266P}, between 91 and 99.99 per cent of the light (respectively) falls within 6 half-light radii of the galaxy's centre.
Ten of our SNe have no visible host galaxies within this distance, and so we label them \textquoteleft hostless\textquoteright\ (namely SNSDF0503.14, SNSDF0503.18, SNSDF0702.06, SNSDF0705.20, SNSDF0705.21, SNSDF0705.24, SNSDF0806.04, SNSDF0806.30, SNSDF0806.49, and SNSDF0806.53).
The probable host galaxy of SNSDF0806.51 appears exclusively in the \B\ and \R\ bands of epoch 1.
Given that our photometric redshift estimate requires at least three photometric bands for its calculation, and that even the \B\ and \R\ detections are barely above the limiting magnitudes in those bands, we treat this SN as hostless as well.
The most probable explanation is that these SNe occurred in galaxies fainter than the limiting magnitudes in all the photometric bands of epoch 1.

Other possibilities to consider are that these candidates are high-$z$ AGNs or flaring Galactic M dwarfs.
The fact that these hostless SN candidates are detected in only a single epoch over a period of 3 years argues against the AGN option, as follows.
Among the 481 objects identified in the Morokuma AGN catalogue, fewer than 1 per cent display detectable variability in only one of our four search-epoch difference images.
50 of the SN candidates in our sample lie within 0.2 arcsec (or 1 pixel) of their respective host-galaxy nuclei, and so could potentially be AGNs.
Together with the above 11 hostless SNe, the predicted number of contaminating AGNs in our sample is $61 \times 0.01 \approx 0.6$.
The Poisson probability of having at least one AGN in the sample is then $\sim 45$ per cent, which is consistent with our having found one such object.
The probability of finding two or more such objects drops to $\sim 12$ per cent (see SNSDF0705.17 in Section~\ref{subsub:0705.17} and SNSDF0705.30 in Section~\ref{subsub:0705.30}).

As to the second possibility, M-dwarf optical flares consist of a fast rise followed by a decay lasting typically of order an hour or less, with the distribution of flare durations steeply falling at longer durations \citep{2011AJ....141...50W}.
The longest known flares last $\sim 10$~hrs \citep{2010ApJ...714L..98K}, and these constitute $<1$ per cent of all flares (E. Hilton, S. Hawley, private communication). 
With such variation timescales, M-star flare events would be filtered out in our nightly image averaging, or would at least show a decline between consecutive half-night averages.
None of the hostless candidates show such a decline. 
We note, further, that flaring activity is limited to the younger M dwarfs in the Milky Way disk that are within a height of $Z<300$~pc above the disk. 
Activity in older dwarfs, which have had time to be scattered to larger heights, is exceedingly rare \citep{2008AJ....135..785W, 2009AJ....138..633K, 2011AJ....141...50W}. 
Any M dwarfs below the SDF detection limits in quiescence, and that had flared into visibility during our observations, would necessarily be at distances $\gtsim 50$~kpc, i.e., they would belong to the Galactic halo, and hence would be even older and less active than the $Z>300$~pc disk stars.
We therefore deem it highly unlikely that any of our hostless SN candidates are optical flares of Galactic M dwarfs.

\subsection{Host redshifts}
\label{subsec:zebra}

From our spectroscopy, detailed in Section~\ref{subsec:spectro}, we derived spectroscopic redshifts (spec-$z$) for 24 of the SN host galaxies.
Of these 24 SN host galaxies, hSDF0705.18 has the highest spec-$z$, at $z=1.412$.
The seven new spectra obtained on 2010 February 15 appear in Fig.~\ref{fig:spectra}.
For the majority of our SN host galaxies, which are too faint for spectroscopy, we derive photometric redshifts, as in P07b, using the Zurich Extragalactic Bayesian Redshift Analyzer (ZEBRA; \citealt{2006MNRAS.372..565F}). 
We calibrated ZEBRA in the manner described by P07b, but with a larger training set of 431 galaxies, of which 150 are in the range $1<z<2$. 
This training set consisted of 123 galaxies imaged in the Keck runs detailed by P07b, along with data from other surveys that had been conducted in the SDF (e.g., \citealt{2003AJ....125...53K, 2006ApJ...637..631K, 2006PASJ...58..313S}; and a new sample obtained by N. Kashikawa in 2008 with DEIMOS on the Keck~II telescope).
ZEBRA was allowed to run over the redshift range $0<z<3$.

Since ZEBRA does not, at the moment, offer an adequate treatment of upper limits, but rather deals with them as with any other photometry measurement, we decided (at the suggestion of R. Feldmann, private communication) to halve the $1\sigma$ {\it FUV} and {\it NUV} flux limits, and treat them as measurements with relative uncertainties of 100 per cent, thus requiring ZEBRA's fit to pass through the region $[0,f_{1\sigma}]$.
If no {\it GALEX} signal existed that could be clearly associated with the optical galaxy, we used the UV flux limit (as described above) as an extra band in the ZEBRA fit, thus constraining the SEDs to those with fluxes lower than the UV flux limit.
These upper limits on the UV flux were particularly useful for constraining the redshifts of galaxies having \textquoteleft Lyman breaks\textquoteright\ due to absorption by neutral hydrogen in the
intergalactic medium (IGM).
If, on the other hand, there was a {\it GALEX} detection, but due to the large {\it GALEX} PSF we could not clearly associate the UV signal with the optical SN host galaxy, we did not use the {\it GALEX} data at all.
For larger samples, where more galaxies have clear signals in the UV, one could treat the UV signal as a lower limit, in similar fashion to our use of nondetections as upper limits.

\begin{figure}
 \begin{center}
 \includegraphics[width=0.5\textwidth]{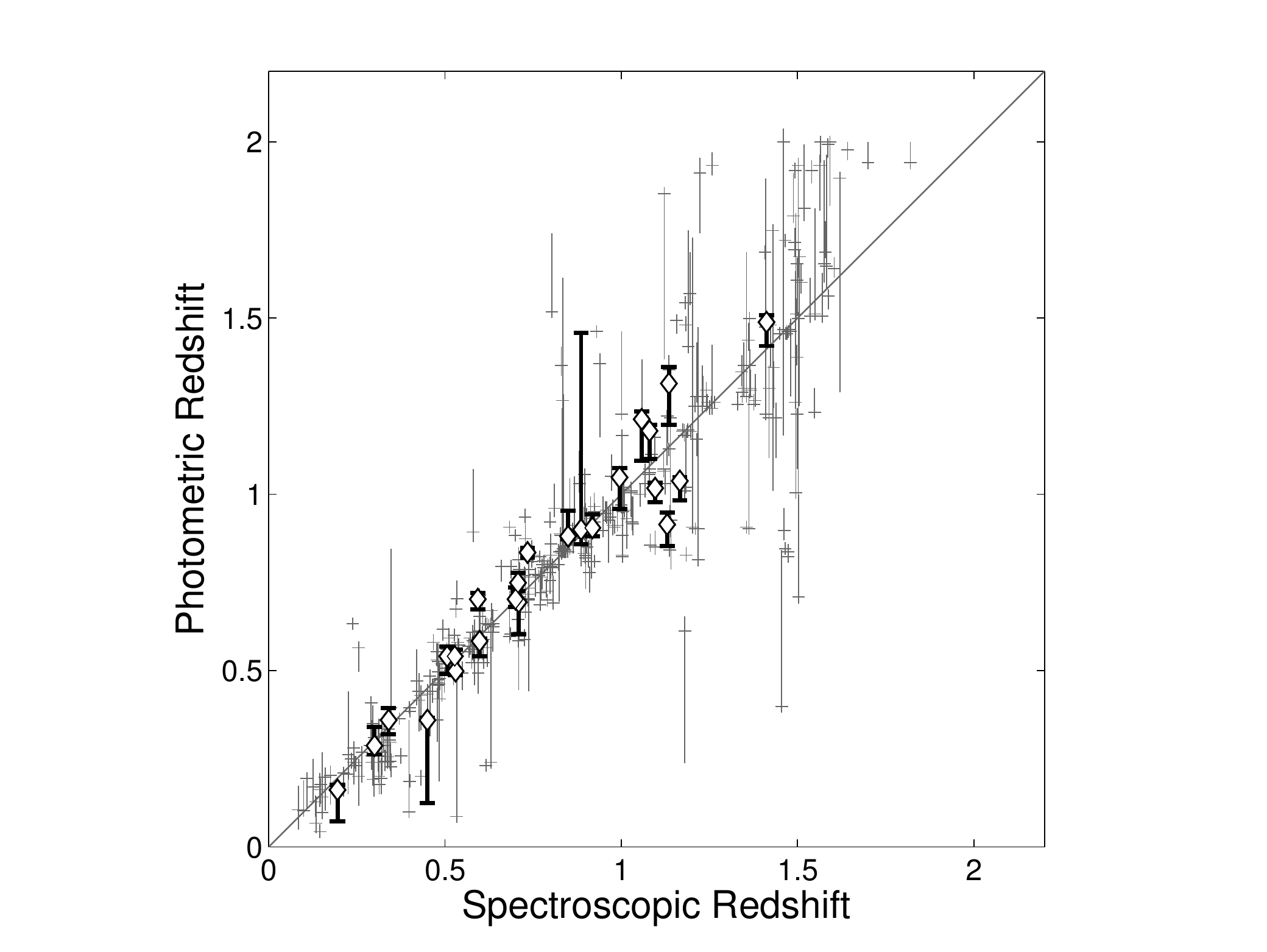}
 \caption{Comparison of the photometric redshifts derived with ZEBRA and the corresponding spectroscopic redshifts for the 431 galaxies in our training set (grey crosses) and for 24 SN host galaxies (empty diamonds). Error bars are the $1\sigma$ confidence limits from the $z$-PDF of each galaxy. The rms scatter of the data is $\sigma_{\Delta{z}/(1+z_s)} = 0.065$ for the training set and $\sigma_{\Delta{z}/(1+z_s)} = 0.028$ for the SN host galaxies.}
 \label{fig:zebra}
\end{center}
\end{figure}

Fig.~\ref{fig:zebra} displays the ZEBRA photo-$z$ values for our training-set galaxies, compared to their spec-$z$ values.
The training set of 431 galaxies has a root-mean square (rms) scatter of $\sigma_{\Delta{z}}/(1+z_s) = 0.075$ (where $\Delta{z} = z_s-z_p$) in the range $0<z_p<2$, after rejecting six $4\sigma$ outliers.
This is consistent with the accuracy achieved by P07b, $\sigma_{\Delta{z}}/(1+z_s) = 0.08$, for 296 galaxies in the range $0<z_p<1.8$ and after rejection of five $4\sigma$ outliers.
The rms scatter for our 24 SN host galaxies is smaller: $\sigma_{\Delta{z}}/(1+z_s) = 0.044$.
There were no $4\sigma$ outliers among these host galaxies.

Of the various end products computed by ZEBRA, we use the redshift probability distribution function ($z$-PDF) of each SN host galaxy that results from marginalizing the full posterior distribution over all templates.
In this manner the uncertainties in the determination of the photo-$z$ are propagated into the classification stage.
While most of the $z$-PDFs display a single, narrow peak, some are more structured, a result of degeneracies between the different combinations of redshifts and normalization constants (i.e., a certain galaxy may fit the same template if it is bright and distant, or if it is faint and nearby) or of a dearth of information. 
For example, the optical continuum shape of late-type galaxies can be approximated with a power law, and so its shape is weakly affected by redshift (see, e.g., Fig.~\ref{fig:3_30_zebra}).
In such cases the UV data can be useful; a clear signal (whether a detection or a nondetection) in the {\it NUV} band would decide among the redshift values.
In order to take the uncertainty introduced by the shape of the $z$-PDFs into account, we use the full $z$-PDFs in the classification stage (see Section~\ref{sec:classify}).

For 23 of the 24 SN host galaxies with spectral redshifts, the spec-$z$ and photo-$z$ values are almost identical, with $\Delta{z}/(1+z) < 0.08$, while for hSDF0503.24 the difference is only $\Delta{z}/(1+z) = 0.10$.
For these galaxies we do not take the $z$-PDF computed by ZEBRA as input for the SNABC, but rather use a Gaussian $z$-PDF centred on that galaxy's spec-$z$, with a width $w_z=0.01$.
For the 11 hostless SNe, we use a $z$-PDF which is the sum of the $z$-PDFs of all the other host galaxies. 
A different composite $z$-PDF, the average of the $z$-PDFs of all the galaxies in the SDF, was also tested for these SNe, and produced the same results.
Given the resulting redshifts, the host galaxies of the hostless SNe would have to be fainter than between $-15.8$ and $-17.0$ (absolute observed \I-band magnitude) to be undetected in the \I-band master images.
This is consistent with these SNe having occured in low luminosity dwarf galaxies (see, e.g., \citealt{2010ApJ...721..777A}). 

\begin{figure*}
\begin{minipage}{\textwidth}
\vspace{0.2cm}
\begin{tabular}{cc}
\includegraphics[width=0.5\textwidth]{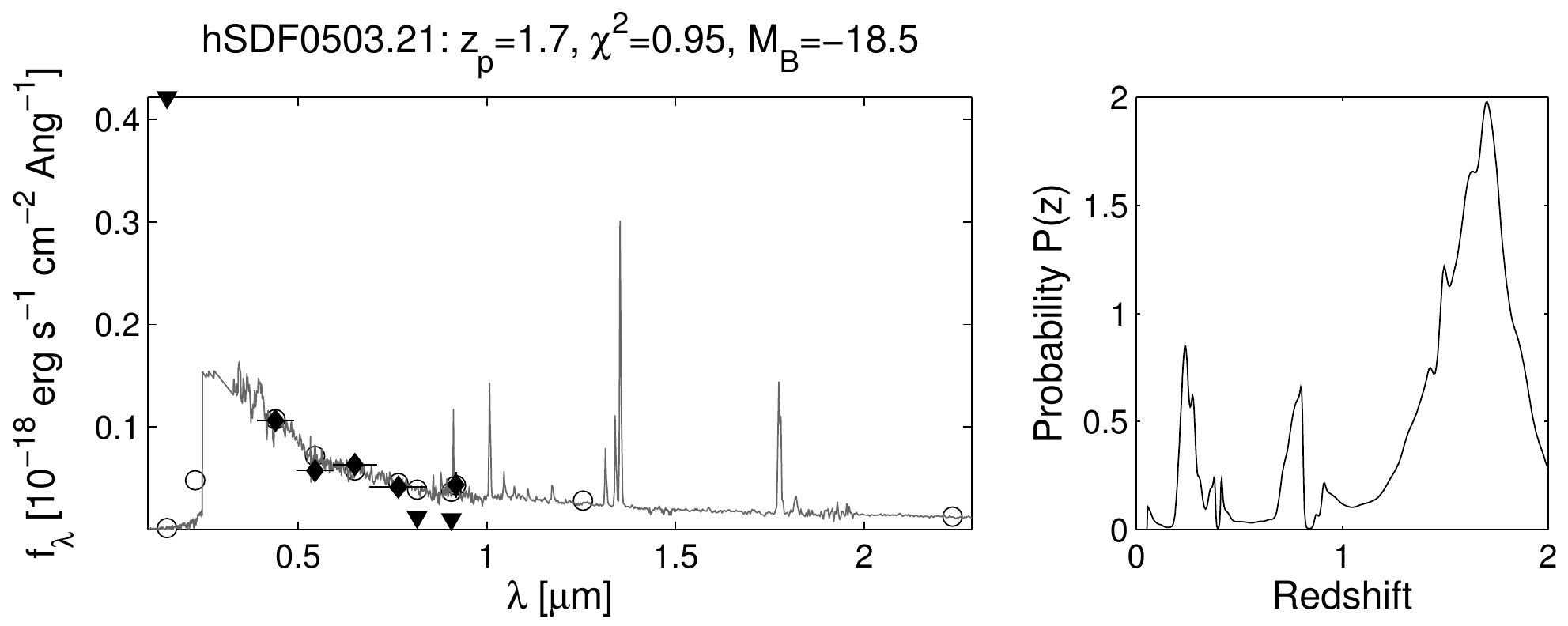} & 
\includegraphics[width=0.5\textwidth]{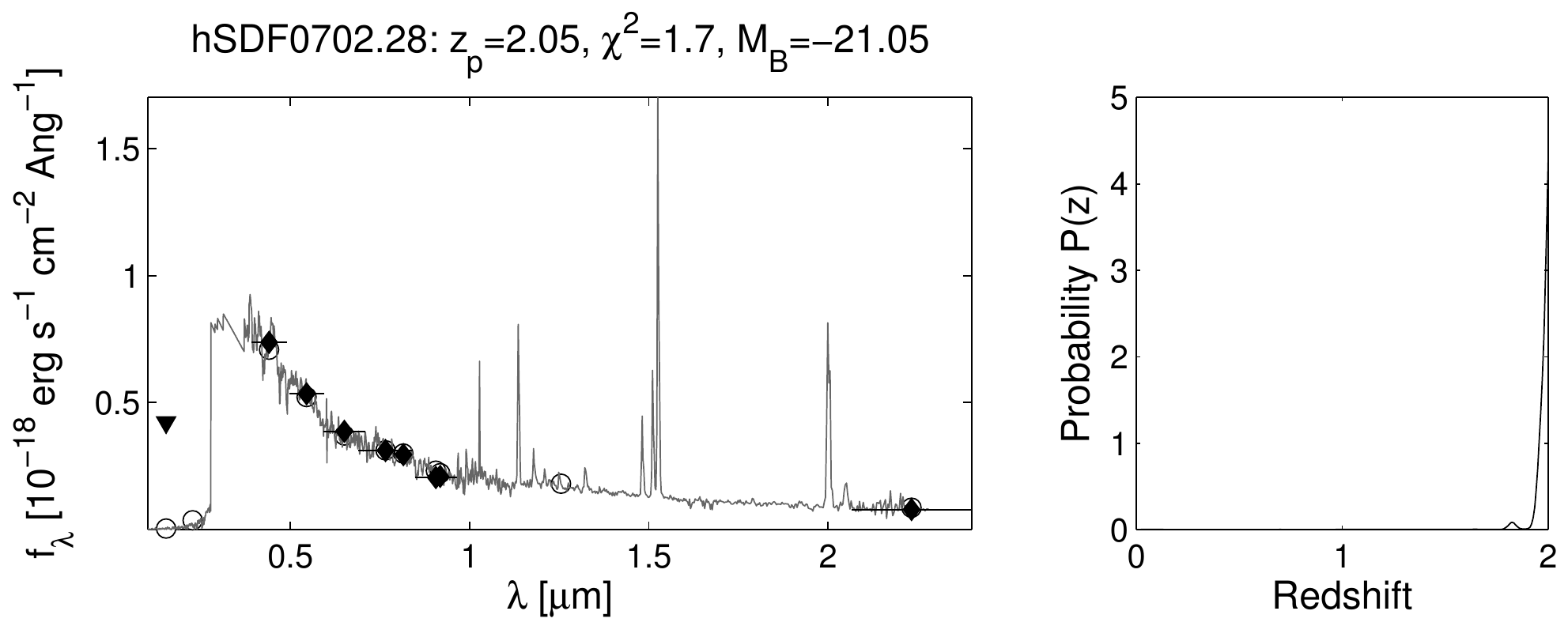} \\

\includegraphics[width=0.5\textwidth]{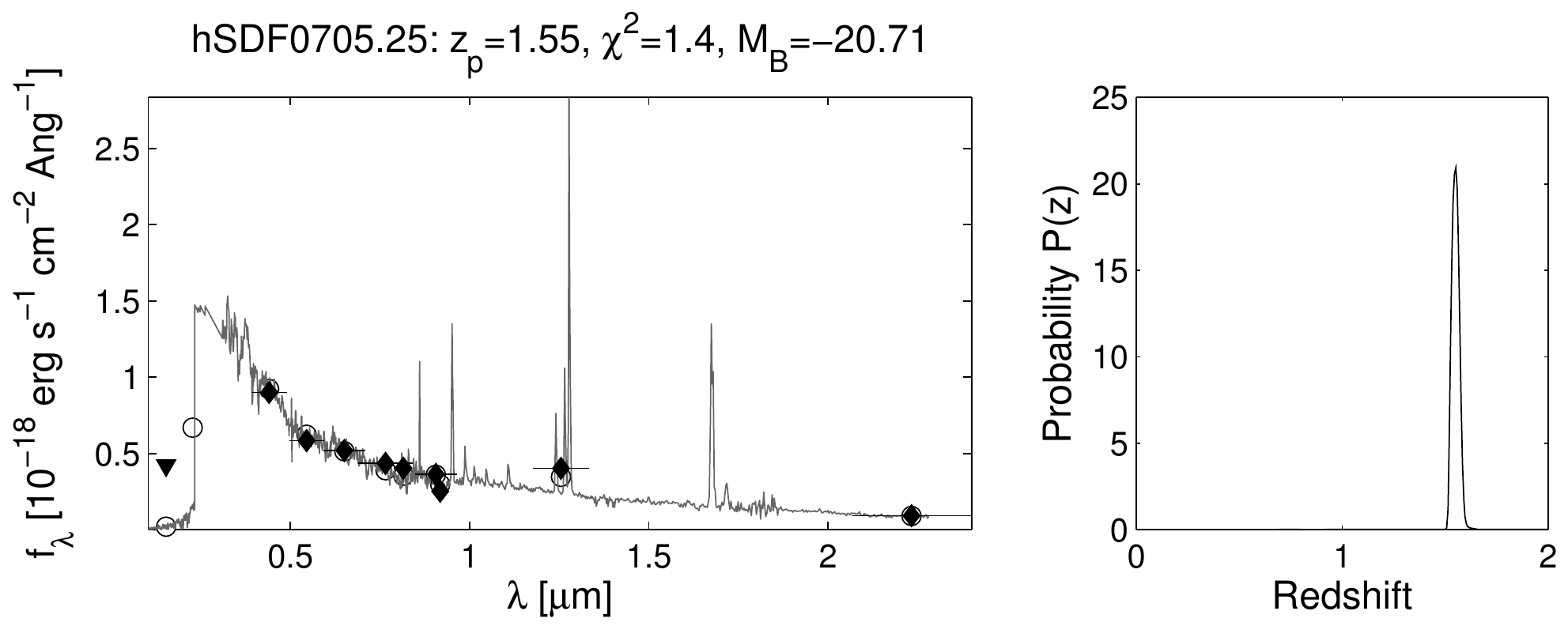} &
\includegraphics[width=0.5\textwidth]{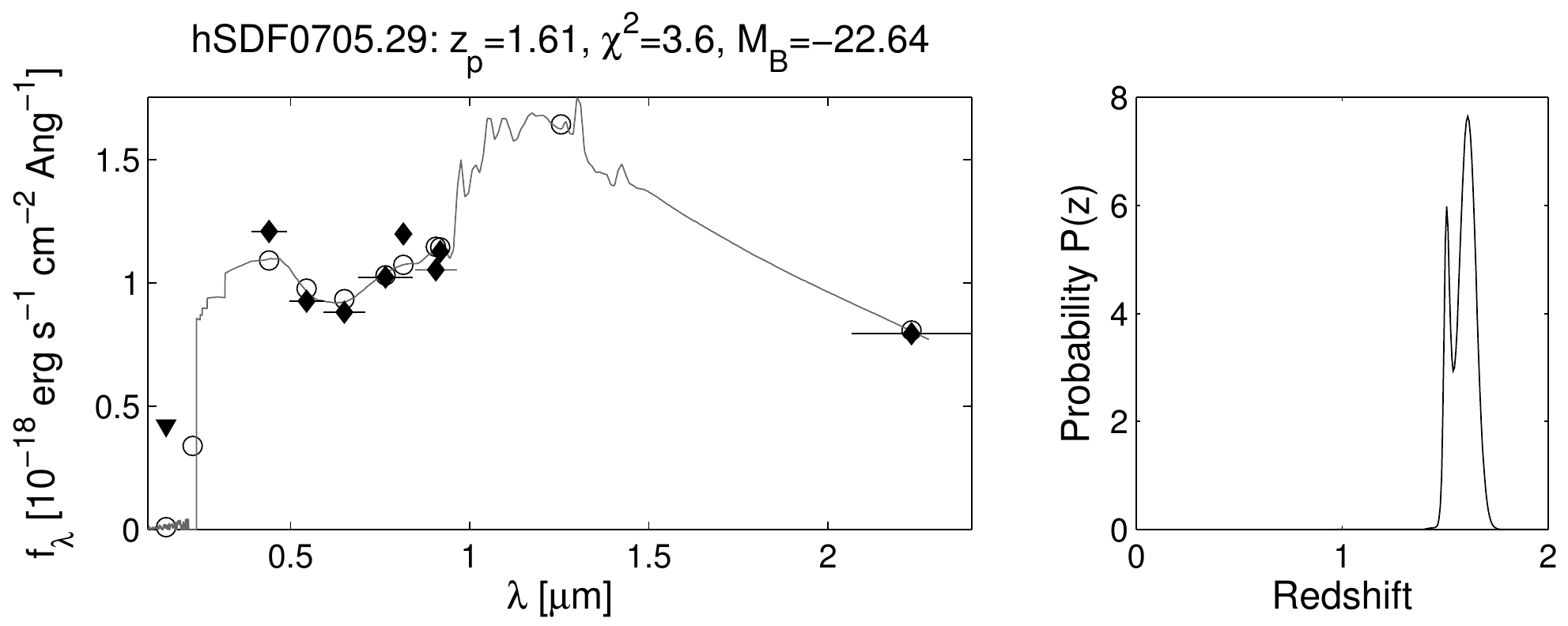} \\

\includegraphics[width=0.5\textwidth]{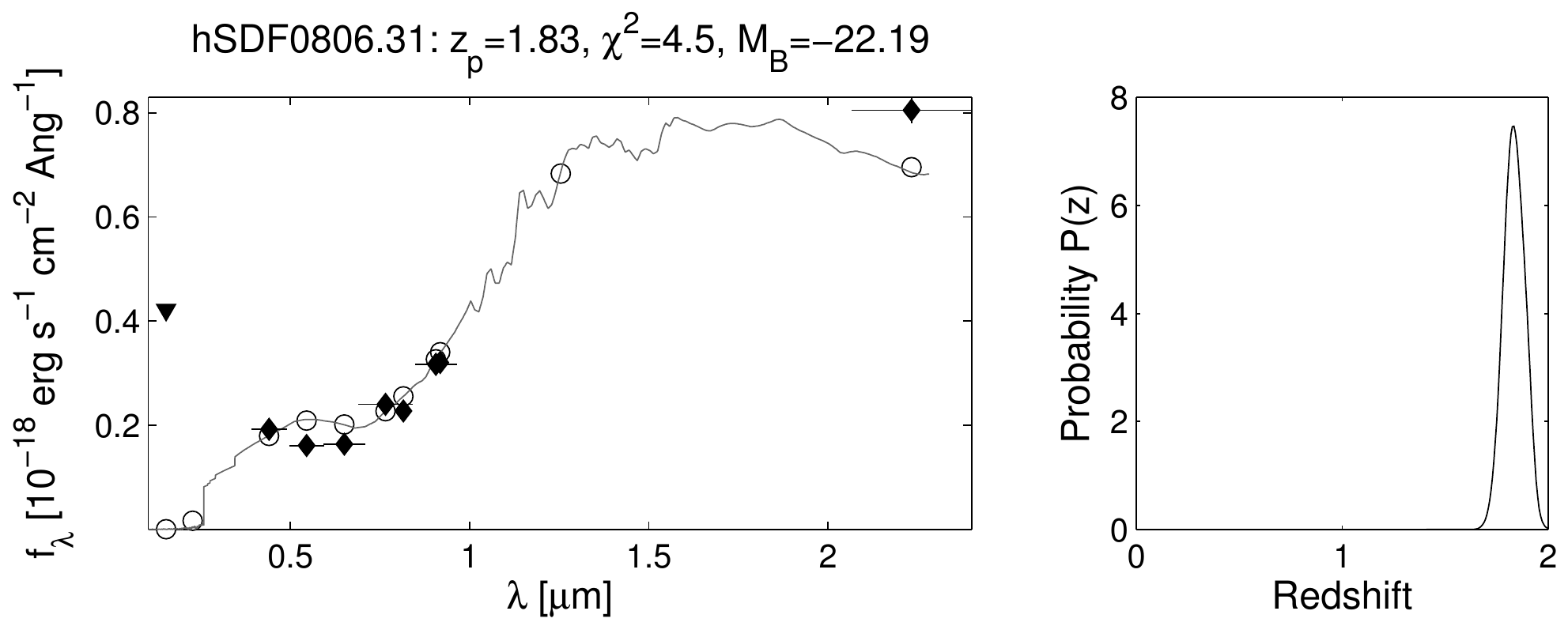} & 
\includegraphics[width=0.5\textwidth]{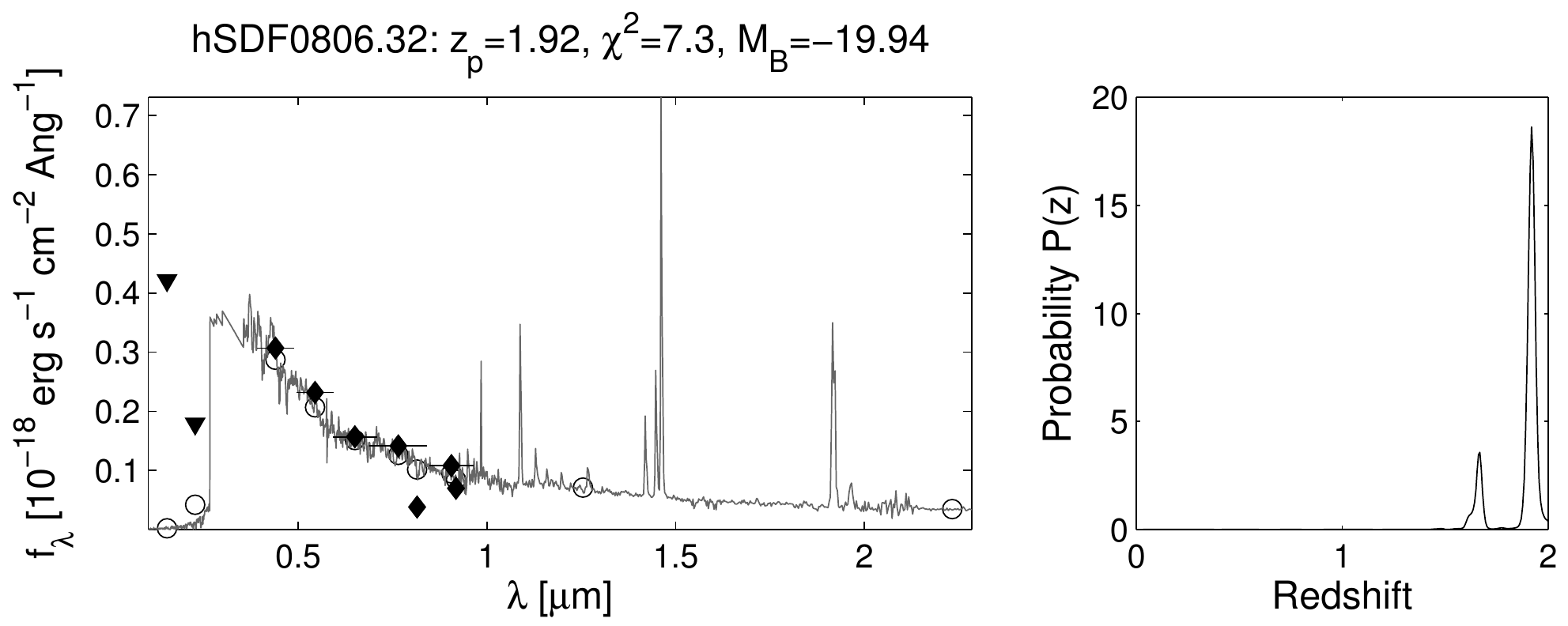} \\

\includegraphics[width=0.5\textwidth]{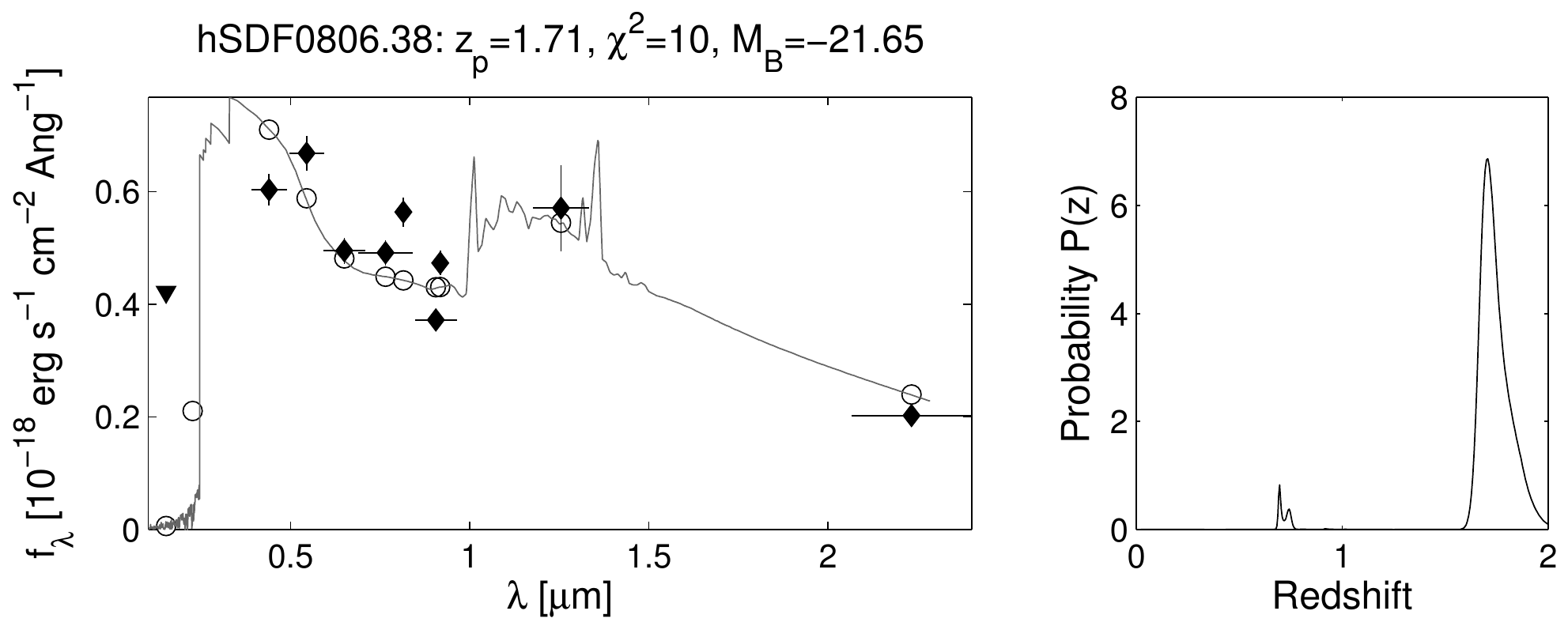} &
\includegraphics[width=0.5\textwidth]{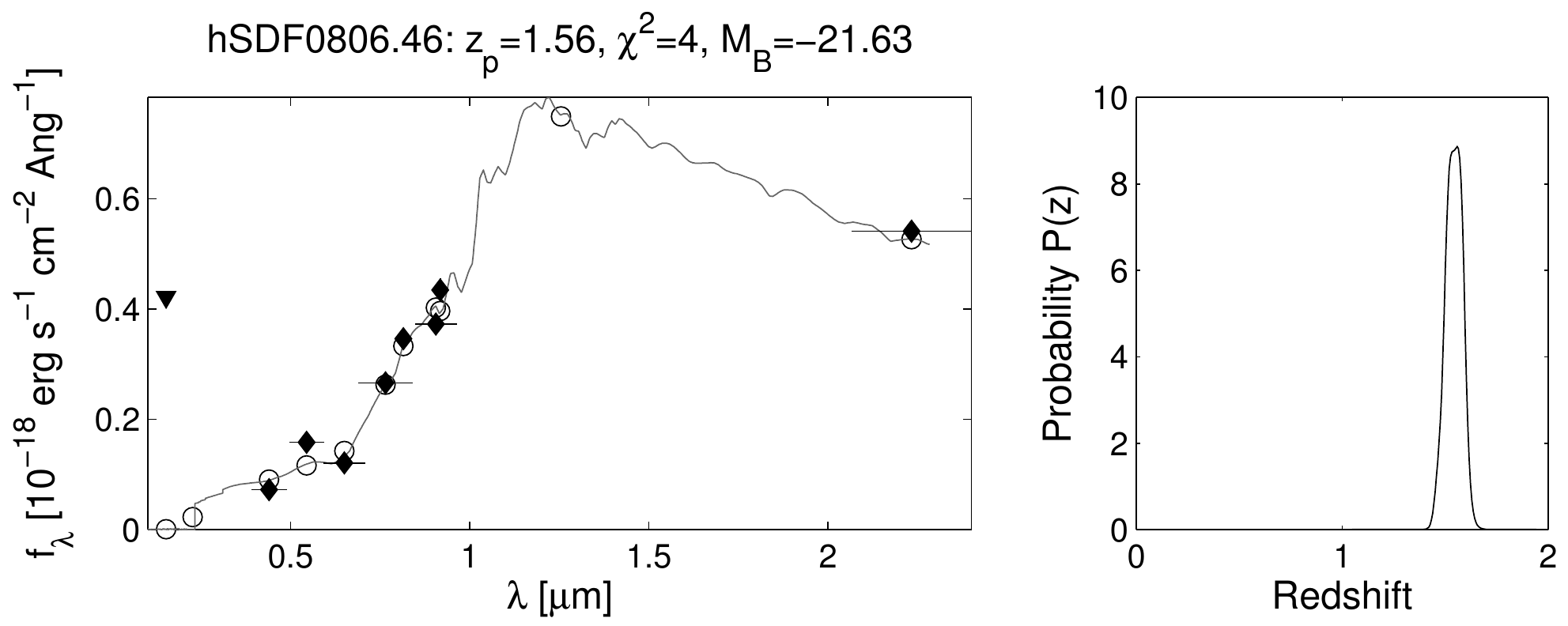} \\

\includegraphics[width=0.5\textwidth]{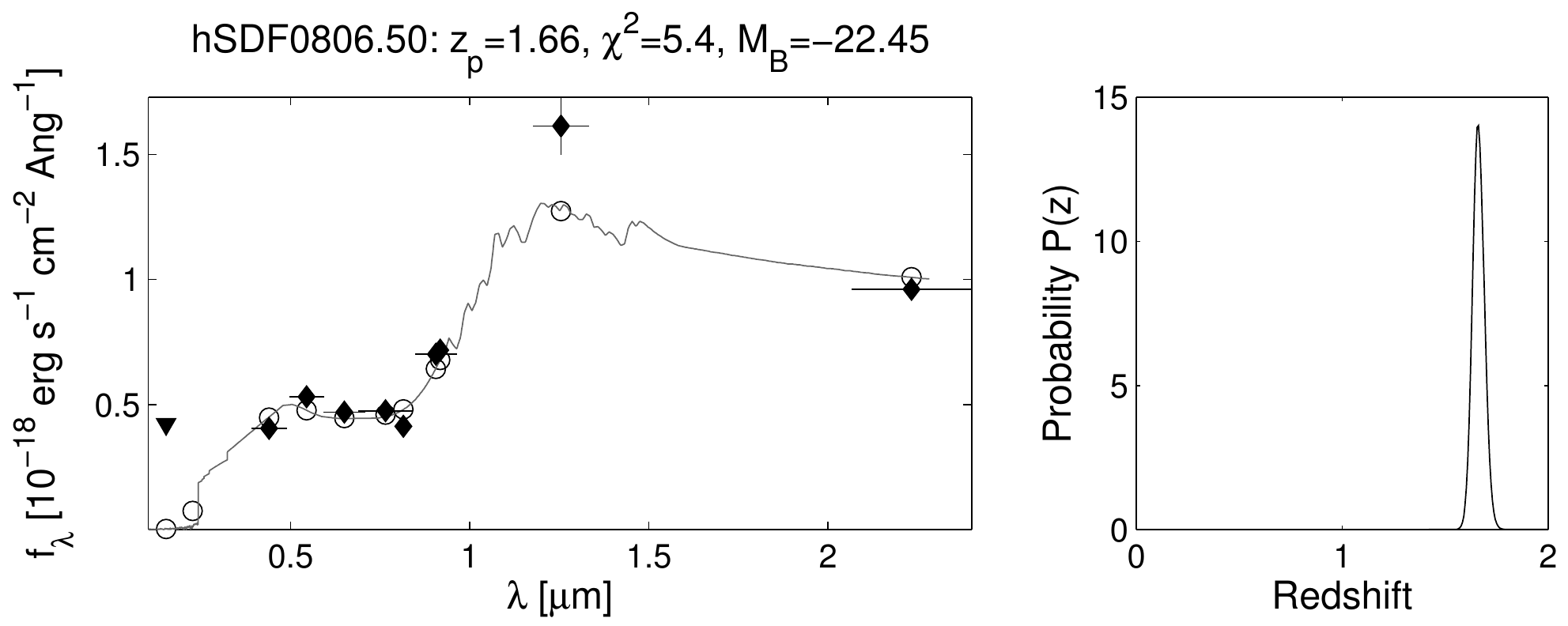} & 
\includegraphics[width=0.5\textwidth]{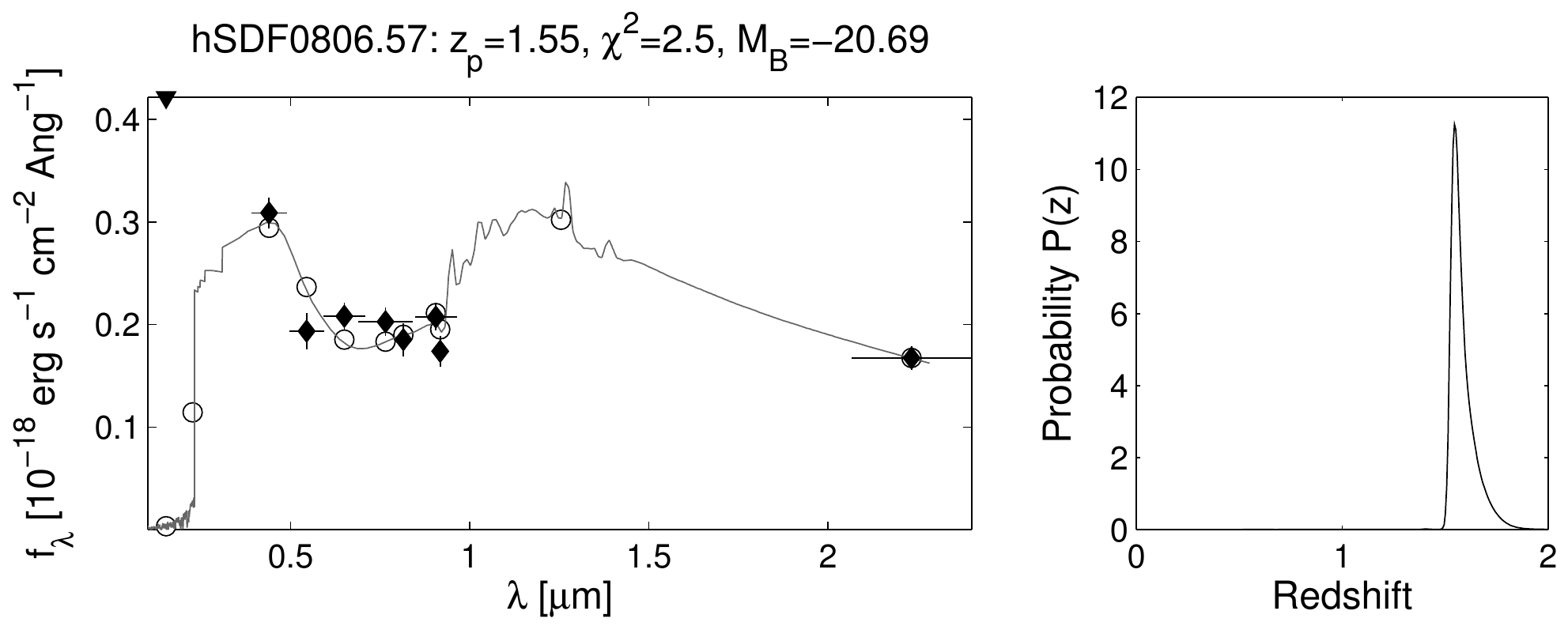}

\end{tabular}
\caption{ZEBRA fits and resultant redshift PDFs of the $1.5<z<2.0$ SN~Ia host galaxies. The left panel of every pair shows the actual photometry (filled circles), the best-fitting galaxy template (solid line), and its synthetic photometry (empty circles). The vertical error bars denote the photometric uncertainty, and the horizontal error bars show the width of the filter. The header gives the designation of the SN host galaxy, most probable photo-$z$ ($z_p$), the spec-$z$ ($z_s$, if such a measure exists for the specific object), the $\chi^2$ per degree of freedom of the fit, and the absolute \B-band magnitude the object would have at $z_p$. The right panel of every pair shows the resultant $z$-PDF. If a spec-$z$ exists for the SN host galaxy, it appears as a cross.} 
\label{fig:zebfits}
\end{minipage}
\end{figure*}


\section{SUPERNOVA CLASSIFICATION}
\label{sec:classify}

\begin{figure*}
\begin{minipage}{\textwidth}
\center
\includegraphics[scale=0.9]{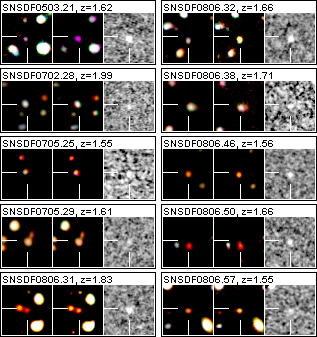}
\caption{SNe~Ia and host galaxies at $1.5 < z < 2.0$. North is up and east is left. All tiles are 10 arcsec on a side. The left-hand tiles show the SN host galaxies as imaged in epoch 1, whereas the centre tiles display the SN host galaxy as imaged in epochs 2 through 5. \R-, \I-, and \z-band images were combined to form the blue, green, and red channels (respectively) of the color composites. The right-hand tiles show the subtraction in the \z\ band. Whereas the stretch of the colour images differs from panel to panel in order to highlight host properties, the greyscale for all difference images is identical. The header of each panel gives the designation of the SN~Ia and its redshift. Similar images of the full sample of SNe are available in the electronic version of the paper.}
\label{fig:high_z_sne}
\end{minipage}
\end{figure*}

\begin{table}
\center
\caption{SN luminosity functions, presented as \B-band absolute magnitudes (Vega) at maximum light, and Gaussian width.}
 \begin{tabular}{l|c|c|l}
 \hline
 \hline
 {Type} & {$M_B$} & {$\sigma$} & {Source} \\
 \hline
 Ia   & $-19.37$ & 0.47 & \citet{2006ApJ...641...50W} \\
 II-P & $-16.98$ & 1.00 & \citet{2002AJ....123..745R} \\
 Ib/c & $-17.60$ & 0.90 & \citet{2010arXiv1011.4959D} \\
 IIn  & $-18.55$ & 1.00 & \citet{2010arXiv1010.2689K} \\
 \hline
 \end{tabular}
\label{table:LF}
\end{table}

We classify our SNe into SNe~Ia and CC~SNe using the SNABC algorithm of P07a.
Briefly, the SNABC receives as input the photometry and $z$-PDF of a SN candidate.
Using the above inputs, the SNABC then compares the colours of the SN candidate to the synthetic colours derived from a set of SN spectral templates of different types, ages, redshifts, host-galaxy and Galactic extinctions (based on the spectral templates of \citealt*{2002PASP..114..803N}, hereafter N02),\footnote{http://supernova.lbl.gov/$\sim$nugent/nugent\textunderscore templates.html} and to the rest-frame \B-band luminosity functions (LFs) of the different SN types.
In this work we used the LFs quoted by B08 for type Ia and II-P SNe, the LF measured by \citet{2010arXiv1011.4959D} for Ib/c SNe, and the LF measured by \citet{2010arXiv1010.2689K} for type IIn SNe. 
\citet{2010arXiv1011.4959D} measured peak magnitudes of $M_R=-17.9 \pm 0.9$ for SNe~Ib and $M_R=-18.3 \pm 0.6$ for SNe~Ic.
We take the weighted average of these magnitudes and get $M_R=-18.2 \pm 0.9$ mag.
Based on the N02 spectral template for SNe~Ib/c, we apply a colour correction of $(B-R)=0.6$ and arrive at $M_B=-17.6 \pm 0.9$ for SNe~Ib/c.
In a similar vein, we apply a colour correction of $(B-V)=-0.15$ to the LF measured by \citet{2010arXiv1010.2689K} for SNe~IIn, and arrive at a peak magnitude of $M_B=-18.55 \pm 1.00$.
The LFs and their sources are listed in Table~\ref{table:LF}.
The host-galaxy extinction was allowed to vary in the range $A_V=$ 0--3 mag, which spans the full range of possible extinctions that we consider (see Section~\ref{sec:debias} for a discussion of the extinction model we use).

The SNABC, as described by P07a, uses only the SN~Ia and SN~II-P spectral templates for classification.
P07a describe how using more templates, such as SN~IIn and SN~Ib/c, allows for better classification of CC~SNe, but at the same time significantly increases the number of SNe~Ia misclassified as CC~SNe, thus lowering the overall classification accuracy. 
We note that the goal of the current survey is not to discover and classify all types of SNe in the SDF, but rather to determine the rates of SNe~Ia statistically. 
The SNABC was designed and discussed specifically with the SDF survey, and its statistical approach, in mind.

The SNABC computes the likelihood of each comparison, and then marginalises over age, redshift, and extinction to arrive at the \textquoteleft evidence\textquoteright\ that the candidate is of a certain type: $E(\textrm{Ia})$ and $E(\textrm{CC})$.
The evidence is then used to derive the probability that the candidate is either a SN~Ia or CC~SN, according to
\begin{equation}\label{eq:PIa}
P(\textrm{Ia}) = \frac{E(\textrm{Ia})}{E(\textrm{Ia})+E(\textrm{CC})}.
\end{equation}

In addition to $P(\textrm{Ia})$, for each SN type the SNABC also produces a posterior $z$-PDF, which is constrained by the prior $z$-PDF input from ZEBRA. 
The SNABC also produces a $\chi^2$ value that indicates how well the SN's colours compared with those of the best-fitting spectral template.
A high $\chi^2$ value may imply the SN is a peculiar type of SN, an AGN, or a subtraction residual which was not rejected earlier.
An event is considered a SN~Ia if $P(\textrm{Ia})>0.5$ (and a CC~SN if $P(\textrm{Ia})<0.5$).
P07a have shown that $P(\textrm{Ia})$ can also be viewed as a confidence estimator: the closer it is to unity (zero), the safer the classification of the candidate as a SN~Ia (CC~SN).
P07a also found that for the sake of classification, most CC~SNe resemble SNe~II-P more than SNe~Ia.
Thus, while SN~Ia classifications usually result in small $\chi^2$ values ($\chi^2<1$), CC~SN classifications may result in higher values, since SNe~IIn or SNe~Ib/c are forcibly compared to SN~II-P spectral templates.

The posterior redshift assigned to each SN by the SNABC usually matches the prior redshift assigned by ZEBRA to within 5 per cent.
In those cases where the difference between the two exceeds 5 per cent, we check the shape of the $z$-PDF.
A wide or multi-peaked $z$-PDF implies that the colours of the SN provided either more information than the $z$-PDF itself, or enough information to break the degeneracy between the different peaks in the $z$-PDF.
In such instances (20 of the 150 SNe in our sample), we use the posterior redshift computed by the SNABC.
For example, SNSDF0806.32 has a posterior redshift of 1.66, even though this value corresponds to the weaker of the two peaks in the $z$-PDF of hSDF0806.32, as shown in Fig.~\ref{fig:zebfits}.

Table~\ref{table:SNe} lists the SNe in our sample, along with their redshifts and classifications.
Of the 150 SNe in our sample, 26 were found in the $z<0.5$ bin, of which 5 were classified as SNe~Ia and 21 as CC~SNe.
The $0.5<z<1.0$ bin contains 86 SNe, of which 50 were classified as SNe~Ia and 36 as CC~SNe.
The $1.0<z<1.5$ and $1.5<z<2.0$ bins contain 26 and 12 SNe, respectively, all of which were classified as SNe~Ia.
Two of the 12 SNe in the $1.5<z<2.0$ bin have high $\chi^2$ values, and are dealt with individually in Section~\ref{subsub:0705.30}.
The remaining 10 high-$z$ SNe~Ia are shown in Fig.~\ref{fig:high_z_sne}.

\subsection{Notes on individual supernovae}
\label{subsec:pec-sne}
The high $\chi^2$ values ($>10$) of some of the 163 transients in our sample prompted their reevaluation and, in some cases, rejection.
The final sample, after such rejections, includes 150 SNe.
All $\chi^2_r$ values quoted are per degree of freedom.

\subsubsection{SNSDF0503.25}
\label{subsub:0503.25}
While SNSDF0503.25 was classified as a CC~SN [$P(\rm{Ia})=0.06$] with a high $\chi^2_r$ value of 32, it is displaced from the nucleus of its spiral host by $0.63 \pm 0.07$ arcsec.
This, together with the absence of the object at other epochs, argues against its being an AGN, though it could be a variable background quasar.
Since the SNABC compares all candidates to SN~Ia and SN~II-P spectral templates, it is, in effect, forcibly comparing all subtypes of CC~SNe to SNe~II-P.
This leads us to believe that this SN is, in fact, a non-II-P CC~SN.
A similar situation is encountered for SNSDF0806.14.

\subsubsection{SNSDF0702.01}
\label{subsub:0702.01}
SNSDF0702.01 was classified as a SN~Ia [$P(\rm{Ia})=1$], but with $\chi^2_r=13$.
At a separation of $3.61 \pm 0.02$ arcsec, this $z=0.18$ transient is well offset from the centre of its spiral host galaxy, and so precludes the possibility of an AGN (though it could be a variable background quasar).
The high $\chi^2$ value arises from this object's $\R-\I$ colour, which does not fit the SN~Ia template.
As its absolute \R-band magnitude is $M_R=-17.01$, we checked whether this could be a SN 1991bg-like SN~Ia by comparing its photometry to the N02 SN 1991bg template.
While the \z-band magnitude matches the template, the $\R-\I$ and $\I-\z$ colours do not.
Though the \z-band magnitude and $\I-\z$ colour raise the possibility that this is an early SN~II-P, it is still too blue in the \R\ band.
We also checked whether the excess flux in the \R\ band might be the result of a SN caught during shock breakout, by comparing the \R-band photometry in our half-night stacks, but there was no discernible difference between the \R-band flux in the first two nights and in the second two.
At this point we conclude that this object is too faint and too blue to be a SN~Ia, and it might be either a very blue SN~II-P, or a peculiar SN of a different kind.
As detailed in Section~\ref{sec:debias}, since this object is at $z=0.18$, it enters neither the SN~Ia nor the CC~SN rate calculations.

\begin{figure}
\center
\includegraphics[width=0.48\textwidth]{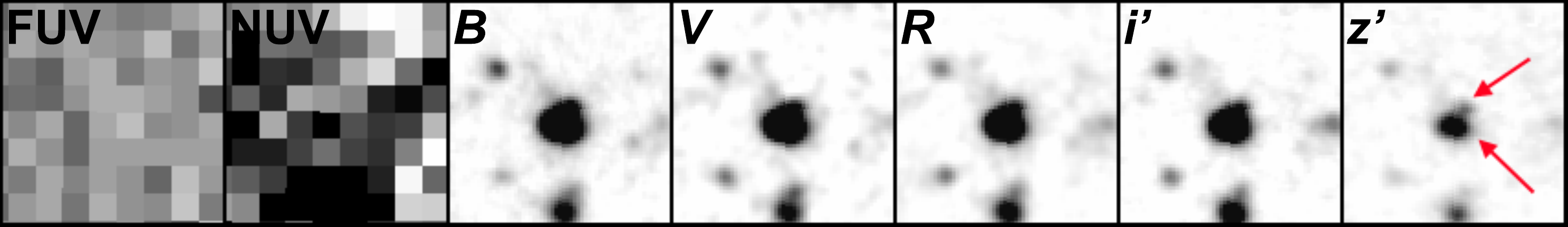}
\caption{The possible host galaxies of SNSDF0702.30. The arrows in the \z-band image point to the two possible hosts: hSDF0702.30a is the resolved galaxy, while hSDF0702.30b is a compact source to the NW (above and to the right) of the latter. While there may be an ambiguous detection in the {\it NUV} band, both galaxies are clearly undetected in the {\it FUV} band. All tiles are 10 arcsec on a side.}
\label{fig:3_30}
\end{figure}

\begin{figure}
\center
\includegraphics[width=0.5\textwidth]{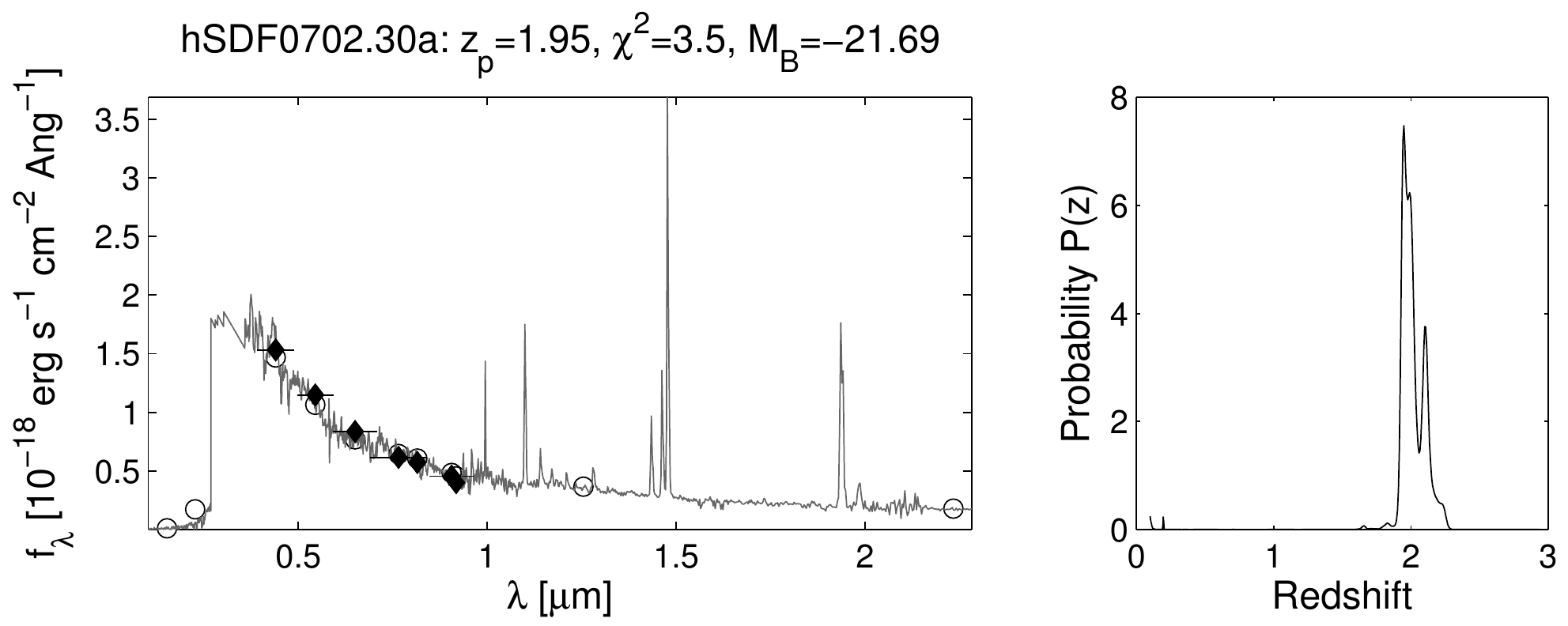} \\
\includegraphics[width=0.5\textwidth]{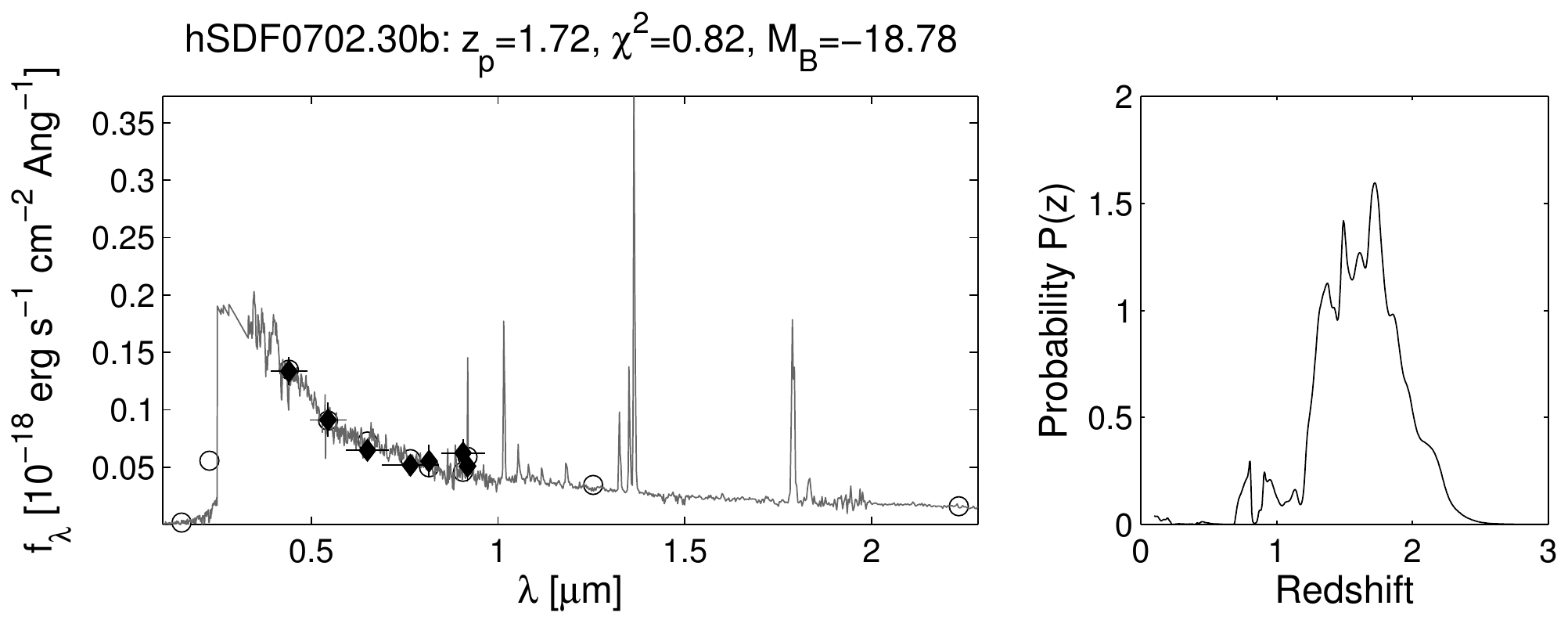}
\caption{Photometric redshift derivation for the two possible hosts of SNSDF0702.30, shown in Fig.~\ref{fig:3_30}. The galaxy hSDF0702.30a appears on top, while hSDF0702.30b is below. Symbols as in Fig.~\ref{fig:zebfits}. The first $z$-PDF is peaked at $z=1.95$, and the second $z$-PDF is peaked at $z=1.72$.}
\label{fig:3_30_zebra}
\end{figure}

\subsubsection{SNSDF0702.30}
\label{subsub:0702.30}
SNSDF0702.30 has two possible host galaxies, as shown in Fig.~\ref{fig:3_30}: a resolved galaxy designated hSDF0702.30a, and a compact galaxy to the NW (upper right; hSDF0702.30b).
We used the software GALFIT\footnote{http://users.obs.carnegiescience.edu/peng/work/galfit/galfit.html} \citep{2010AJ....139.2097P} to fit and subtract the larger galaxy, thus enabling us to perform photometry of each galaxy on its own.
The resulting photometry and best-fitting ZEBRA SEDs are shown in Fig.~\ref{fig:3_30_zebra}.
Both galaxies agree well with the power-law SED of a star-forming galaxy at a high redshift (hSDF0702.30a at $z=2.0$ with $\chi^2=3.5$, and hSDF0702.30b at $z=1.7$ with $\chi^2=0.8$).
While the fit in Fig.~\ref{fig:3_30_zebra} does not utilize UV data, the results agree with the nondetections observed in the {\it FUV} band, as seen in Fig.~\ref{fig:3_30}.

Using the resulting $z$-PDF of hSDF0702.30a as a prior, the SNABC classifies this SN as a CC~SN [$P(\rm{Ia})=0.04$] at redshift $z=1.95$, with $\chi^2_r=37$.
The $z$-PDF of hSDF0702.30b, on the other hand, yields a different classification: [$P(\rm{Ia})=0.68$] at redshift $z=0.8$, with $\chi^2_r=0.4$.
In this case, the SNABC chooses the smaller $z$-PDF peak at $z \approx 0.8$, instead of the main peak at $z \approx 1.7$, in order to avoid a high $\chi^2_r$ value such as that achieved with the sharply peaked $z$-PDF of hSDF0702.30a.
When run through the SNABC with a flat $z$-PDF, the SN best resembles a CC~SN [$P(\rm{Ia})=0.30$] at $z=0.6$ with $\chi^2_r=1.0$. 
The $z$-PDF constructed from the best-fitting redshifts of the other SN host galaxies does not change this result much; the posterior redshift changes to $z=0.7$, with a lower $\chi^2_r=0.4$.

In this case, the SNABC is dominated by the SN~II-P LF. 
Since the colours of the SN match those of a SN~II-P, it places it at $z<1$, the redshift range where the apparent magnitude of the SN would still match the SN~II-P LF.
This is also the reason it produces a high $\chi^2_r$ value when forced to higher redshifts.
In summary, SNSDF0702.30 may be either a CC~SN at $z=0.6$--0.8, or a non-Ia luminous SN at $z=1.7$--1.95.
The possible observation of overluminous non-Ia SNe at high redshifts in our sample is further discussed in Section~\ref{subsubsec:ccsne}, below.
If it is a low-$z$ CC~SN, it will not be counted in the rates, as it is fainter than the detection limit adopted in Section~\ref{sec:debias}.
Since it may be a high-$z$ non-Ia SN, we do not include this SN in our $1.5<z<2.0$ SN~Ia sample.

\subsubsection{SNSDF0705.17}
\label{subsub:0705.17}
SNSDF0705.17 was classified as a CC~SN [$P(\rm{Ia})=0.02$] at $z=2.87$, with $\chi^2_r=58$.
The offset of the candidate from its host galaxy is $0.15 \pm 0.10$ arcsec, or $\sim 1 \pm 1$ pixel.
If one were to redshift a fiducial SN~Ia template (i.e., at peak, with no stretch) to $z=2.87$, the synthetic observed \z-band magnitude would be $\z=29.5$, which is 3.9 mags fainter than the observed $\z=25.6 \pm 0.2$ of the object.
Thus, at this redshift, the object is too bright to be either a SN~Ia or a normal CC~SN.
This object appears in epoch 4, which is separated from epoch 3 by only $\sim 90$ days in the observer's frame.
In the object's rest-frame, this interval corresponds to $\sim 23$ days.
The high redshift, coupled with the high $\chi^2_r$ value, raises the suspicion that this candidate, even though it shows no variability in other epochs, is still an AGN.
Alternatively, the object might be a hyperluminous SN~IIn, or even a pair-instability SN.
Since both luminous SNe~IIn and pair-instability SNe decay slowly (e.g., \citealt{2002ApJ...573..144D, 2009Natur.462..624G}), if this object were one of the two it would likely have been detected in both epochs, unless it exploded between the two epochs.
Our preferred conclusion is that this is an AGN and as such, we have removed it from our sample.

\subsubsection{SNSDF0705.18}
\label{subsub:0705.18}
SNSDF0705.18 lies $3.06 \pm 0.10$ arcsec, about 3 half-light radii, from the closest (and only probable host) galaxy.
We obtained a spectrum of this galaxy, which places it at $z=1.41$.
If this is indeed the SN's host galaxy, it is classified as a SN~Ia [$P(\rm{Ia})=0.98$], with $\chi^2_r=17$.
The SNABC is sensitive to the $z$-PDF it receives as input, and since for this galaxy the input was a very narrow ($\sigma = 0.01$) Gaussian centred on the measured spec-$z$, we ran this SN through the SNABC once more, this time treating it as a hostless SN.
This resulted in a classification as a CC~SN [$P(\rm{Ia})=0.39$] at a posterior redshift of 0.7, with a much better $\chi^2_r=0.2$.
At this redshift, the synthetic photometry derived from redshifting the SN~II-P template, at peak, would be $\z=25.37$ mag.
This is consistent with the measured $\z=25.6 \pm 0.2$ mag.
The \z-band master image of epoch 4 has a limiting magnitude of 27.24 mag.
At $z=0.7$, a galaxy would have to be fainter than $-15.9$ mag so as not to be detected.
This could mean that the candidate is indeed a CC~SN that went off in a dwarf galaxy undetected in the SDF (see, e.g., \citealt{2010ApJ...721..777A}).
Since the fit to a CC~SN at $z=0.7$ is much better than the earlier SN~Ia classification, we treat this SN as a \textquoteleft hostless,\textquoteright\ intermediate redshift CC~SN.
As this SN is fainter than the detection limit adopted for this redshift bin (see Section~\ref{sec:debias}, below), it will not be counted in the rates.
To account for the possibility that this is a SN~Ia in the range $1.0<z<1.5$, we add a systematic uncertainty of $+1$ to the number of SNe~Ia in this bin.

\subsubsection{SNSDF0705.30 and SNSDF0806.35}
\label{subsub:0705.30}
SNSDF0705.30 and SNSDF0806.35 are both classified as SNe~Ia [$P(\rm{Ia})=0.90$ and $P(\rm{Ia})=0.99$, respectively] at high redshifts ($z=1.93$ and $z=1.94$, respectively), but with high $\chi^2_r$ values (34 and 22, respectively).
While these SNe are both offset from the cores of their host galaxies (by $0.5 \pm 0.1$ and $0.7 \pm 0.1$ arcsec, respectively), they are much bluer than any of the SN~Ia or CC~SN spectral templates.
SNSDF0806.35 has $\R-\I$ and $\I-\z$ colours consistent with those of the $z=1.189$ pulsational pair-instability SN SCP~06F6 \citep{2009ApJ...690.1358B, 2009arXiv0910.0059Q}, redshifted to $z=1.94$.
SNSDF0705.30, on the other hand, is even bluer.
It might be a very blue non-Ia SN, or a background variable quasar.
As both of these SNe are clearly not SNe~Ia, we exclude them from our $1.5<z<2.0$ bin.


\section{DEBIASING: DERIVATION OF INTRINSIC SUPERNOVA TYPE AND REDSHIFT DISTRIBUTIONS}
\label{sec:debias}

The success rate of the SNABC depends on the intrinsic parameters of the SNe (e.g., type, age, redshift, and extinction). 
P07a have found that degeneracies between these parameters lead to misclassifications, which in this work may introduce biases in the SN rate calculations (i.e., if an appreciable number of SNe~Ia are misclassified as CC~SNe, the SN~Ia rates will be systematically lower).
In order to correct for potential misclassifications, we follow the debiasing procedure described by P07a and P07b.
We use the spectral templates from N02 to simulate a sample of 40,000 SN light curves, divided into four subtypes: Ia, II-P, Ib/c, and IIn.
These templates have been normalised so that the \B-band absolute magnitude at maximum luminosity, for a stretch $s=1$ \citep{1999ApJ...517..565P} SN~Ia, is zero, in the Vega magnitude system.
In order to construct the light curves in our sample, we follow the recipe outlined by \citet{2006AJ....131..960S}.
For SNe~Ia, the light curves are constructed according to:
\begin{equation}\label{eq:LC}
m=m_{z=0,s=1}+M_B+\mu-\alpha(s-1),
\end{equation}
and
\begin{equation}\label{eq:LC1}
t_s=t_{s=1}\alpha,
\end{equation}
where $m_{z=0, s=1}$ is the basic light curve, at $z=0$ and with $s=1$, constructed from the spectral templates; $M_B$ is the peak brightness in the \B\ band, drawn from a Gaussian centred on $-19.37$ mag, with a dispersion of $\sigma=0.17$ mag, mimicking the intrinsic SN~Ia dispersion in peak brightness \citep{1995AJ....109....1H, 1996AJ....112.2398H, 1999AJ....118.1766P}; $\mu$ is the distance modulus; $\alpha=1.52\pm0.14$ \citep{2006A&A...447...31A}; $s$ is the stretch parameter of the SN, which is modeled as a Gaussian centred on $s=1$ with a dispersion of $\sigma=0.25$, and allowed to vary in the range $0.7\leq s \leq 1.3$ \citep{2006AJ....131..960S}; and $t_s$ is the age of the stretched-light-curve SN.

The dispersion in $s$, taken from \citet{2006AJ....131..960S}, is larger than the observed dispersion among normal SNe~Ia (e.g., \citealt{2007ApJ...667L..37H}), in order to include both the very subluminous and overluminous SNe~Ia.
The above recipe results in a LF that is consistent with those assumed by N06 and \citet{2006AJ....131..960S}, and measured by \citet{dilday2008}. 
Recently, \citet{li2011LF} measured a larger fraction of subluminous SNe~Ia than is represented here, which means our subsequent SN~Ia rates may be underestimated. 
However, since the \citet{li2011LF} LF is not corrected for extinction, nor is it in a standard magnitude system, we cannot use it to estimate how many subluminous SNe~Ia may be unaccounted for in our calculations.

The CC~SN light curves are constructed in much the same way, but without any stretching.
Host extinction is added using the \citet*{cardelli1989} extinction law, with $R_V=3.1$, and $A_V$ values drawn from the extinction model of N06: the positive side of a Gaussian centred on $A_V=0$ mag, with a dispersion of $\sigma=0.62$ for SNe~Ia and $\sigma=0.93$ mag for CC~SNe \citep{2006AJ....131..960S}.

As with our observed SN sample, one sixth of the simulated sample is assigned a random spec-$z$ in the form of a Gaussian $z$-PDF with $\sigma=0.01$.
The rest of the SNe in the sample are randomly assigned a redshift from the $z$-PDF of the entire SDF, out to $z=3$. 
Each simulated SN is assigned a \textquoteleft real\textquoteright\ redshift and a \textquoteleft measured\textquoteright\ redshift drawn from its $z$-PDF.
This mimics the ZEBRA redshift determinations.
While the simulated light curves are redshifted according to the real redshift, we keep the entire $z$-PDFs for the classification stage.

The resulting light curves are \textquoteleft observed\textquoteright\ at a random day, and each measurement is assigned an uncertainty according to the photometric uncertainties measured in our survey.
At redshifts $z \leq 1$, the light curves do not cover the full time period during which SNe could have been detected by the depth of our survey.
One way to overcome this problem would be to extrapolate the light curves, but this might introduce systematic errors that are difficult to quantify.
Instead, we chose to impose a flux limit on the SNe found in these bins; by raising the detection limit we narrow the time period during which the SNe could have been observed, thus ensuring that we stay within the bounds of the observed light curves.

In the $0.5<z<1.0$ bin, the detection limit was raised to 25.0 mag in the \z\ band for all epochs.
This reduces the number of SNe in this bin from 85 to 29, of which 26 are classified as SNe~Ia and 3 as CC~SNe. In the $z<0.5$ bin, the necessary flux limit leaves no SNe to work with; we thus cannot compute the SN rate in this bin.
We note, however, that rates at $z<1$ are much better measured by wider and shallower surveys that obtain light curves and spectroscopic confirmation for each SN (e.g., SDSS-II, SNLS). 
Our survey is designed specifically for detecting SNe at $z>1$, and for classifying them with single-epoch photometry.

We measure the success fractions of the SNABC in each epoch of observations by selecting only those SNe that would have been detected by our survey (i.e., those SNe which are brighter in the \z\ band than 26.3, 26.6, 26.4, and 26.7 mag for epochs 2 through 5, respectively, in the $z>1$ bins, and brighter than 25.0 mag in all epochs for the $0.5<z<1.0$ bin), leaving 3,000 SNe from each subtype.
The surviving SNe are then classified by the SNABC, and their redshift is determined as in Section~\ref{sec:classify}. 
Next, the SNe are distributed into three redshift bins ($0.5<z<1.0$, $1.0<z<1.5$, and $1.5<z<2.0$), and the success fraction in each bin is calculated by dividing the number of correctly classified SNe by the total number of SNe in that bin.

The resulting success fractions are used to calculate the probability of classifying a SN of any subtype as a SN~Ia, as a function of the intrinsic distribution of SN subtypes (e.g., 10 per cent SN~Ia, 40 per cent SN~II-P, 20 per cent SN~Ib/c, and 30 per cent SN~IIn).
Using steps of 2.5 per cent, there are 12,341 possible distributions.
In each redshift bin, and for each possible distribution, the SN~Ia success fraction is computed by summing the fraction of SNe~Ia that were classified correctly, together with the fractions of CC~SNe that were misclassified as SNe~Ia.
Each possible distribution is weighted according to the number of combinations in which the different CC~SN subtypes may be distributed for a given fraction of SNe~Ia (i.e., if the fraction of SNe~Ia is 50 per cent, there are many different combinations of CC~SN fractions, whereas if the SN~Ia fraction is 100 per cent, there is only one possible combination).

After weighting the different distributions, we marginalise over all of the different combinations for a specific SN~Ia fraction, and are left with the probability of classifying any SN as a SN~Ia, as a function of the intrinsic SN subtype distribution.
Using binomial statistics, this probability is used to answer the following question: Given the number of SNe classified by the SNABC as SNe~Ia in a given redshift bin, the total number of SNe in that bin, and the probability of classifying any SN as a SN~Ia, at a given intrinsic distribution, what is the most probable fraction of SNe~Ia in our sample?
From the resulting PDF we select the most probable value as the true fraction of SNe~Ia in each redshift bin, and define the $1\sigma$ uncertainty as the region that includes 68.3 per cent of the probability density.
To this classification uncertainty we add, in quadrature, the statistical uncertainty, defined as the $1\sigma$ Poisson uncertainty of the debiased number of SNe~Ia in the redshift bin (or the Poisson uncertainty of the number of debiased CC~SNe for the CC~SN uncertainty).

\begin{figure}
\center
\includegraphics[width=0.5\textwidth]{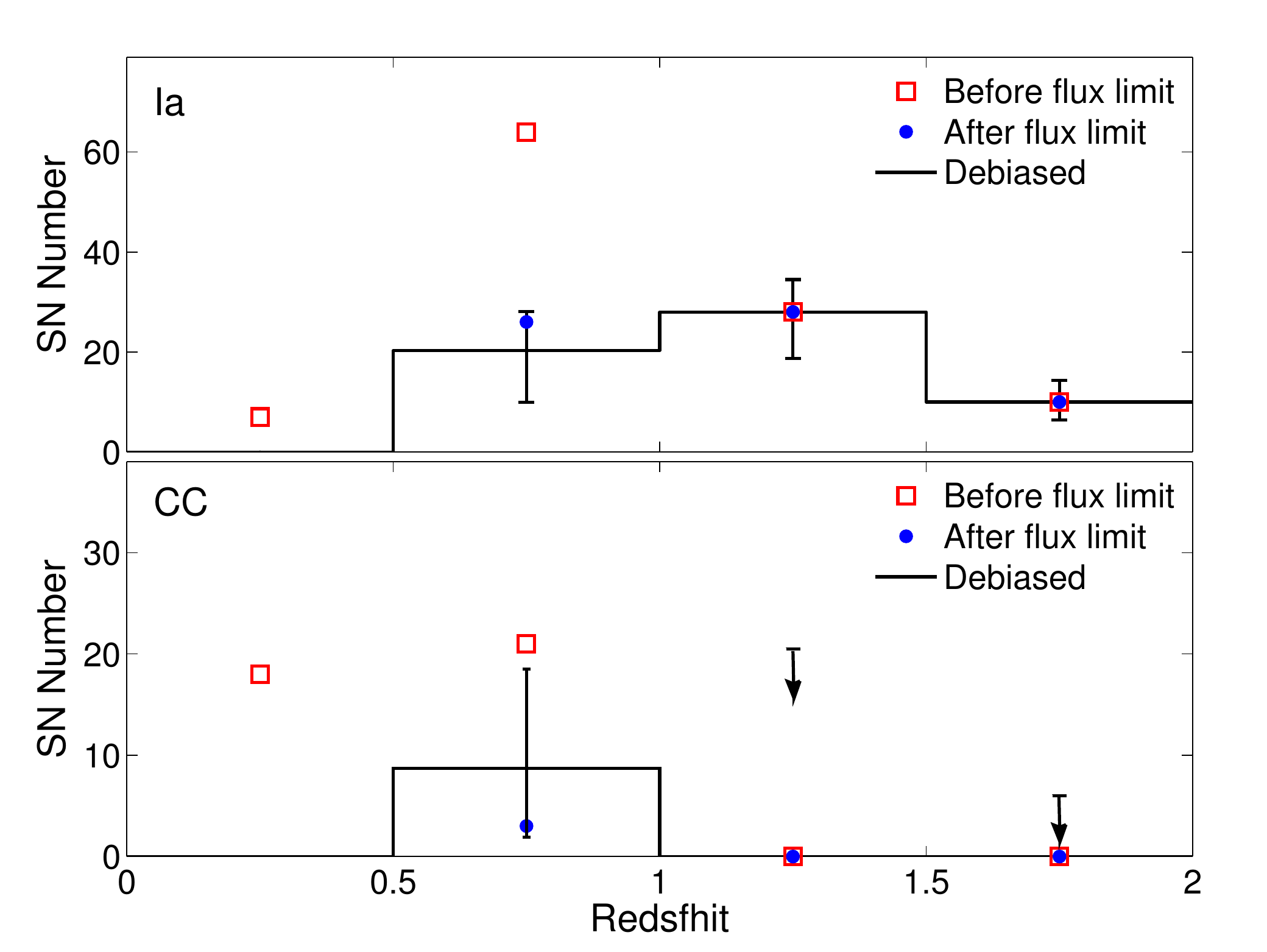}
\caption{Observed (empty squares), flux-limited (filled circles), and debiased (solid line) SDF SN~Ia and CC~SN numbers. Filled squares denote the number of SNe in the $z<1$ bins before application of the flux limit. SN~Ia error bars are $1\sigma$ Poisson and classification uncertainties, added in quadrature. CC~SN $0.5<z<1.0$ debiased error bar is $1\sigma$ Poisson and classification uncertainties, added in quadrature, and $z>1$ debiased numbers are $2\sigma$ upper limits (arrows).}
\label{fig:debias}
\end{figure}

The raw and debiased distributions of SNe~Ia and CC~SNe are presented in Fig.~\ref{fig:debias}.
The debiased number of SNe~Ia is the same as the raw number in the two $z>1$ bins, where our survey is mostly insensitive to CC~SNe.
The possibility that the $z>1$ bins have been contaminated by luminous CC~SNe (e.g., SNe~IIn) is taken into account in the lower systematic uncertainty of the debiased number of SNe Ia in these bins: $28.0^{+6.4,+1.0}_{-5.3,-7.6}$ at $1.0<z<1.5$ and $10.0^{+4.3,+0.0}_{-3.1,-1.7}$ in the $1.5<z<2.0$ bin.
In the $0.5<z<1.0$ bin, the number of SNe~Ia falls to $20.3^{+5.2,+5.8}_{-4.7,-9.3}$.
The post-debiasing number of CC~SNe, on the other hand, rises to $8.7^{+3.2,+9.3}_{-3.5,-5.8}$.
The errors for the above SN numbers are $1\sigma$ Poisson and classification uncertainties, respectively.


\section{SUPERNOVA RATES}
\label{sec:rates}

\begin{figure}
\center
\includegraphics[width=0.5\textwidth]{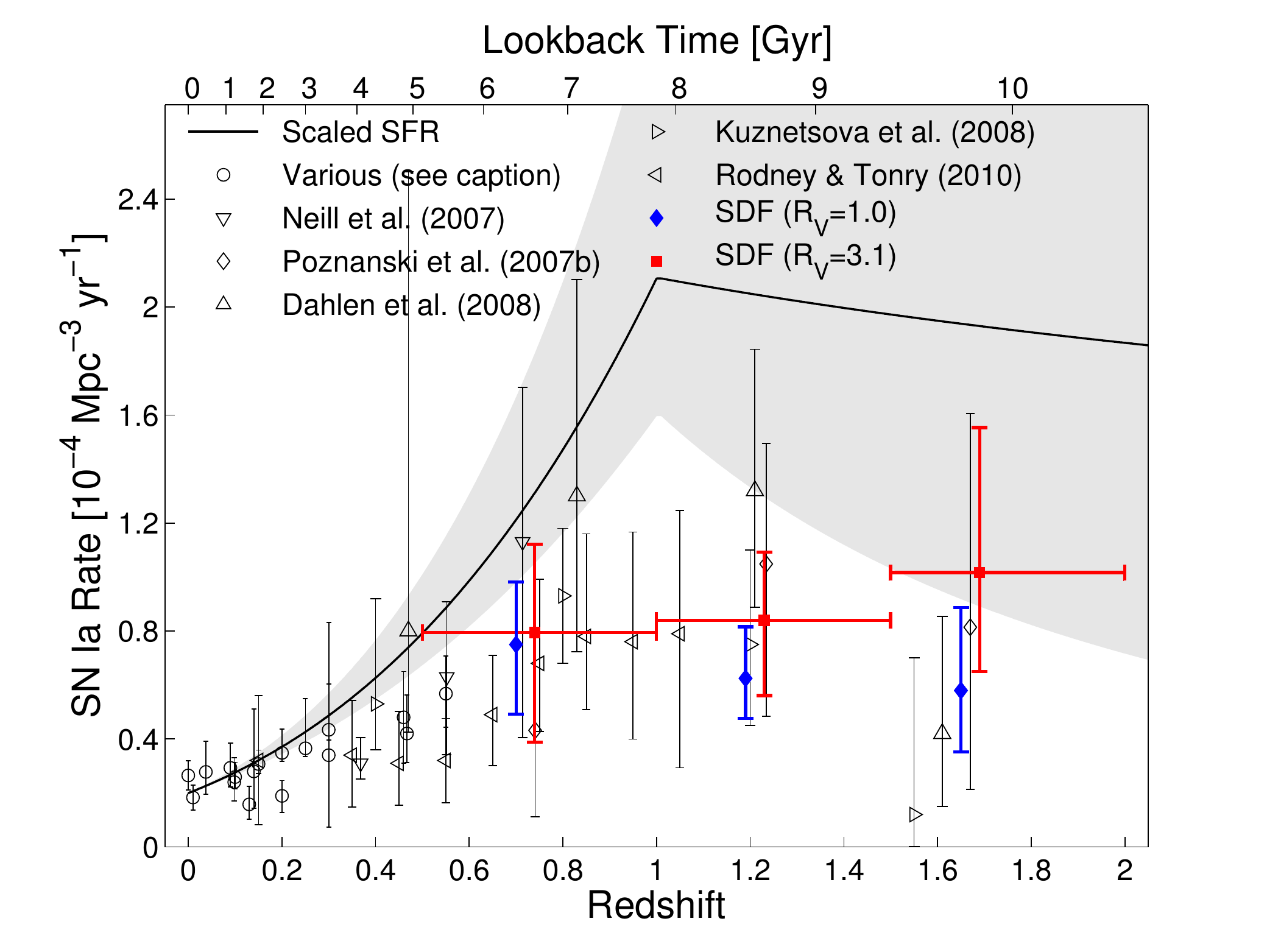}
\caption{SN~Ia rates from the SDF (filled squares) compared to rates from the literature. Circles denote low-$z$ data from \citet{cappellaro1999}, \citet{hardin2000}, \citet{2002ApJ...577..120P}, \citet{2003ApJ...599L..33M}, \citet{2003ApJ...594....1T}, \citet{blanc2004}, \citet{neill2006}, \citet{botticella2008}, \citet{dilday2008}, \citet{horesh2008}, \citet{dilday2010a}, and \citet{li2011rates}. Downturned triangles are for \citet{neill2007}. The corrected IfA Deep Survey rates from \citet{2010ApJ...723...47R} are left-facing triangles. The GOODS rates from \citet{dahlen2008} are denoted by upturned triangles. Right-facing triangles are for \citet{kuznetsova2008goods}. Our initial SDF results \citep{poznanski2007sdf} are shown as diamonds. Filled squares (circles) denote the SDF rates, derived with an extinction law with $R_V=3.1$ ($R_V=1$).The cosmic SFH from Y08, has been scaled to fit the low-$z$ data. The shaded area denotes the plausible range of SFHs with power-law slopes between 3 and 4, out to $z=1$, and between $-2$ and 0 for $z>1$. All vertical error bars include statistical and systematic uncertainties added in quadrature. Horizontal error bars indicate the SDF redshift bins.}
\label{fig:rates_Ia_upperlim}
\end{figure}

In this section, we use the debiased distributions of SNe~Ia and CC~SNe to derive the SN~Ia and CC~SN rates in three redshifts bins: $0.5<z<1.0$, $1.0<z<1.5$, and $1.5<z<2.0$.
Our rates are summarized in Table~\ref{table:rates}, and comparisons to the literature are given in Tables~\ref{table:rates_lit_Ia} and \ref{table:rates_lit_CC}, and in Figs.~\ref{fig:rates_Ia_upperlim} and~\ref{fig:rates_CC}.
All rates from the literature have been converted to $h=0.7$. 
In cases where they are originally reported in SNuB (SNe per century per $10^{10}\, \rm{L}_{\odot,B}$), we have converted them to volumetric rates using the redshift-dependent luminosity density function from B08:
\begin{equation}\label{eq:Ldense}
j_B(z)=(1.03+1.76\, z)\times 10^8~\rm{L}_{\odot,B}~\textrm{Mpc}^{-3}.
\end{equation}

\subsection{The type Ia supernova rate}
\label{subsec:rates-Ia}

The volumetric SN~Ia rate is
\begin{equation}\label{eq:rateIa}
R_{\textrm{Ia}}(\langle z \rangle_i)=\frac{N_{\textrm{Ia},i}}{\int t_v(z)~\frac{dV}{dz}~dz},
\end{equation}
where $\langle z \rangle_i$ is the effective redshift of each redshift bin $i$, $N_{\textrm{Ia},i}$ is the number of debiased SNe~Ia in bin $i$, and $t_v(z)$ is the survey visibility time, integrated over the comoving survey volume element $dV$, at all redshifts $z$ within bin $i$.

The visibility time is the total amount of time we could have observed a SN, given the parameters of our survey.
At a given redshift, we need to consider the dispersion in light curves that originates in three separate effects: the intrinsic dispersion in peak magnitude, the stretch-luminosity relation, and the host-galaxy extinction.
To account for these different effects, we calculate the visibility time of each possible light curve, weight it by its probability (which is just the product of the probabilities of the separate effects), and sum over all possible combinations.

As in the previous section, we construct each possible light curve according to Equation~\ref{eq:LC}. 
We construct light curves with all the possible combinations of peak magnitude, stretch, and extinction, where $M_B$ is allowed to vary as a Gaussian in the $2\sigma$ range around $-19.37$ mag (where $1\sigma=0.17$); the stretch parameter $s$ is allowed to vary as a Gaussian centred on $s=1$ with a dispersion of 0.25 in the range $0.7 \leq s \leq 1.3$, with $\alpha=1.52$; and $A_V$ ranges between 0 and 3 mag according to the N06 model.

Each point in the light curve is multiplied by the appropriate detection efficiency taken from the functions in Section~\ref{subsec:fakes_eff}, and the entire light curve is then summed over the time it lies above the detection efficiency limit (the 50 per cent detection efficiency limits for the $z>1$ bins, and 25.0 mag in the $0.5<z<1.0$ bin).
Finally, we sum over the different epochs (since for each epoch the detection efficiency limit is different), and end up with the visibility time of our entire survey.
Symbolically,
\begin{equation}\label{eq:vistime}
\begin{split} 
t_v(z)=\sum_{\rm{epoch}}{ \iiint dM_B~ds~dA_V~p(M_B)~p(s)~p(A_v)} \\
\qquad\qquad\qquad\qquad\qquad\qquad\quad \times \int\limits_{m>m_{1/2}} \epsilon[m_z(t)]dt.
\end{split}
\end{equation}
\noindent
We take the weighted average of the redshifts in a bin as the bin's effective redshift, where the weight is the visibility time integrated over the volume element within that bin:
\begin{equation}\label{eq:redshift}
\langle z \rangle_i = \frac{\int t_v~z~dV}{\int t_v~dV}.
\end{equation}

\begin{table*}
\center
\hspace{1.5in}\parbox{5.8in}{\caption{SN~Ia and CC~SN numbers and rates}\label{table:rates}}
\begin{tabular}{l|l|l|l|l}
\hline
\hline
{Subsample} & {$0.0<z<0.5$} & {$0.5<z<1.0$} & {$1.0<z<1.5$} & {$1.5<z<2.0$} \\
\hline
Total & 25 & 85 & 28 & 12 \\
\hline
SN host galaxies with spec-$z$ & 4 & 13 & 7 & 0 \\ 
Hostless SNe & 0 & 11$^a$ & 1 & 0 \\
\hline
SNe~Ia (raw) & 7 & 64 & 28 & 10$^b$ \\
SNe~Ia (after flux limit) & 0 & 26 & 28 & 10 \\
SNe~Ia (debiased)$^c$ & ... & $20.3^{+5.2,+5.8}_{-4.7,-9.3}$ & $28.0^{+6.4,+1.0}_{-5.3,-7.6}$ & $10.0^{+4.3,+0.0}_{-3.1,-1.7}$ \\
SN~Ia rate (10$^{-4}$ yr$^{-1}$ Mpc$^{-3}$) & ... & $0.79^{+0.33}_{-0.41}$ & $0.84^{+0.25}_{-0.28}$ & $1.02^{+0.54}_{-0.37}$ \\
SN~Ia rate without host-galaxy extinction & ... & $0.60^{+0.23}_{-0.31}$ & $0.62^{+0.14}_{-0.21}$ & $0.45^{+0.20}_{-0.16}$ \\
Effective redshift & ... & 0.74 & 1.23 & 1.69 \\
\hline
CC~SNe (raw) & 18 & 21 & 0 & 0 \\
CC~SNe (after flux limit) & 0 & 3 & 0 & 0 \\
CC~SNe (debiased)$^d$ & ... & $8.7^{+3.2,+9.3}_{-3.5,-5.8}$ & $<3.8,+20.2$ & $<3.8,+4.7$ \\
CC~SN rate (10$^{-4}$ yr$^{-1}$ Mpc$^{-3}$) & ... &  $6.9^{+9.9}_{-5.4}$ & {...} & {...} \\
CC~SN rate without host-galaxy extinction & ... & $1.8^{+2.0}_{-1.4}$ & {...} & {...} \\
Effective redshift & ... & 0.66 & {...} & {...} \\
\hline
\multicolumn{5}{l}{$^a$This includes SNSDF0705.18, which is treated as a hostless SN, as detailed in Section~\ref{subsub:0705.18}.} \\
\multicolumn{5}{l}{$^b$Two of the 12 SNe in this bin are clear non-Ia transients.} \\
\multicolumn{5}{l}{$^c$Errors are $1\sigma$ Poisson and classification uncertainties, respectively.} \\
\multicolumn{5}{l}{$^d$Errors in the $0.5<z<1.0$ bin are $1\sigma$ Poisson and classification uncertainties, respectively.} \\
\multicolumn{5}{l}{The $z>1$ rates are upper limits. Errors are $2\sigma$ Poisson and classification uncertainties, respectively.}
\end{tabular}
\end{table*}

\begin{table*}
\center
\hspace{1.5in}\parbox{5.5in}{\caption{SN~Ia rate measurements}\label{table:rates_lit_Ia}}
\begin{tabular}{c|c|c|l}
\hline
\hline
Redshift & $N_{\textrm{Ia}}$ & Rate [$10^{-4}$ yr$^{-1}$ Mpc$^{-3}$]  & Reference \\
\hline

0.01 & 70 & $0.183\pm 0.046$ & \citet{cappellaro1999}$^b$ \\

$<0.019^a$ & 274 & $0.265^{+0.034,+0.043}_{-0.033,-0.043}$ & {\citet{li2011rates}} \\

0.0375 & 516$^c$ & $0.278^{+0.112,+0.015}_{-0.083,-0.000}$ & \citet{dilday2010a} \\

0.09 & 17 & $0.29^{+0.09}_{-0.07}$ & \citet{dilday2008} \\

0.098 & 19 & $0.24^{+0.12}_{-0.12}$ & \citet{2003ApJ...599L..33M}$^b$ \\

0.1 & 516$^c$ & $0.259^{+0.052,+0.018}_{-0.044,-0.001}$ & \citet{dilday2010a} \\

0.13 & 14 & $0.158^{+0.056,+0.035}_{-0.043,-0.035}$ & \citet{blanc2004}$^b$ \\

0.14 & 4 & $0.28^{+0.22,+0.07}_{-0.13,-0.04}$ & \citet{hardin2000}$^b$ \\

0.15 & 516$^c$ & $0.307^{+0.038,+0.035}_{-0.034,-0.005}$ & \citet{dilday2010a} \\

0.15 & 1.95 & $0.32^{+0.23,+0.07}_{-0.23,-0.06}$ & \citet{2010ApJ...723...47R} \\

0.2 & 17 & $0.189^{+0.042,+0.018}_{-0.034,-0.015}\pm 0.42$ & \citet{horesh2008} \\

0.2 & 516$^c$ & $0.348^{+0.032,+0.082}_{-0.030,-0.007}$ & \citet{dilday2010a} \\

0.25 & 1 & $0.17\pm 0.17$ & \citet{2006ApJ...637..427B} \\

0.25 & 516$^c$ & $0.365^{+0.031,+0.182}_{-0.028,-0.012}$ & \citet{dilday2010a} \\

0.3 & 31.05$^d$ & $0.34^{+0.16,+0.21}_{-0.15,-0.22}$ & \citet{botticella2008} \\

0.3 & 516$^c$ & $0.434^{+0.037,+0.396}_{-0.034,-0.016}$ & \citet{dilday2010a} \\

0.35 & 5 & $0.530\pm 0.024$ & \citet{2006ApJ...637..427B} \\

0.35 & 4.01 & $0.34^{+0.19,+0.07}_{-0.19,-0.03}$ & \citet{2010ApJ...723...47R} \\ 

0.368 & 17 & $0.31^{+0.05,+0.08}_{-0.05,-0.03}$ & \citet{neill2007} \\

0.40 & 5.44 & $0.53^{+0.39}_{-0.17}$ & \citet{kuznetsova2008goods} \\

0.45 & 9 & $0.73\pm 0.24$ & \citet{2006ApJ...637..427B} \\

0.45 & 5.11 & $0.31^{+0.15,+0.12}_{-0.15,-0.04}$ & \citet{2010ApJ...723...47R} \\

0.46 & 8 & $0.48\pm 0.17$ & \citet{2003ApJ...594....1T} \\

0.467 & 73 & $0.42^{+0.06,+0.13}_{-0.06,-0.09}$ & \citet{neill2006} \\

0.47 & 8 & $0.80^{+0.37,+1.66}_{-0.27,-0.26}$ & \citet{dahlen2008} \\

\hline

0.55 & 38 & $0.568^{+0.098,+0.098}_{-0.088,-0.088}$ & \citet{2002ApJ...577..120P}$^a$ \\

0.55 & 29 & $2.04\pm 0.38$ & \citet{2006ApJ...637..427B} \\

0.55 & 6.49 & $0.32^{+0.14,+0.07}_{-0.14,-0.07}$ & \citet{2010ApJ...723...47R} \\

0.552 & 41 & $0.63^{+0.10,+0.26}_{-0.10,-0.27}$ & \citet{neill2007} \\

0.65 & 23 & $1.49\pm 0.31$ & \citet{2006ApJ...637..427B} \\

0.65 & 10.09 & $0.49^{+0.17,+0.14}_{-0.17,-0.08}$ & \citet{2010ApJ...723...47R} \\

0.714 & 42 & $1.13^{+0.19,+0.54}_{-0.19,-0.70}$ & \citet{neill2007} \\

0.74 & 5.5 & $0.43^{+0.36}_{-0.32}$ & \citet{poznanski2007sdf} \\ 

\textbf{0.74} & \textbf{20.3} & $\mathbf{0.79^{+0.33}_{-0.41}}$ & \textbf{SDF (this work)} \\

0.75 & 28 & $1.78\pm 0.34$ & \citet{2006ApJ...637..427B} \\

0.75 & 14.29 & $0.68^{+0.21,+0.23}_{-0.21,-0.14}$ & \citet{2010ApJ...723...47R} \\

0.80 & 18.33 & $0.93^{+0.25}_{-0.25}$ & \citet{kuznetsova2008goods} \\

0.83 & 25 & $1.30^{+0.33,+0.73}_{-0.27,-0.51}$ & \citet{dahlen2008} \\

0.85 & 15.43 & $0.78^{+0.22,+0.31}_{-0.22,-0.16}$ & \citet{2010ApJ...723...47R} \\

0.95 & 13.21 & $0.76^{+0.25,+0.32}_{-0.25,-0.26}$ & \citet{2010ApJ...723...47R} \\

\hline

1.05 & 11.01 & $0.79^{0.28,+0.36}_{-0.28,-0.41}$ & \citet{2010ApJ...723...47R} \\

1.20 & 8.87 & $0.75^{+0.35}_{-0.30}$ & \citet{kuznetsova2008goods} \\

1.21 & 20 & $1.32^{+0.36,+0.38}_{-0.29,-0.32}$ & \citet{dahlen2008} \\

1.23 & 10.0 & $1.05^{+0.45}_{-0.56}$ & \citet{poznanski2007sdf} \\ 

\textbf{1.23} & \textbf{28.0} & $\mathbf{0.84^{+0.25}_{-0.28}}$ & \textbf{SDF (this work)} \\

\hline

1.55 & 0.35 & $0.12^{+0.58}_{-0.12}$ & \citet{kuznetsova2008goods} \\

1.61 & 3 & $0.42^{+0.39,+0.19}_{-0.23,-0.14}$ & \citet{dahlen2008} \\

1.67 & 3.0 & $0.81^{+0.79}_{-0.60}$ & \citet{poznanski2007sdf} \\

\textbf{1.69} & \textbf{10.0} & $\mathbf{1.02^{+0.54}_{-0.37}}$ & \textbf{SDF (this work)} \\

\hline

\multicolumn{4}{l}{Note -- Redshifts are means over the redshift intervals probed by each survey. $N_{\rm{Ia}}$ is the number} \\
\multicolumn{4}{l}{of SNe~Ia used to derive the rate. Where necessary, rates have been converted to $h=0.7$.} \\
\multicolumn{4}{l}{Where reported, the statistical errors are followed by systematic errors, and separated by commas.} \\
\multicolumn{4}{l}{The uncertainties of the SDF results are statistical and systematic, added in quadrature.} \\
\multicolumn{4}{l}{$^a$\citet{li2011rates} consider SNe~Ia within 80 Mpc.}  \\
\multicolumn{4}{l}{$^b$Rates have been converted to volumetric rates using Equation~\ref{eq:Ldense}.} \\
\multicolumn{4}{l}{$^c$\citet{dilday2010a} compute their rates using 516 SNe~Ia in the redshift range $z<0.5$.} \\
\multicolumn{4}{l}{$^d$\citet{botticella2008} found a total of 86 SN candidates of all types. See their section 5.2} \\
\multicolumn{4}{l}{for details on their various subsamples and classification techniques.}

\end{tabular}
\end{table*}

\begin{table*}
\hspace{1.5in}\parbox{5.5in}{\caption{CC~SN rate measurements}\label{table:rates_lit_CC}}
\begin{tabular}{c|c|c|l}
\hline
\hline
Redshift & $N_{\textrm{CC}}$ & Rate [$10^{-4}$ yr$^{-1}$ Mpc$^{-3}$]  & Reference \\
\hline

$<0.0066^a$ & 92 & $>0.96$ & \citet{smartt2009mnras} \\

0.01 & 67 & $0.43\pm 0.17$ & \citet{cappellaro1999}$^b$ \\

$<0.014^a$ & 440 & $0.62^{+0.07,+0.17}_{-0.07,-0.15}$ & \citet{li2011rates} \\

0.21 & 44.95$^c$ & $1.15^{+0.43,+0.42}_{-0.33,-0.36}$ & \citet{botticella2008} \\

0.26 & 31.2$^d$ & $1.88^{+0.71}_{-0.58}$ & \citet{cappellaro2005}$^b$ \\

0.3 & 17 & $2.51^{+0.88,+0.75}_{-0.75,-1.86}$ & \citet{dahlen2004} \\

0.3 & 117 & $1.63^{+0.34,+0.37}_{-0.34,-0.28}$ & \citet{bazin2009} \\

\hline

\textbf{0.66} & \textbf{8.7} & $\mathbf{6.9^{+9.9}_{-5.4}}$ & \textbf{SDF (this work)} \\

0.7 & 17 & $3.96^{+1.03,+1.92}_{-1.06,-2.60}$ & \citet{dahlen2004} \\

\hline

\multicolumn{4}{l}{Note -- Where reported, the statistical errors are followed by systematic errors, and separated by commas.} \\
\multicolumn{4}{l}{The uncertainties of the SDF results are statistical and systematic, added in quadrature.} \\
\multicolumn{4}{l}{$^a$\citet{smartt2009mnras} and \citet{li2011rates} consider CC~SNe within 28 and 60 Mpc, respectively.} \\
\multicolumn{4}{l}{$^{b,c}$Same as in Table~\ref{table:rates_lit_Ia}.} \\
\multicolumn{4}{l}{$^d$Total number of CC~SNe and SNe~Ia detected throughout the survey.}

\end{tabular}
\end{table*}

The uncertainties of the rates are the classification and Poisson uncertainties of the debiased SN~Ia numbers, propagated and added in quadrature.
To test how the uncertainties in the detection efficiency functions, as plotted in Fig.~\ref{fig:fakes_eff}, affect the rates, we ran 500 Monte Carlo simulations in which each efficiency measurement was varied according to its uncertainty.
This produced variations in the detection efficiency limits of $\pm 0.1$ mag.
This propagates to a $1\sigma$ dispersion in the SN~Ia rates that is between one and two orders of magnitude smaller than our main uncertainties, thus having a negligible effect on the resulting rates.
The SN~Ia rates, with and without taking host-galaxy extinction into account, are shown in Table~\ref{table:rates}.

\subsubsection{High-redshift dust}
\label{subsubsec:dust}
As star formation increases with redshift, so does injection of dust into the interstellar medium, leading to an expected increase of extinction with redshift (e.g., \citealt*{2007MNRAS.377.1229M}).
This effect should lead to a decrease in the number of observed SNe at high redshifts.
\citet{2007MNRAS.377.1229M} have shown that at high redshifts ($z>1$) a large fraction of SNe, both CC~SNe and SNe~Ia, would be missed in optical surveys, due to extinction by dust in massive starburst galaxies, which make up a larger fraction of the galaxy population at higher redshifts \citep{2005ApJ...632..169L, 2005ApJ...631L..13D, 2005ApJ...630...82P}.
Using the \citet{2006MNRAS.370..773M} DTD model, \citet{2007MNRAS.377.1229M} calculated that in the range $1.0<z<2.0$, 15 to 35 per cent of SNe~Ia would be missed.
Assuming a power-law DTD model with a slope of $-1$ (see Section~\ref{sec:DTD}, below), the fraction of missing SNe~Ia would be 5--13 per cent in the above redshift range (F. Mannucci, private communication).
The corrected SN~Ia rates are shown in Fig.~\ref{fig:rates_Ia_upperlim} and in Table~\ref{table:rates_lit_Ia}.

\subsubsection{Different extinction laws}
\label{subsubsec:rv}
Throughout our classification, debiasing, and subsequent derivation of the SN~Ia rates, we have assumed a \citet{cardelli1989} extinction law with the Galactic average $R_V=3.1$.
However, several SN surveys have discovered SNe~Ia that underwent extinction best fit with lower values of $R_V$, from 1.7 to 2.5 \citep{tripp1998,2009ApJ...700..331H,2009ApJ...699L.139W}.
To gauge the systematic effect of lower $R_V$ values on our rates, we reran the classification, debiasing, and rates derivation stages assuming an extinction law with $R_V=1$.
The resultant rates are shown as filled diamonds in Fig.~\ref{fig:rates_Ia_upperlim}. 
They are consistent with the rates derived with $R_V=3.1$, but in the three redshift bins are systematically lower by 6, 26, and 43 per cent, respectively.

Whereas it is possible that the extinction law in the immediate vicinity of SNe~Ia is different from the Galactic average, recent studies (e.g., \citealt{guy2010,2011ApJ...729...55F}) have raised the possibility that SNe~Ia have an intrinsic colour scatter, which together with dust extinction is responsible for their overall reddening.
\citet{chotard2011} have found that once they corrected for an intrinsic scatter in the Si and Ca features of 76 SNe~Ia spectra, they recovered a \citet{cardelli1989} extinction law with $R_V=2.8 \pm 0.3$, consistent with the Galactic average value.
We do not add the systematic uncertainty produced by different values of $R_V$ to our final quoted SN~Ia rates.
However, in Section~\ref{sec:DTD} we do take this systematic uncertainty into account when deriving the SN~Ia DTD.

\subsubsection{Contamination from high-$z$ non-Ia transients}
\label{subsubsec:ccsne}

While our survey is largely insensitive to CC~SNe at $z>1$, there remains the possibility of contamination by non-Ia luminous SNe (e.g., \citealt{2007ApJ...666.1116S, 2007ApJ...668L..99Q, 2009ApJ...690.1358B}).
As described in Section~\ref{subsec:pec-sne}, we have discovered two luminous non-Ia SNe in the $1.5<z<2.0$ bin.
This ratio of 2:10 non-Ia~SNe to SNe~Ia is consistent with the 1:11 ratio found by \citet{2010arXiv1010.5786B}, who found one non-Ia transient (SCP~06F6) and 11 field SNe~Ia in the redshift range $z=0.6$--1.3.

As for contamination by AGNs, the extremely blue colours of SNSDF0705.30 and the classification of SNSDF0705.17 as a CC~SN at $z=2.87$ hint that these objects might be variable background quasars, as discussed in Section~\ref{subsec:pec-sne}.
This is consistent with the expected number of contaminating AGNs in our sample, as detailed in Section~\ref{subsec:gal_phot}.
In summary, beyond the non-Ia objects we have identified, contamination of the $1.5<z<2.0$ SN~Ia sample is unlikely.

\subsubsection{Biases in the photometric redshifts}
\label{subsub:photz}
As shown in Fig.~\ref{fig:zebra}, there is a bias in our photo-$z$ method towards overestimation of the redshift at high redshifts. 
This is caused by the dearth of spectroscopic redshifts at $1.5<z<2.0$ (only 24, or $\sim 6$ per cent, of the training-set galaxies), and the inherent difficulty of measuring the redshift of late-type galaxies (see Section~\ref{subsec:zebra}). 
We have taken two steps to overcome this bias. 
First, the measured colours of the SNe were compared to those predicted by the templates of SNe~Ia, SNe~II-P, SNe~Ib/c, and SNe~IIn, at different redshifts, spanning the entire $0<z<2$ range. 

Eight out of the ten $1.5<z<2.0$ SNe agree within $2\sigma$ with the fiducial SN~Ia template colors one would observe at their host galaxies' photo-$z$ (namely, SNSDF0705.21, SNSDF0806.31, SNSDF0806.46, SNSDF0806.50, .25, SNSDF0705.29, SNSDF0806.38, and SNSDF0806.57). 
SNSDF0702.28 may be an example of the bias seen in Fig.~\ref{fig:zebra}; whereas its late-type host galaxy has a photo-$z$ of $\sim 2$, the SN colours favour a lower redshift of $\sim 1.6$--1.7. 
Finally, the colours of SNSDF0806.32 favour the SN~IIn template over the entire $1.2<z<2.0$ redshift range. 
The possibility that this SN has been misclassified as a SN~Ia is taken into account in the systematic uncertainty of the SN~Ia rate in this bin, as quoted in Table~\ref{table:rates}.

To further investigate the issue of the redshifts of the candidate $z>1.5$ SNe, and to check whether any of them are contaminating AGNs, we are pursuing {\it HST} and ground-based spectroscopic observations of these hosts.

\subsubsection{Probing the UV part of the SN spectrum}
\label{subsub:UV}
From a theoretical standpoint, the spectra of SNe~Ia at high redshifts may differ from their low-redshift counterparts due to changes in, for example, progenitor metallicity.
Such differences are expected to show up in the UV part of the SN~Ia spectrum, introducing a possible systematic uncertainty into any survey (such as the current work) which probes this part of the spectrum (\citealt*{1998ApJ...495..617H}; \citealt{2000ApJ...530..966L, 2008MNRAS.391.1605S}).
Several recent surveys have found evidence for such differences between low- and high-redshift SNe~Ia (e.g., \citealt{2009ApJS..185...32K, 2009ApJ...704..687C, 2010arXiv1010.2749F}), which might provide an additional explanation for the high $\chi^2$ values of the two peculiar SNe in our $1.5<z<2.0$ sample.

\subsection{The core-collapse supernova rate}
\label{subsec:rates-CC}

Since our survey is insensitive to normal CC~SNe at redshifts higher than 1, we do not use the debiased results to derive the rates in the $1.0<z<1.5$ and $1.5<z<2.0$ redshift bins.
We now proceed to derive the CC~SN rate in the $0.5<z<1.0$ redshift bin.

To account for the division of CC~SNe into subtypes, in the calculation of the visibility time we have weighted the contribution of each subtype according to its fraction of the total CC~SN population, and then summed the different contributions.
The CC~SN subtype fractions were taken from the volume-limited sample of \citet{li2011LF}, with two alterations: (a) the SN~II-P and SN~II-L fractions have been combined, as the separation between these subclasses is currently ill-defined \citep{2002PASP..114..833P}; and (b) the SN~Ib/c and SN~IIb fractions have also been combined, since their light curves are nearly identical \citep{1994AJ....107.1453B}.
The final volume-limited CC~SN fractions are 60.0 per cent II-P/L, 33.5 per cent Ib/c/IIb, and 6.5 per cent IIn.
We note that \citet{li2011LF, li2011rates} only targeted $\sim\rm{L_*}$ galaxies, and so the CC~SN fractions and rates might be different for an untargeted survey (e.g., \citealt{2010ApJ...721..777A}).
We calculate the flux-limited fractions at the effective redshift of $z=0.66$ as being 37 per cent II-P/L, 44 per cent Ib/c/IIb, and 19 per cent IIn.

\begin{figure}
\center
\includegraphics[width=0.5\textwidth]{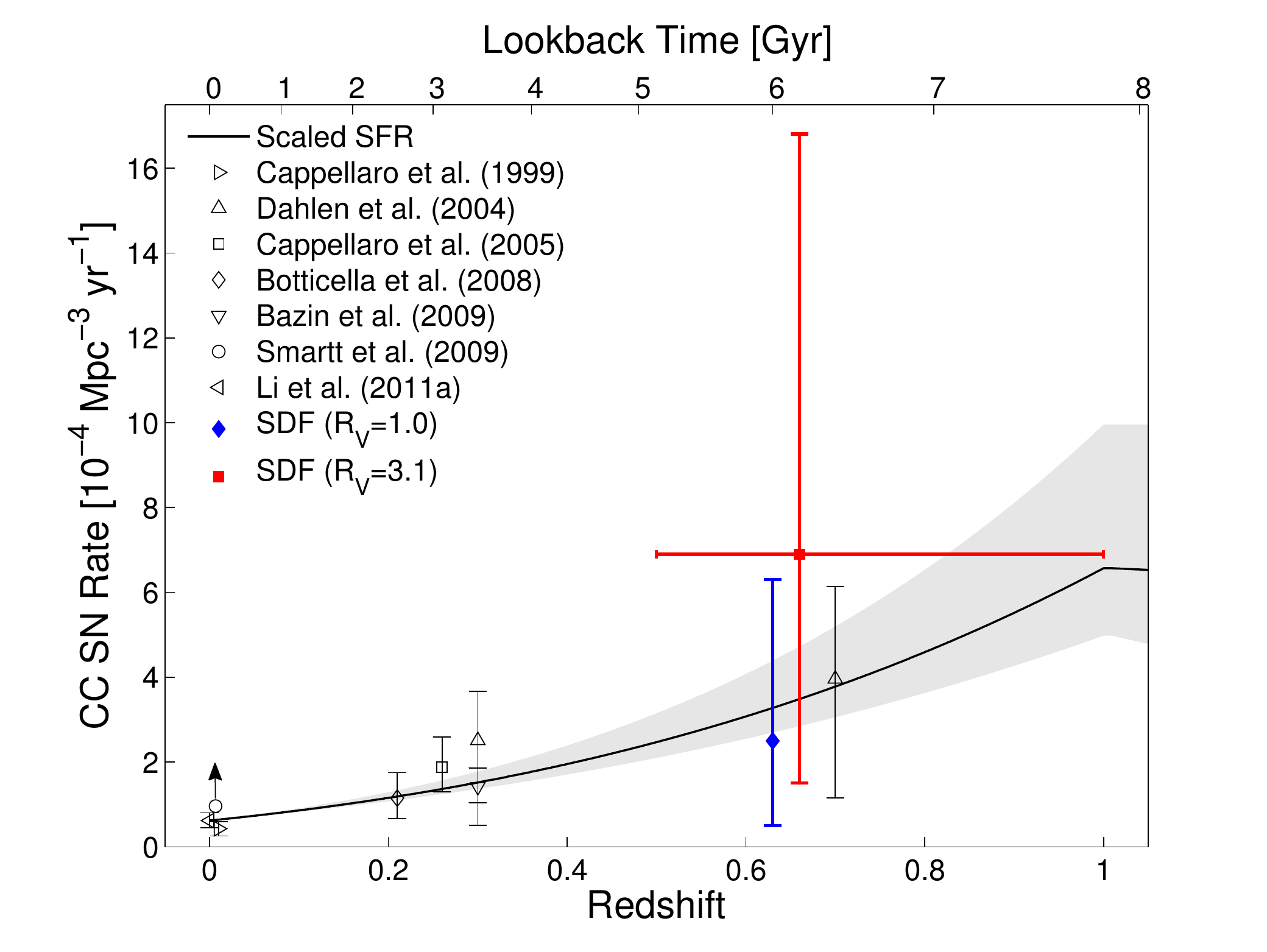}
\caption{CC~SN rate from the SDF (filled square) compared to rates from the literature: \citealt{cappellaro1999} (right-facing triangle), \citealt{dahlen2004} (upward triangles), \citealt{cappellaro2005} (square), \citealt{botticella2008} (diamond), \citealt{bazin2009} (downward triangle), lower limit from \citealt{smartt2009mnras} (circle), and \citealt{li2011rates} (left-facing triangle). As in Fig.~\ref{fig:rates_Ia_upperlim}, the SFH from Y08 has been scaled to fit the low-$z$ data. All vertical error bars from the literature are 1$\sigma$ uncertainties. The horizontal error bar indicates the SDF redshift bin.}
\label{fig:rates_CC}
\end{figure}

As in the previous section, the visibility time of each CC~SN subtype was derived using Equation~\ref{eq:vistime}, but without stretch.
In the present case, $M_B$ was limited to the $2\sigma$ range around the peak magnitude of each subtype.
The probability for $A_V$ was drawn from a one-sided Gaussian PDF centred on 0 with a dispersion of $\sigma=0.93$, and the probability for $M_B$ was drawn from the LF of each subtype.
Without host-galaxy extinction, the rates of each CC~SN subtype (in units of $10^{-4}$ SNe yr$^{-1}$ Mpc$^{-3}$) are 1.3 for SNe~II-P/L, 0.4 for SNe~Ib/c/IIb, and 0.1 for SNe~IIn.
This results in an overall rate of $R_{\textrm{CC}}(\langle z \rangle = 0.66) = 1.8^{+2.0}_{-1.4} \times 10^{-4}$ SNe yr$^{-1}$ Mpc$^{-3}$.
Once host-galaxy extinction is added, the rates of each CC~SN subtype (in the same units) become 5.8 for SNe~II-P/L, 0.9 for SNe~Ib/c/IIb, and 0.2 for SNe~IIn.
This yields an overall rate of $R_{\textrm{CC}}(\langle z \rangle = 0.66) = 6.9^{+7.7}_{-5.4} \times 10^{-4}$ SNe yr$^{-1}$ Mpc$^{-3}$.
After correcting for the fraction of CC~SNe missed due to high-redshift dust \citep{2007MNRAS.377.1229M}, the final CC~SN rate is $R_{\textrm{CC}}(\langle z \rangle = 0.66) = 6.9^{+9.9}_{-5.4} \times 10^{-4}$ SNe yr$^{-1}$ Mpc$^{-3}$.
This value is consistent with both the rate derived by D04 in this redshift bin and with the scaled Y08 SFH at that redshift, as shown in Fig.~\ref{fig:rates_CC}.
We present a summary of CC~SN rates from the literature, along with our measured rate at $\langle z \rangle = 0.66$, in Table~\ref{table:rates_lit_CC}.

The statistical and systematic uncertainties affecting the SN~Ia and CC~SN rates are summarised in Table~\ref{table:errs}.

\section{THE TYPE IA SUPERNOVA DELAY-TIME DISTRIBUTION}
\label{sec:DTD}

\begin{table}
\begin{minipage}{\textwidth}
\caption{SN rate uncertainty percentages}
\label{table:errs}
\begin{tabular}{l|c|c|c|}
\hline
\hline

Uncertainty & $0.5<z<1.0$ & $1.0<z<1.5$ & $1.5<z<2.0$ \\
\hline
\multicolumn{4}{c}{SN~Ia rates} \\
\hline

Poisson & $+25/-23$ & $+23/-19$ & $+42/-31$ \\
Classification & $+28/-45$ & $+3/-27$ & $+0/-18$ \\
High-$z$ dust & $+3/-0$ & $+6/-0$ & $+9/-0$ \\
Extinction law$^a$ & $+0/-6$ & $+0/-26$ & $+0/-43$ \\
Total & $+41/-51$ & $+29/-33$ & $+51/-36$ \\

\hline
\multicolumn{4}{c}{CC~SN rates} \\
\hline

Poisson & $+37/-40$ &  &  \\
Classification & $+107/-67$ &  &  \\
High-$z$ dust & $+32/-0$ &  &  \\
Extinction law$^a$ & $+0/-52$ &  &  \\
Total & $+145/-78$ &  &  \\

\hline
\multicolumn{4}{l}{All uncertainties are reported as percentages of the rates.} \\
\multicolumn{4}{l}{$^a$This uncertainty is not added to the final quoted rates, but is} \\
\multicolumn{4}{l}{propagated directly into the derivation of the SN~Ia DTD} \\
\multicolumn{4}{l}{(see Section~\ref{subsubsec:rv}).}

\end{tabular}
\end{minipage}
\end{table}

In this section we make use of our measured SN~Ia rates, together with published rates at various redshifts, to recover the DTD.
The different SN~Ia rates used in our fits are presented in Table~\ref{table:rates_lit_Ia}.
Where necessary \citep{cappellaro1999, hardin2000, 2002ApJ...577..120P, 2003ApJ...599L..33M, blanc2004}, rates from the literature have been converted to volumetric rates using the redshift-dependent luminosity density function from B08 (see Equation~\ref{eq:Ldense}).
Furthermore, all rates have been converted to $h=0.7$.
We make use of all the SN~Ia rate measurements in Table~\ref{table:rates_lit_Ia}, except for the \citet{2006ApJ...637..427B} measurements, which have been superseded by \citet{2010ApJ...723...47R}; the \citet{kuznetsova2008goods} measurements, which make use of much the same data as D08; and our initial results reported by P07b, which are superseded by the present results.
In total, there are 36 SN~Ia rate measurements, of which 31 are at $z<1$ and 5 at $z>1$.

\begin{figure}
\includegraphics[width=0.5\textwidth]{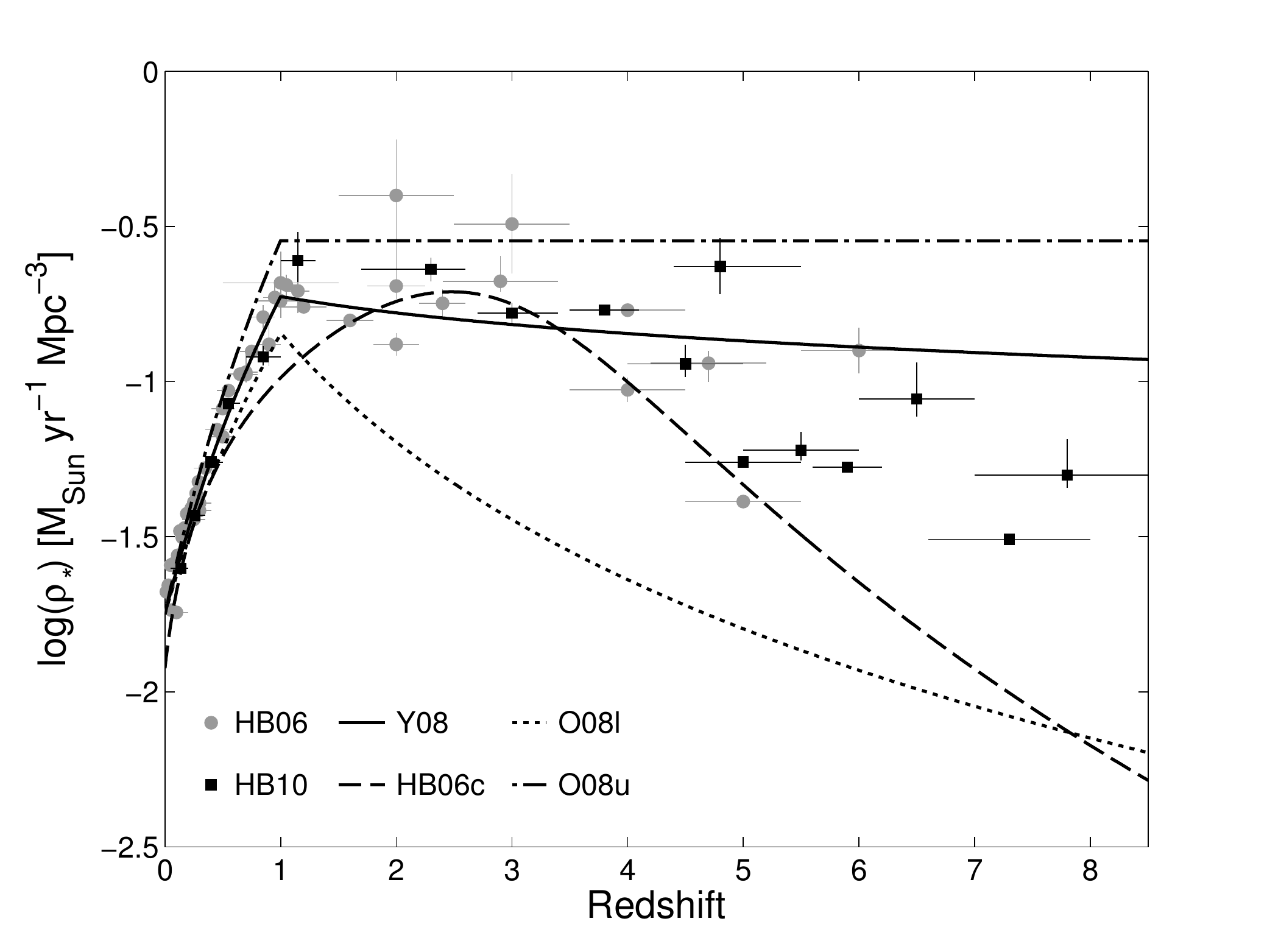}
\caption{SFH measurements and parametrizations. Measurements include the compilation from HB06 (circles) and additional data presented by \citet{horiuchi2010} (squares; here noted as HB10). The dashed line represents the \citet{2001MNRAS.326..255C} parametrization with parameter values from HB06. The solid (Y08), dot-dashed (O08u), and dotted (O08l) lines are power laws with parameter values as detailed in Table~\ref{table:SFH_beta_params}.}
\label{fig:sfh}
\end{figure}

We recover the DTD by convolving different trial DTDs with various SFH fits from the literature, resulting in a model SN~Ia rate evolution.
One such SFH is the one presented in fig.~2 of HB06 (HB06c).
Other recent estimates of the SFH and their parametrizations (e.g., Y08 and O08) can be approximated by broken power laws, with a break at $z=1$, and various power-law indices above and below the break.
To test the systematic uncertainty in our DTD derivation produced by this range of possible SFHs, we parametrize the SFH as being proportional to $(1+z)^\gamma$, with $\gamma$ in the range 3--4 at $z<1$, a break at $z=1$, and $\gamma$ values in the range  $-2$ to 0 at $z>1$.
This range of parametrizations covers most of the SFHs that have been recently proposed. 
The indices, breaks, and normalizations of each SFH at $z=0$ are collected in Table~\ref{table:SFH_beta_params}, and the resulting SFHs are shown in Fig.~\ref{fig:sfh}.
For a given SFH, variations of the normalization will translate to inverse scalings of the
amplitude of the best-fitting DTD, without affecting the DTD shape, which is our main interest here. 
There remains considerable debate among different authors as to the amount and the redshift dependence of extinction corrections in SFH estimates (see, e.g., \citealt{2010arXiv1006.4360B, 2010Natur.468...49R}). 
Different extinction correction choices can shift much or all of a SFH curve up or down by up to a factor of two (F. Mannucci, private communication). 
To account for this uncertainty, we also calculate the range in DTD amplitude that results
when the SFH varies between the extreme case of O08u and the HB06c level.

\begin{figure}
\includegraphics[width=0.5\textwidth]{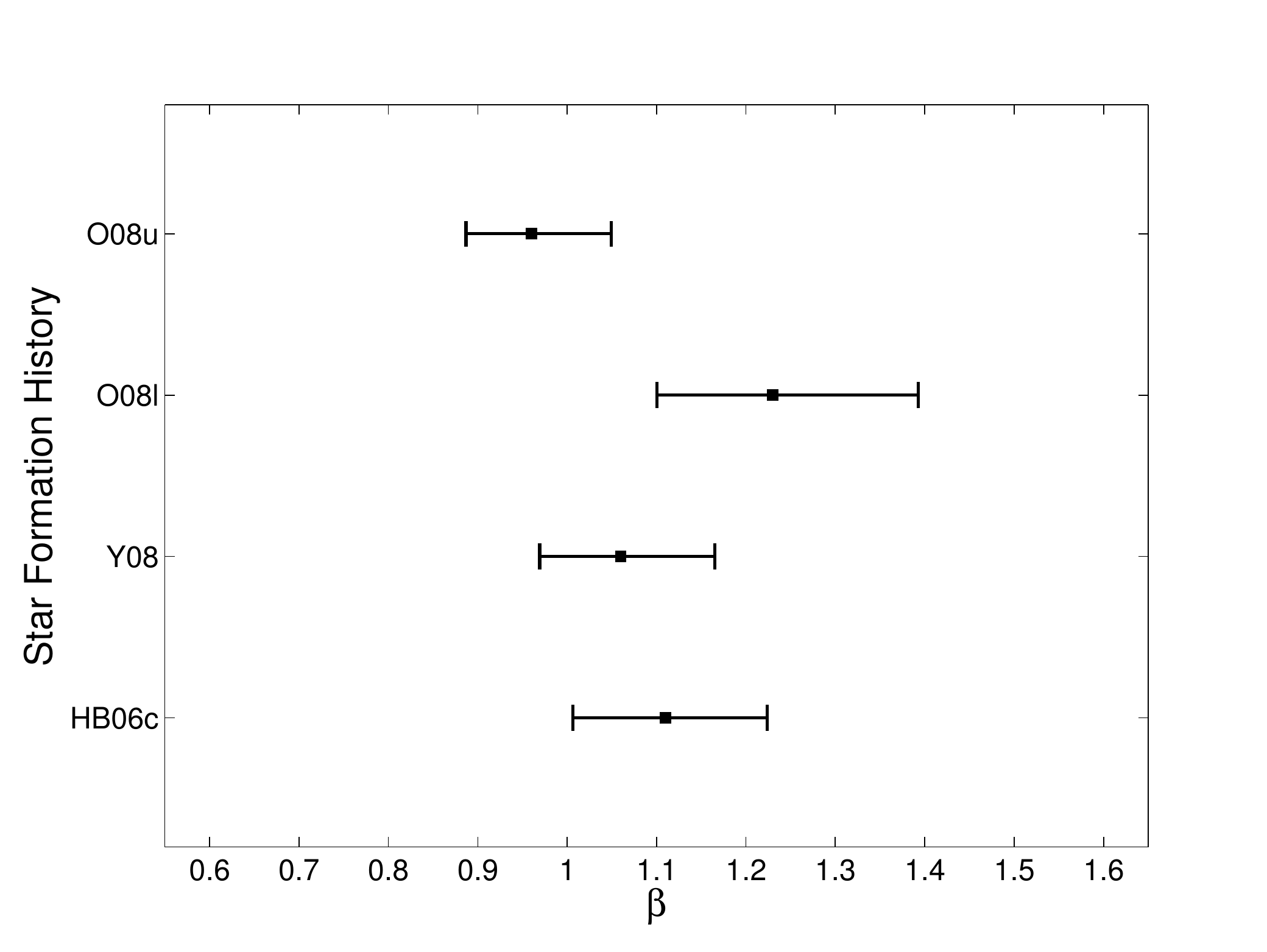}
\caption{Best-fitting values and 68 per cent statistical uncertainties of the slope $\beta$ of a power-law DTD of the form $\Psi(t) = \Psi_1(t/1~\rm{Gyr})^{\beta}$, when convolved with various SFHs, as marked. See Table~\ref{table:SFH_beta_params} for SFH abbreviations and parameters.}
\label{fig:beta}
\end{figure}

\begin{figure}
\includegraphics[width=0.5\textwidth]{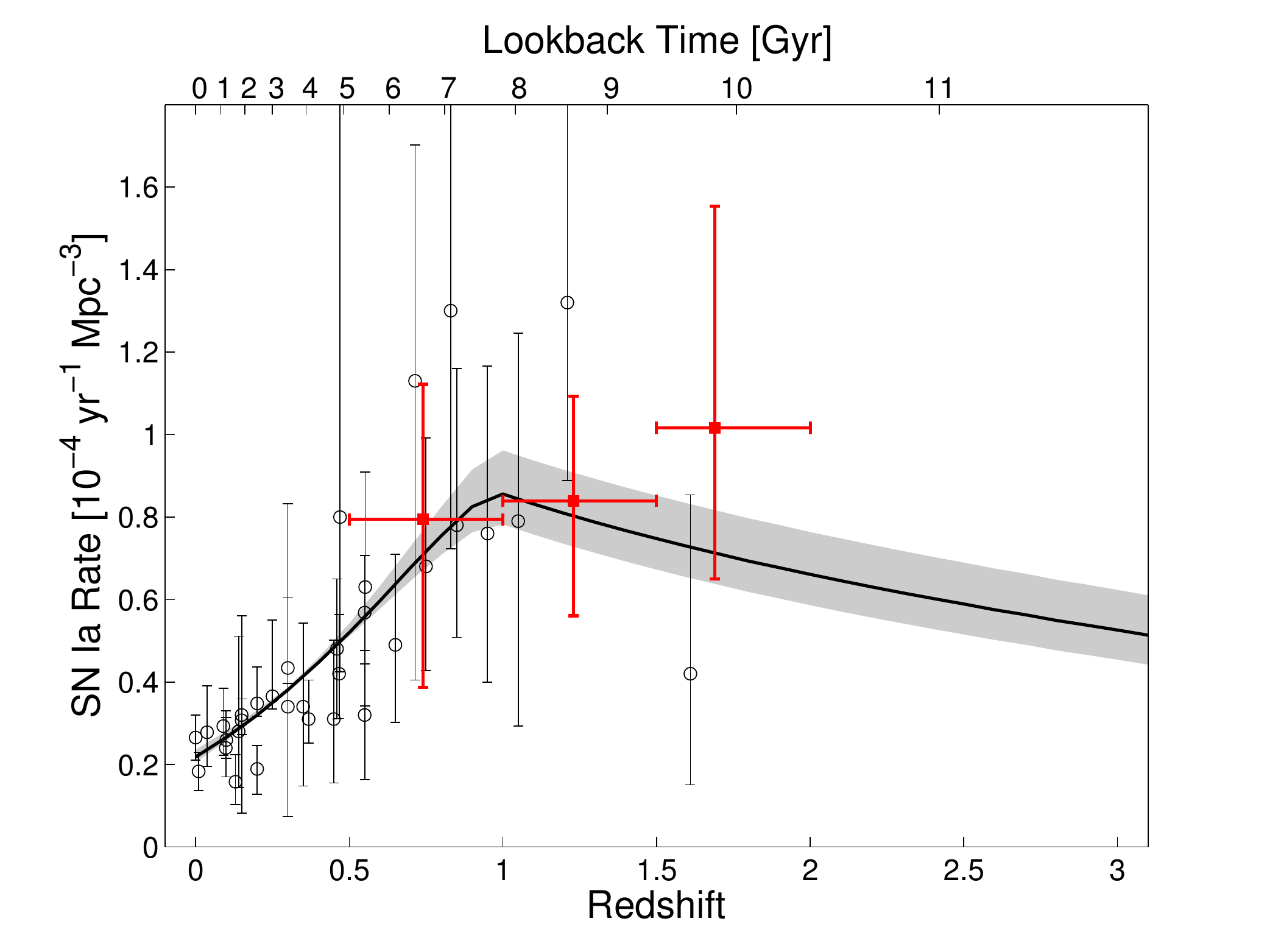} \\
\includegraphics[width=0.5\textwidth]{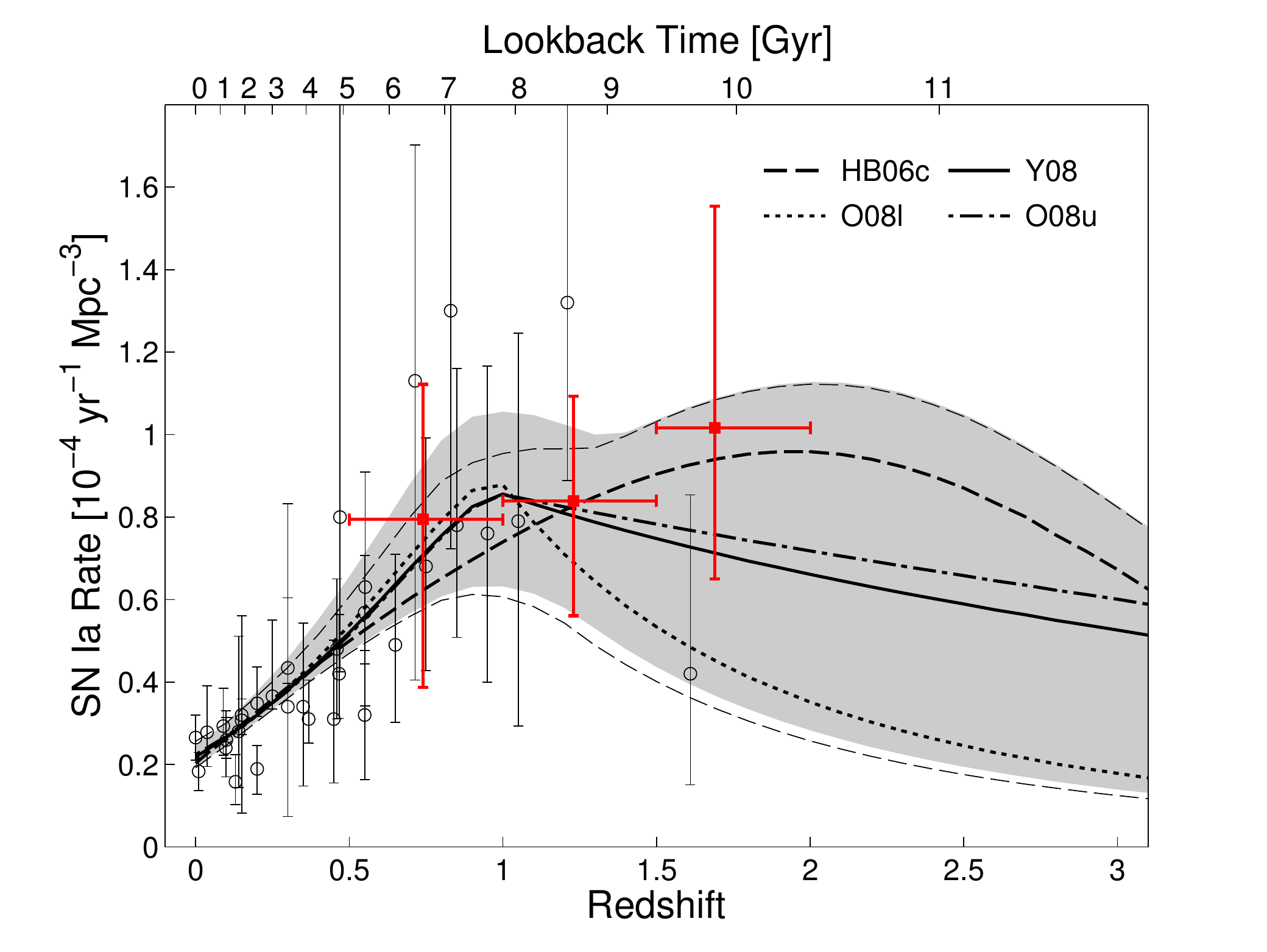}
\caption{Top panel: Observed SN~Ia rates compared to prediction from convolution of the Y08 SFH with a best-fitting power-law DTD of the form $\Psi(t) = \Psi_1(t/1~\rm{Gyr})^{\beta}$ (solid line).
Non-independent measurements, which are therefore excluded from the fits, are not plotted -- \citet{kuznetsova2008goods} and P07b, which are superseded by D08 and this work, respectively.
The shaded area is the confidence region resulting from the 68 per cent statistical uncertainty of $\beta$ from the convolution of the DTD with the Y08 SFH fit. Bottom panel: Same as top panel, but for each of the SFHs in Table~\ref{table:SFH_beta_params}, and showing the combined effect of the 68 per cent statistical uncertainties of $\beta$. The thin dashed lines indicate the 68 per cent uncertainty region produced without the new SDF measurements.}
\label{fig:dtd_power}
\end{figure}

Throughout this derivation we assume a \textquoteleft diet-Salpeter\textquoteright\ initial mass function (IMF; \citealt{2003ApJS..149..289B}).
This IMF assumption means that the SFHs of HB06 and Y08, who assumed a \citet{1955ApJ...121..161S} IMF, are scaled down by a factor of $0.7$.
We then use the $\chi^2$ statistic to find the best-fitting values of the parameters of the DTD, along with their statistical 68 and 95 per cent confidence regions, defined as the projections of the $\Delta\chi^2=1$ contour on the parameter axes.
Below we present reduced $\chi^2$ values, denoted by $\chi^2_r$.
To the statistical uncertainty of the parameters we add the systematic uncertainty that originates in the shapes of the different SFHs.
Finally, for each model we calculate the number of SNe~Ia per formed stellar mass, integrated over a Hubble time.

We first test a power-law DTD of the form $\Psi(t)=\Psi_1(t/1~\rm{Gyr})^{\beta}$.
Such a power law, with $\beta \approx -1$, is generic to the DD scenario, where two WDs merge due to loss of energy and angular momentum to gravitational waves (see, e.g., \citealt{maoz2010clusters}).
Several recent experiments, in different environments and different redshifts, have indeed found best-fitting DTDs consistent with this form \citep{2008PASJ...60.1327T, maoz2010magellan, maoz2010clusters, maoz2010loss}.
The DTD is set to zero for the first 40~Myr, until the formation of the first WDs.
We fit for the normalization $\Psi_1$ and the slope $\beta$.
Based on the Y08 SFH fit, we find best-fitting values of $\beta = 1.1 \pm 0.1 (0.2)$, where the statistical uncertainties are the 68 and 95 (in parentheses) per cent confidence regions, respectively. 
The range of SFHs tested here adds a systematic uncertainty of $^{+0.17}_{-0.10}$.
The best-fitting values of $\beta$ for all four SFHs, with their respective reduced $\chi^2$ values, appear in Table~\ref{table:SFH_beta_params}.
These best-fitting values result in reduced $\chi^2_r$ values of $0.7$ to $0.8$, for 34 degrees of freedom (DOF) for all SFH fits.
The number of SNe~Ia per formed stellar mass, integrated over a Hubble time, lies in the range $N_{\textrm{SN}}/M_*=(0.5$--$1.5) \times 10^{-3}~\rm{M_\odot}^{-1}$, where this range accounts for the statistical uncertainties in $\beta$ and $\Psi_1$.
However, the uncertainty in the normalization of the SFH is such that this range might easily be higher by a factor of two (F. Mannucci, private communication).
The best-fitting values of $\beta$ are presented in Fig.~\ref{fig:beta}, and the resulting SN~Ia rate evolution tracks are presented in Fig.~\ref{fig:dtd_power}.

\begin{figure}
\includegraphics[width=0.5\textwidth]{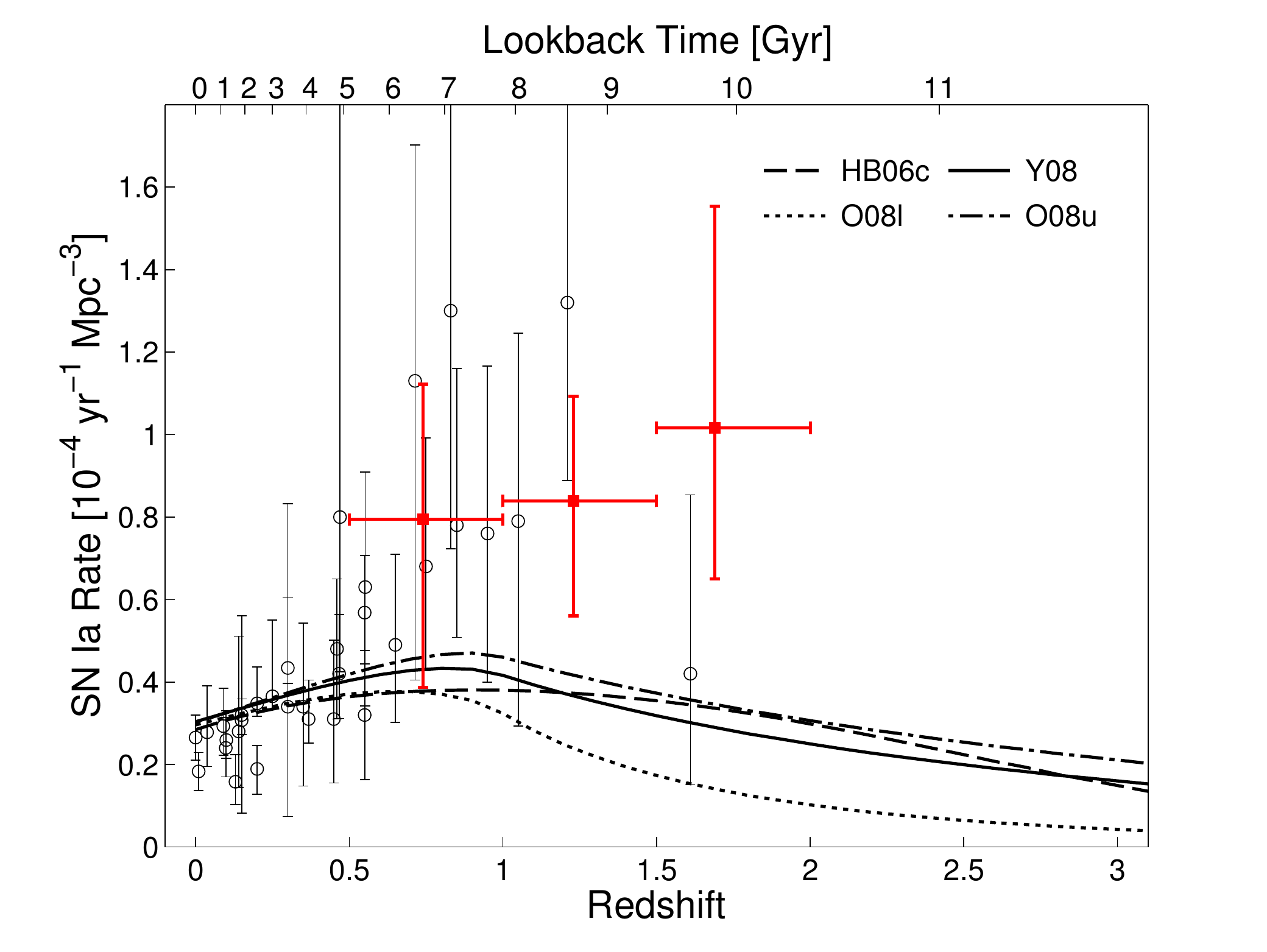} \\ 
\includegraphics[width=0.5\textwidth]{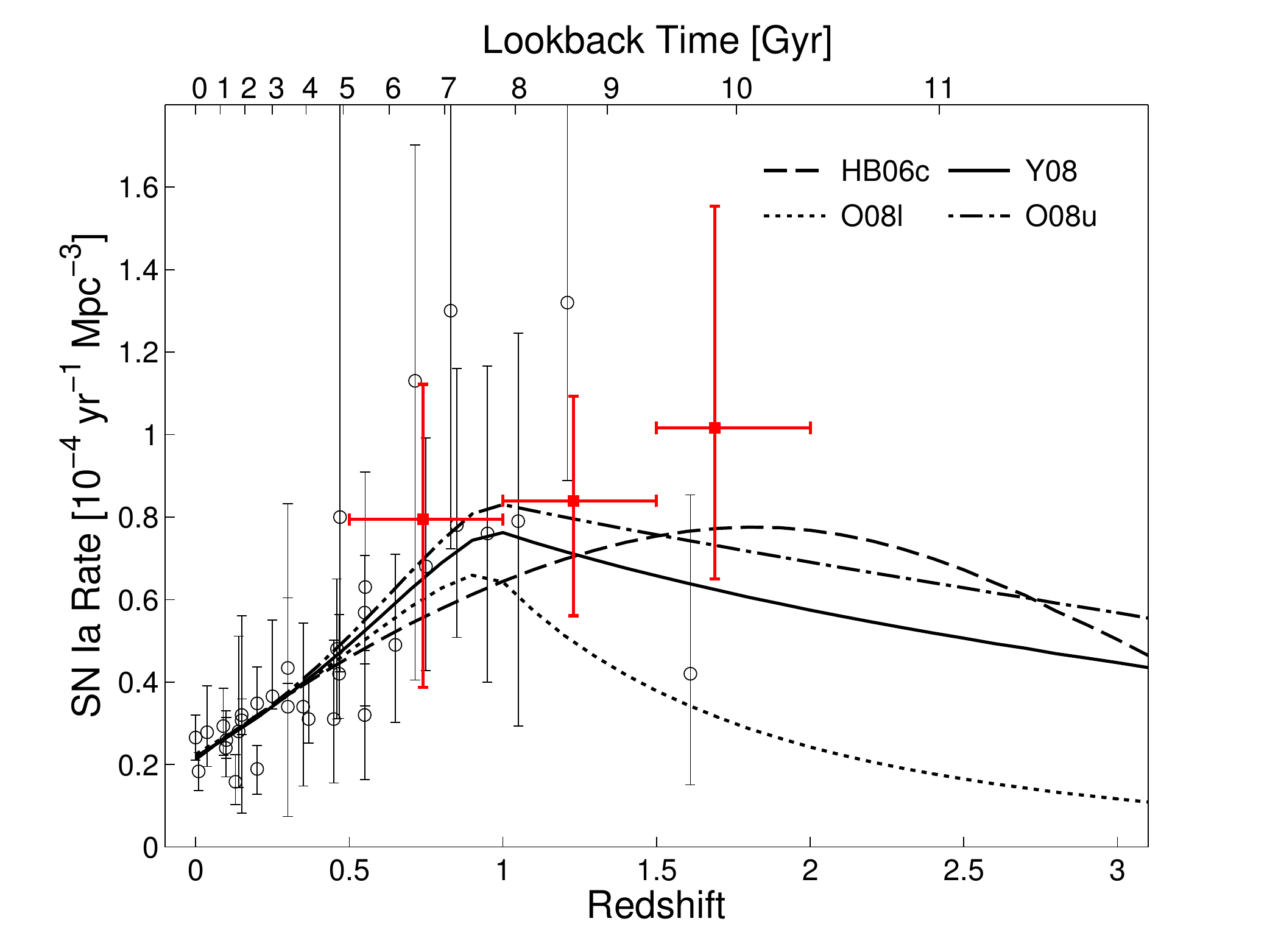} \\  
\includegraphics[width=0.5\textwidth]{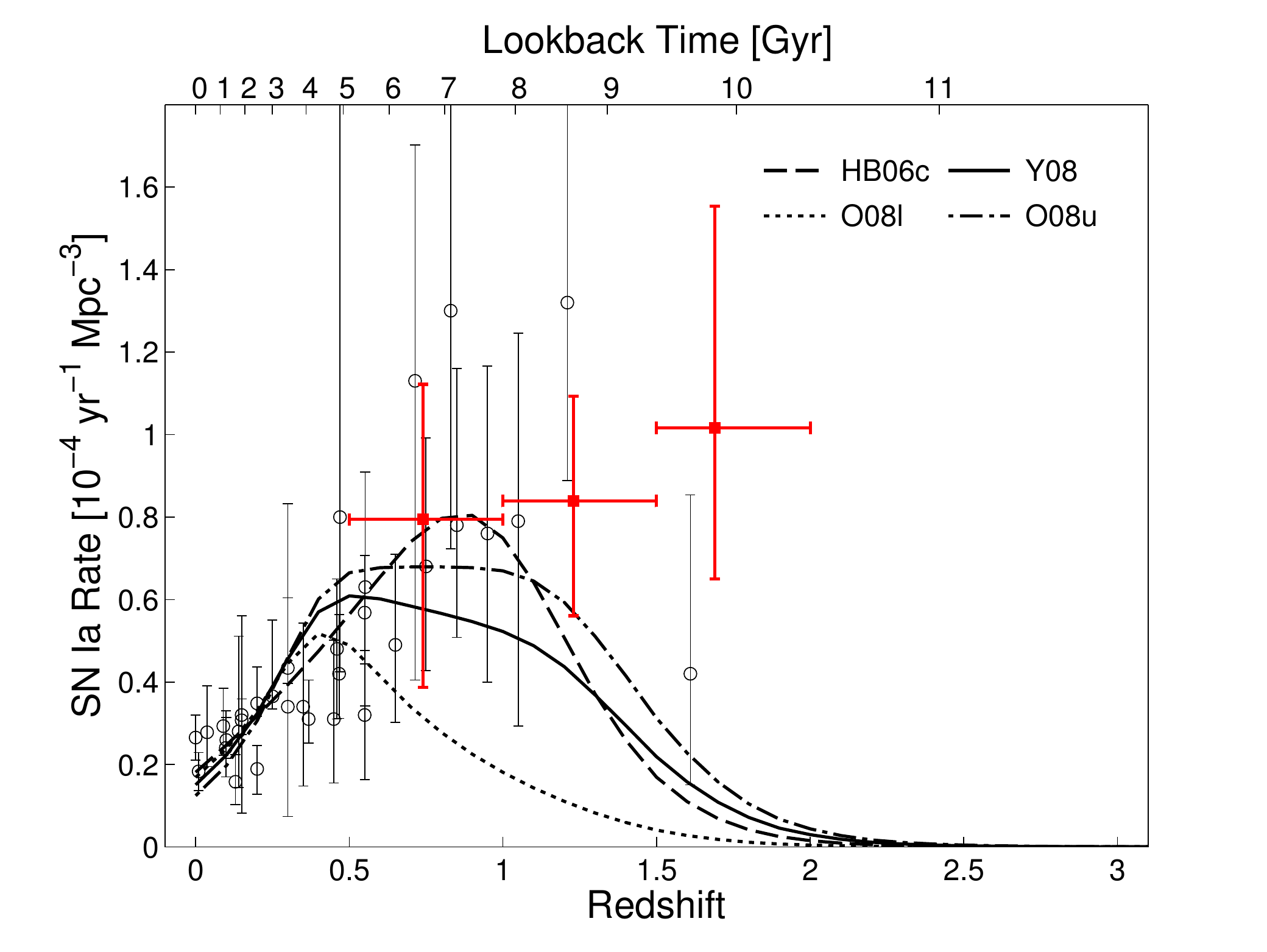}
\caption{Observed SN~Ia rates compared to predictions from convolution of the SFHs in Table~\ref{table:SFH_beta_params} with a best-fitting (top) power-law DTD of the form 
$\Psi(t) = \Psi_1(t/1~\rm{Gyr})^{-1/2}$; (centre) broken power-law DTD of the form 
$\Psi(t) \propto t^{-1/2}$ up to $t_c$, and 
$\Psi(t) \propto t^{-1}$ afterward; and (bottom) D08 Gaussian DTD.
Symbols are as marked.}
\label{fig:dtds}
\end{figure}

\begin{table*}
\begin{minipage}{\textwidth}
\center
\hspace{1.5in}\parbox{6.3in}{\caption{Star formation histories and resultant best-fitting DTD parameters.}\label{table:SFH_beta_params}}
\begin{tabular}{l|l|c|c|c|c|}
\hline
\hline
\multicolumn{2}{c}{SFH} & \multicolumn{2}{c}{Power-law DTD} & \multicolumn{2}{c}{Broken power-law DTD$^a$}\\

\hline

Ref.$^b$ & Parametrization$^c$ & $\beta^d$ & $\chi^2/\rm{DOF}$ & $t_c$ [Gyr]$^e$ & $\chi^2/\rm{DOF}$ \\

\hline
\multicolumn{6}{c}{Galactic dust extinction: $R_V=3.1$} \\
\hline

HB06c & \citet{2001MNRAS.326..255C} with values from HB06 & $1.11^{+0.10(0.24)}_{-0.10(0.20)}$ & 0.7 & $0.05^{+0.14(0.70)}_{-0.01(0.01)}$ & 0.8 \\

Y08 & $S(0)=17.8,~\gamma_1=3.4,~z_b=1,~\gamma_2=-0.3$ & $1.1 \pm 0.1(0.2)$ & 0.7 & $0.05^{+0.17(0.72)}_{-0.01(0.01)}$ & 0.7 \\

O08l & $S(0)=17.8,~\gamma_1=3,~z_b=1,~\gamma_2=-2$ & $1.23^{+0.16(0.40)}_{-0.13(0.25)}$ & 0.8 & $0.05^{+0.06(0.32)}_{-0.01(0.01)}$ & 0.9 \\

O08u & $S(0)=17.8,~\gamma_1=4,~z_b=1,~\gamma_2=0$ & $0.96^{+0.09(0.17)}_{-0.07(0.16)}$ & 0.7 & $0.18^{+0.30(^f)}_{-0.14(0.14)}$ & 0.7 \\

\hline
\multicolumn{6}{c}{Low dust extinction: $R_V=1$} \\
\hline

HB06c & \citet{2001MNRAS.326..255C} with values from HB06 & $1.02^{+0.10(0.20)}_{-0.09(0.19)}$ & 0.8 & $0.04^{+0.43(1.39)}_{-0.00(0.00)}$ & 0.8 \\

Y08 & $S(0)=17.8,~\gamma_1=3.4,~z_b=1,~\gamma_2=-0.3$ & $0.99^{+0.09(0.18)}_{-0.08(0.17)}$ & 0.7 & $0.04^{+0.43(1.24)}_{-0.00(-0.00)}$ & 0.7 \\

O08l & $S(0)=17.8,~\gamma_1=3,~z_b=1,~\gamma_2=-2$ & $1.18^{+0.13(0.28)}_{-0.22(0.26)}$ & 0.7 & $0.05^{+0.06(0.29)}_{-0.01(-0.01)}$ & 0.8 \\

O08u & $S(0)=17.8,~\gamma_1=4,~z_b=1,~\gamma_2=0$ & $0.90^{+0.08(0.15)}_{-0.07(0.15)}$ & 0.7 & $0.55^{+0.21(^f)}_{-0.29(-0.51)}$ & 0.8 \\

\hline

\multicolumn{6}{l}{$^a \Psi(t) \propto t^{-1/2}$ power law at $t<t_c$, and $\Psi(t) \propto t^{-1}$ at $t>t_c$.} \\
\multicolumn{6}{l}{$^b$SFH references: HB06c -- \citet{2006ApJ...651..142H}; Y08 -- \citet{2008ApJ...683L...5Y}; O08l and O08u -- \citet{2008PASJ...60..169O}.} \\
\multicolumn{6}{l}{$^c$Except for HB06c, all other SFHs are parametrized as broken power laws of the form $S(z)=S(0)(1+z)^{\gamma_i}$,} \\
\multicolumn{6}{l}{with $\gamma_1$ at $z<z_b$, and $\gamma_2$ at $z>z_b$. $S(0)$ is in units of $10^{-3}~\rm{M_\odot}$~yr$^{-1}$~Mpc$^{-3}$.} \\
\multicolumn{6}{l}{$^d$The first and second errors (in parentheses) are 68 and 95 per cent confidence regions, respectively, for the slope $\beta$} \\
\multicolumn{6}{l}{of the power-law DTD.} \\
\multicolumn{6}{l}{$^e$The first and second errors (in parentheses) are 68 and 95 per cent confidence regions, respectively, for $t_c$, the break} \\
\multicolumn{6}{l}{between a $t^{-1/2}$ and a $t^{-1}$ power law.} \\
\multicolumn{6}{l}{$^f$As the O08u SFH was found to be compatible with a $t^{-1/2}$ power-law DTD at all times, there is no 95 per cent} \\
\multicolumn{6}{l}{upper limit for this measurement.}

\end{tabular}
\end{minipage}
\end{table*}

Whereas the power law discussed above extends all the way back to $t=40$ Myr, it is possible that at early times the DTD is dictated not by the WD merger rate, but rather by the supply of progenitor systems. 
\citet*{2008ApJ...683L..25P} have suggested a $t^{-1/2}$ power-law DTD, which is proportional to the formation rate of WDs.
A pure $t^{-1/2}$ power law, convolved with the HB06c, Y08, and O08l SFHs, produces fits with a minimal $\chi^2_r>1.5$ for 35 DOF, ruling out this model at the 95 per cent confidence level.
The O08u SFH results in a fit with a minimal $\chi^2_r$ value of 1.4, which is marginally acceptable.
\citet{2009A&A...501..531M} also argue against this model, as it does not reproduce the observed G-dwarf metallicity distribution in the solar vicinity (see their fig.~7).
The resulting SN~Ia rate evolution tracks are presented in the top panel of Fig.~\ref{fig:dtds}.

Another possibility is that the DTD is controlled by the WD formation rate up to some characteristic time $t_c$, beyond which newly formed WDs no longer have the combined mass to constitute SN~Ia progenitors; from this point on only the merger rate sets the DTD.
The \citet{2005A&A...441.1055G} DD3-close model, for example, is such a broken power law, with $t^{-1/2}$, $t^{-1.3}$, and a break at $t_c=0.4$ Gyr.
This value for $t_c$ corresponds to the lifetime of $3~\rm{M_\odot}$ stars.
A larger value of $t_c$ would imply that WD binaries with a smaller primary mass can explode as SNe~Ia, and ultimately contribute to the observed SN~Ia rate.
We therefore investigate whether the SN~Ia rate data may be fit by a broken power law behaving as $t^{-1/2}$ at $t<t_c$, and as $t^{-1}$ thereafter.
Fitting for $t_c$ and the normalization $\Psi_1$, we find that $t_c$ lies in the 68 per cent confidence range 0.04--0.48 Gyr.
As a $t^{-1/2}$ power-law DTD at all times is still an acceptable option for the O08u SFH, we cannot constrain $t_c$ at the 95 per cent confidence level for that SFH.
However, the other SFHs suggest that $t_c$ may be lower than $\sim 0.8$ Gyr, at the 95 per cent confidence level.
The best-fitting parameters result in reduced $\chi^2_r$ values of 0.7--0.9, for 34 DOF for all SFH fits.
The integrated number of SNe~Ia per stellar mass formed resulting from this DTD lies in the range $N_{\textrm{SN}}/M_*=(0.5$--$1.0) \times 10^{-3}~\rm{M_\odot}^{-1}$, where the uncertainty derives from the normalizations of the SFHs and from the statistical uncertainty $t_c$. 
This range is similar to that obtained with the single power-law DTD.
The best-fitting parameters, along with reduced $\chi^2$ values, are presented in Table~\ref{table:SFH_beta_params}, and the resulting SN~Ia rate evolution tracks are presented in the centre panel of Fig.~\ref{fig:dtds}.

Finally, D04, D08, and \citet{2004ApJ...613..200S, 2010ApJ...713...32S} advocate a Gaussian DTD with parameters $\tau=3.4$ Gyr and $\sigma=0.2\tau$.
D04 used the SFH determined by \citet{2004ApJ...600L.103G} in order to derive the parameters of the Gaussian DTD.
As we use different SFHs, we leave the normalization of the Gaussian, $\Psi_{\rm{G}}$, as a free parameter. 
The best-fitting value, derived with the HB06c SFH fit, has a minimal $\chi^2_r=1.1$ for 35 DOF.
However, all of the other SFHs result in best-fitting values with minimal $\chi^2_r>1.5$, ruling out this model at the 95 per cent confidence level.
The resulting SN~Ia rate evolution tracks are plotted in the bottom panel of Fig.~\ref{fig:dtds}.

We propagate the systematic uncertainty brought about by the possibility that the extinction law in the immediate vicinity of SNe is different from the Galactic average by fitting the different DTDs to the SN~Ia measurements derived with $R_V=1$, as detailed in Section~\ref{subsubsec:rv}.
The \citet{2008ApJ...683L..25P} $t^{-1/2}$ power-law and D08 Gaussian DTDs are still excluded, at the 95 per cent confidence level, when using the same SFHs as detailed above.
The resulting best-fitting parameter for the $t^{-1}$ power-law and broken DTDs are presented in Table~\ref{table:SFH_beta_params}.
The lower $R_V$ value adds a systematic uncertainty of $^{+0.00}_{-0.07}$ to the best-fitting value of $\beta$ for the Y08 SFH.
The overall best-fitting value of $\beta$ for the Y08 SFH is thus $\beta = 1.1 \pm 0.1 (0.2) ~\rm{(statistical)} \pm 0.17 ~\rm{(systematic)}$, where the statistical errors are the 68 and 95 per cent (in parentheses) confidence regions, respectively. 
The upper 68 per cent limit on $t_c$ for the broken-power-law DTD rises to 0.76 Gyr, and the 95 per cent upper limit afforded by the Y08, HB06c, and O08l SFHs rises to 1.43 Gyr.


\section{THE TYPE IA SUPERNOVA RATE AT REDSHIFT $>2$ AND COSMIC IRON ACCUMULATION}
\label{sec:iron}

Our analysis, above, has provided the most precise determination to date of the SN~Ia rate at $1<z<2$.
As seen in the bottom panel of Fig.~\ref{fig:dtd_power}, the best-fitting power-law DTD, convolved with each SFH, can also be used to predict the SN~Ia rate at $z>2$. 
The shaded regions in the figure show the uncertainty regions produced by the statistical and systematic uncertainties of the DTD slope $\beta$, where the statistical uncertainties result from the SN~Ia rate measurements, and the systematic uncertainties result from the uncertainty in the slope of the SFHs at $z<z_b$.

\begin{figure*}
\begin{minipage}{\textwidth}
\center
\begin{tabular}{cc}
\includegraphics[width=0.5\textwidth]{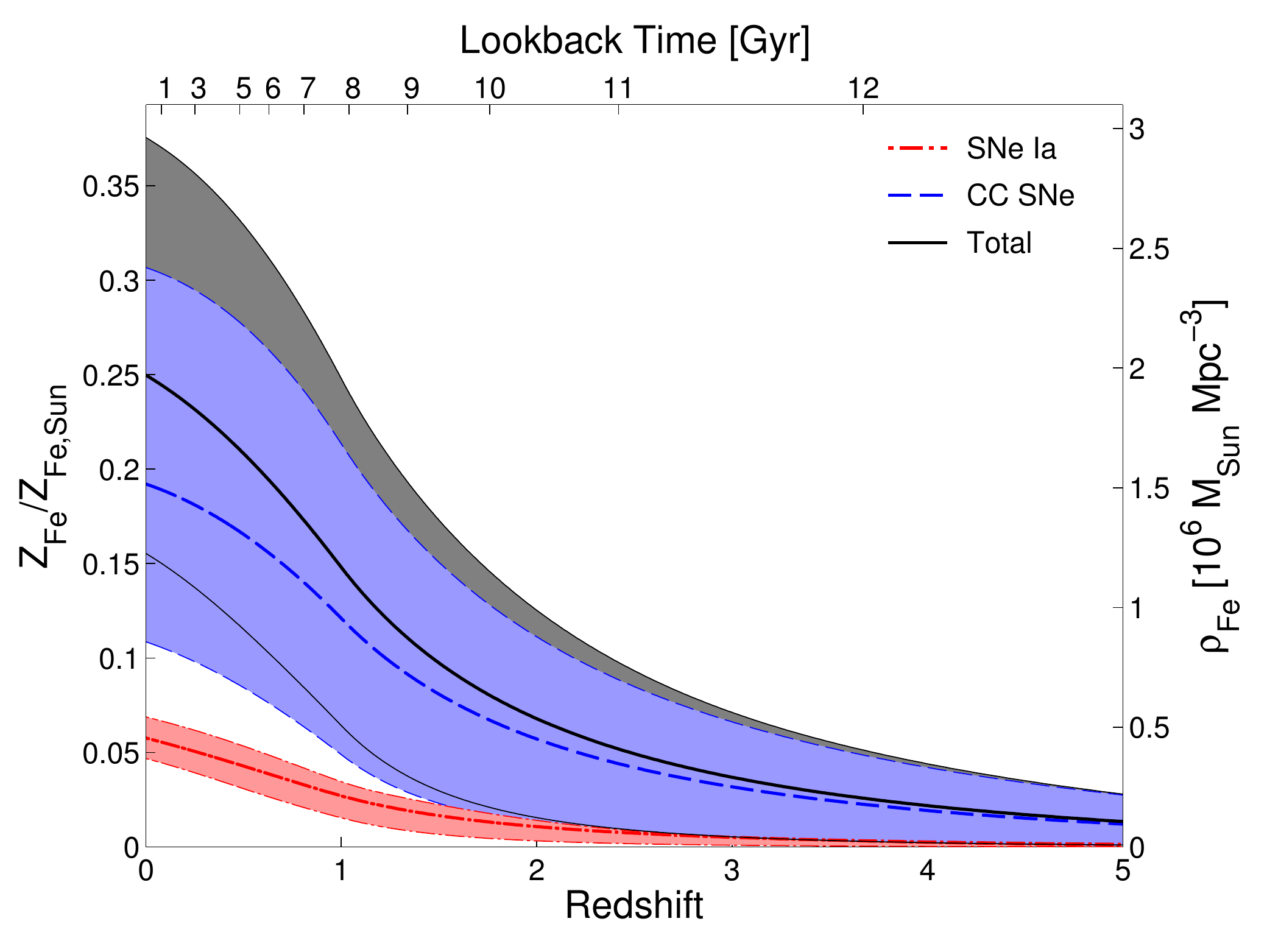} & 
\includegraphics[width=0.5\textwidth]{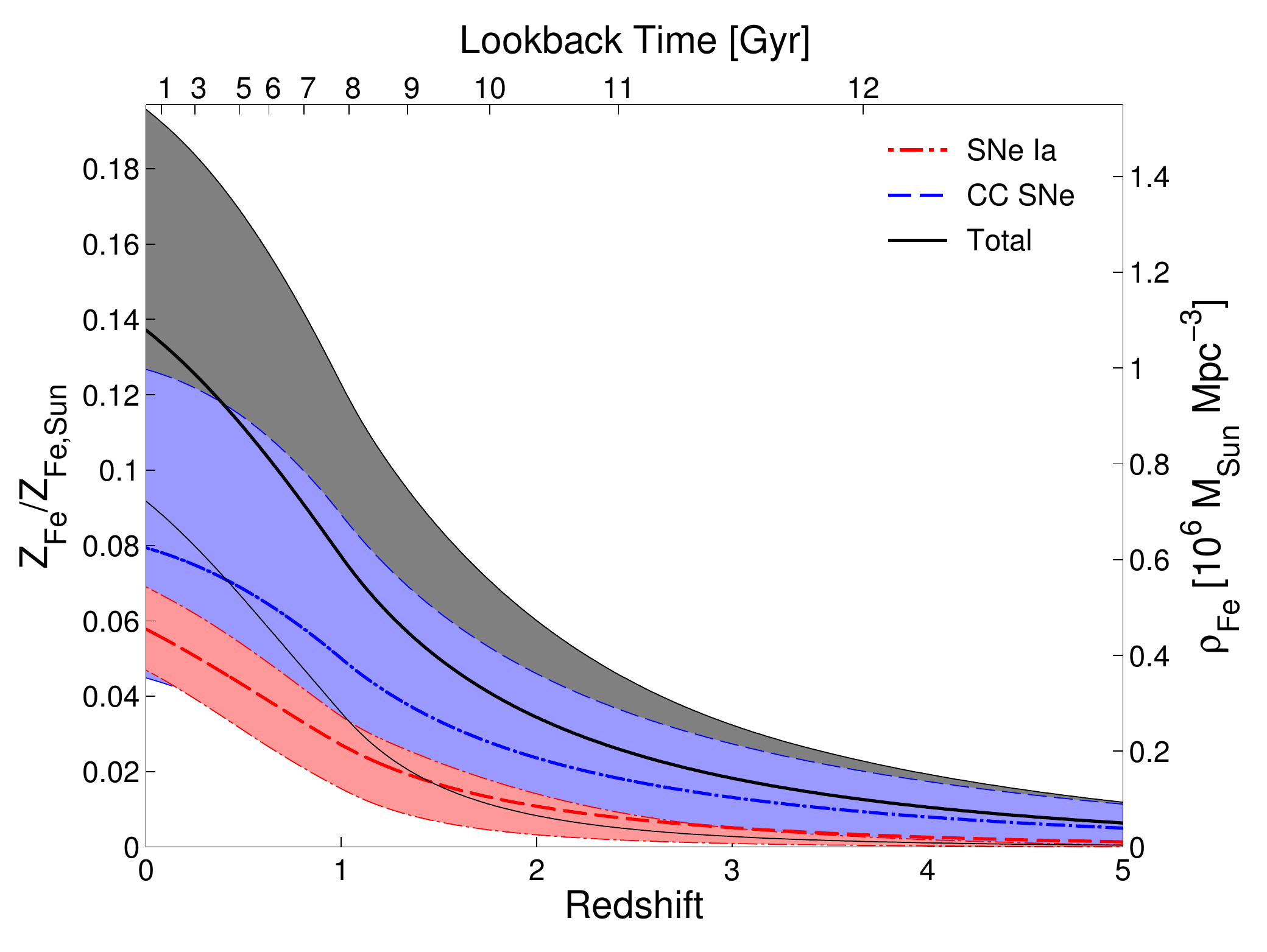}
\end{tabular}
\caption{Cosmic iron density as a function of redshift. In both panels the SN~Ia contribution is denoted by the dot-dashed line, the CC~SN contribution by the dashed line, and the total amount of iron by the solid line. The dark region around the SN~Ia contribution is the systematic and 68 per cent statistical uncertainty introduced by the SFH fits and the SN~Ia rate measurements, respectively. 
The shaded region around the CC~SN contribution is the result of the systematic uncertainty in the SFH fits alone. The dark region around the total iron density curve is the uncertainty introduced by both SN components. Thin lines delineate the uncertainty regions of each component.
Left: assuming 1 per cent of the CC~SN progenitor mass is converted into iron. 
Right: assuming each CC~SN, on average, produces 0.066 $\rm{M_\odot}$ of iron.}
\label{fig:iron}
\end{minipage}
\end{figure*}

Following \citet{2008NewA...13..606B}, we can use our results to calculate the mean cosmic accumulation of iron.
A typical SN~Ia produces $\sim 0.7~\rm{M_\odot}$ of iron (e.g., \citealt{2007ApJ...670..592M}).
We integrate over the SN~Ia rate evolution derived from convolving the power-law DTD described in the previous section with the Y08 SFH, multiplied by the above iron yield, to derive the amount of iron produced by SNe~Ia.
The uncertainty in the amount of iron contributed by SNe~Ia is calculated by integrating the upper and lower bounds of the shaded area in Fig.~\ref{fig:dtd_power}, multiplied by the above iron yield.
This takes into account both the spread in the SN~Ia rate measurements, and the plausible range of SFH shapes.
We calculate the amount of iron produced by CC~SNe by integrating over the Y08 SFH fit.
Using the \citet{1955ApJ...121..161S} IMF (as assumed by Y08), we calculate either the number of stars with masses $8<M<50~\rm{M_\odot}$ or the mass in such stars. 
If we assume that 1 per cent of the CC~SN progenitor mass is converted into iron (as in \citealt{maoz2010clusters}), then the present-day ratio of iron mass produced by SNe~Ia to that produced by CC~SNe is 1:4.
If, on the other hand, we assume that each CC~SN produces, on average, 0.066 $\rm{M_\odot}$ of iron (as in \citealt{2008NewA...13..606B}, based on CC~SN samples from \citealt{2003astro.ph.10057Z} and \citealt{2003ApJ...582..905H}), then the ratio increases to 1:1.
As the major source of uncertainty in the amount of iron contributed by CC~SNe is the normalization of the SFH, we integrate over the O08u and HB06c SFHs to derive upper and lower bounds on the uncertainty region.
Finally, we sum the lower (upper) uncertainty bounds of the separate SN~Ia and CC~SN contributions to arrive at lower (upper) limits on the total cosmic density of iron.

Both scenarios are presented in Fig.~\ref{fig:iron}.
The mean cosmic iron abundance in solar units, marked on the left ordinate axis, is calculated assuming $\Omega_b=0.0445$ for the baryon density in units of the critical closure density \citep{2011ApJS..192...18K}, and $Z_{\rm{Fe,\odot}}=1.3\pm0.1 \times 10^{-3}$ for the solar iron abundance \citep{1998SSRv...85..161G}.
We see that the predicted present-day mean cosmic iron abundance lies in the range (0.09--0.37)~$Z_{\rm{Fe,\odot}}$.
Between $z=0$ and 2, for a given choice of SFH, the abundance behaves roughly linearly as
\begin{equation}\label{eq:iron_param}
\begin{array}{ll}
Z_{\rm{Fe},L} \approx 0.36-0.10(1+z), \\
Z_{\rm{Fe},R} \approx 0.20-0.06(1+z),
\end{array}
\end{equation}
for the best-fitting solid black curves in the left (L) and right (R) panels of Fig.~\ref{fig:iron}, respectively.
The choice of SFH propagates to a dominant systematic uncertainty in the CC~SN contribution to the iron abundance.

From the figures above, we see that the bulk of the predicted IGM enrichment occurs at $z<2$. 
At these epochs, most of the IGM (holding the majority of baryons in the Universe) is in the warm-hot intergalactic medium (WHIM) phase. 
The WHIM has yet to be detected clearly in the X-ray absorption lines of intermediate elements, let alone of iron, which is extremely challenging. 
However, it is conceivable that future X-ray missions, such as the International X-ray Observatory \citep{2011arXiv1102.2845B} or the recently cancelled EDGE, having large effective areas and high spectral resolution, could detect FeXVII absorption at $\lambda \approx 17$\AA, and eventually lead to an actual low-$z$ iron abundance measurement \citep{2008SSRv..134..405P}.
Such a measurement can then be compared with these predictions to constrain both the integrated iron production of CC~SNe and the efficiency with which metals produced by SNe are ejected into the IGM.


\section{CONCLUSIONS}
\label{sec:discuss}

By surveying four deep epochs of the 0.25 deg$^2$ SDF, we have assembled a sample of 150 SNe, of which 26 are SNe~Ia at $1.0<z<1.5$, and 10 are SNe~Ia at $1.5<z<2.0$.
This is the largest sample of SNe~Ia at such high redshifts to date.
The number of SNe~Ia in our $1.0<z<1.5$ bin is comparable to that of D08 in the same range, but our $1.5<z<2.0$ sample is 2.5 times as large.
While we may have discovered some non-Ia transients in the redshift range $1.5<z<2.0$, we have argued that further contamination of our high-$z$ SN~Ia sample is unlikely.
Through various tests, we have shown that the high-$z$ SNe in our sample are securely associated with galaxies at these redshifts, and since our survey is mostly insensitive to CC~SNe, they must be SNe~Ia.
The SN~Ia rates derived from our sample are consistent with those of D08, but are 2--3 times more precise, with uncertainties of 30--50 per cent.
Our measurements indicate that, following the rise at $0<z<1$, the SN~Ia rate appears to level off after $z \approx 1$, but there is no evidence for a decline in the SN~Ia rate evolution of the form advocated by D08.

Based on these rates and on a growing number of accurate measurements at $z<1$, and combined with different SFHs, we find that a power-law DTD of the form $\Psi(t)=\Psi_1(t/1~\rm{Gyr})^\beta$ fits the data well, with $\beta=-1.1 \pm 0.1 (0.2)$ (68 and 95 per cent statistical confidence, respectively) $\pm 0.17$ (systematic).
This form is consistent with the DTDs found by most of the recent SN~Ia surveys, in a variety of environments, at different redshifts, and using different methodologies \citep{2008PASJ...60.1327T, maoz2010loss, maoz2010clusters, maoz2010magellan}.
A $t^{-1/2}$ power law at all delay times, as proposed by \citet{2008ApJ...683L..25P}, is marginally consistent with the data.
DTDs consisting of broken power laws are also acceptable, as long as $t_c$, the time at which the DTD transitions from a $t^{-1/2}$ power law to a $t^{-1}$ power law, is less than $\sim 0.8$ Gyr (68 per cent confidence).
The Gaussian DTD proposed by D04, D08, and \citet{2004ApJ...613..200S, 2010ApJ...713...32S} is ruled out by all but one of the SFHs tested here.
Overall, these results are suggestive of the DD progenitor scenario, for which a power law with $\beta \approx -1$ is a generic prediction.
In contrast, DTDs predicted by SD models have a variety of forms, but as a rule, they fall off steeply or cut off completely beyond delays of a few Gyr (e.g., \citealt*{2011arXiv1105.5265M}).
The DD channel may not be the only one that produces SNe~Ia, but it appears that a large fraction of SNe~Ia are formed in this way, or in some other way that mimics the DTD predictions of the DD channel.

Using the best-fitting power-law DTD, we have reconstructed how the mean iron abundance of the universe has evolved with cosmic time, and predict it is now in the range (0.09--0.37)~$Z_{\rm{Fe,\odot}}$.
This prediction is consistent with those of \citet{2004ApJ...616..643F} and \citet{2008NewA...13..606B}, but is now based on the most recent and accurate SN~Ia rate measurements, the full range of plausible cosmic SFHs, and the current DTD estimations.

The time-integrated number of SNe~Ia per unit mass derived from the best-fitting power-law DTD, assuming a \textquoteleft diet-Salpeter\textquoteright\ IMF \citep{2003ApJS..149..289B}, is in the range $N_{\textrm{SN}}/M_*=(0.5$--$1.5) \times 10^{-3}~\rm{M_\odot}^{-1}$, though it might easily be higher if the normalization of the SFH is found to be lower than currently assumed. 

The CC~SN rate at $\langle z \rangle = 0.66$ is $6.9^{+9.9}_{-5.4} \times 10^{-4}$ yr$^{-1}$ Mpc$^{-3}$.
This value is consistent with the only other measurement in this redshift range (D04), and shows that, as expected, the CC~SN rate tracks the cosmic SFH out to $z \approx 1$. 

Our survey in the SDF has reached the point where the systematic uncertainties in the SN rates are comparable to the statistical uncertainties.
The 1.5 deg$^2$ Hyper-Suprime Cam \citep{2010SPIE.7740E..83F}, soon to be installed on the Subaru Telescope, could allow discovery of larger numbers of SNe per epoch and thus a further reduction in the statistical uncertainties.
A new SN survey in a well-studied field, such as the SDF or the SXDF, but with cadences designed to probe the light curves of the SNe, could permit classification of the SNe at a higher level of accuracy, thus reducing the systematic uncertainties as well.
This will also apply to future massive surveys such as the Large Synoptic Survey Telescope \citep{2004AAS...20510802S} or the Synoptic All-Sky Infrared Survey \citep{2009arXiv0905.1965B}, for which traditional spectroscopic followup observations will be impossible, but to which the approach we have adopted here is optimally suited.

Two {\it HST} Treasury programmes --- CLASH (GO12065-12069, GO12100-12104) and CANDELS (GO12060-12061) --- have recently begun deep IR observations utilizing the {\it F125W} and {\it F160W} filters on the Wide Field Camera 3.
These filters, similar to \J\ and $H$, will probe the optical part of the SN spectrum out to $z \approx 1.5$, and the near-UV part of the spectrum out to $z \approx 2.7$.
By observing the optical part of the spectrum in the observer-frame IR, one can reduce the uncertainties due to high-redshift dust, thus lowering the systematic uncertainty of the SN rates in the redshift range $1.0<z<1.5$.
Ultimately, these programmes will provide independent measurements of the SN~Ia rate in the $1.0<z<2.0$ range probed by this work, as well as extend our knowledge of the SN~Ia rate evolution out to $z \approx 2.7$.
Based on the results presented here, as seen in Fig.~\ref{fig:dtd_power}, we predict that CLASH (CANDELS) will observe 10--24 (9--19) SNe~Ia at $1.0<z<2.0$, and 0--4 (2--7) SNe~Ia at $2.0<z<2.7$.

\section*{Acknowledgments}

We thank Mamoru Doi for his contributions to this project, Robert Feldmann, Suzanne Hawley, Eric Hilton, Weidong Li, and Lucianne Walkowicz for helpful discussions and comments, and Masao Hayashi, Nobunari Kashikawa, Chun Ly, Matt Malkan, and Tomoki Morokuma for sharing their data.
The referee is thanked for many thoughtful comments that improved the presentation.
O.G. thanks Joshua Bloom for hosting him during a month-long visit to the University of California, Berkeley.
This work was based on data collected at the Subaru Telescope, which is operated by the National Astronomical Observatory of Japan.
Additional data presented here were obtained at the W. M. Keck Observatory, which is operated as a scientific partnership among the California Institute of Technology, the University of California, and the National Aeronautics and Space Administration; the Observatory was made possible by the generous financial support of the W. M. Keck Foundation. 
The authors wish to recognize and acknowledge the very significant cultural role and reverence that the summit of Mauna Kea has always had within the indigenous Hawaiian community. 
We are most fortunate to have the opportunity to conduct observations from this mountain.
This research has made use of NASA's Astrophysics Data System (ADS) Bibliographic Services.

D.M. acknowledges support by a grant from the Israel Science Foundation (ISF).
D.P. is supported by an Einstein Fellowship, and by the US Department of Energy Scientific Discovery through Advanced Computing (SciDAC) programme under contract DE-FG02-06ER06-04.
A.V.F. is grateful for the financial support of NSF grant AST-0908886, the TABASGO Foundation, and Department of Energy grant DE-FG0-08ER41563.
R.J.F. is supported by a Clay fellowship.
A.G. is supported by an FP7/Marie Curie IRG fellowship and a grant from the ISF.



\newpage
\begin{landscape}
\begin{table}
\caption{$1.5<z<2.0$ SNe discovered in the SDF. The full table, including the entire sample, is available in the electronic version of the paper.}
\begin{tabular}{l|c|c|c|c|c|c|c|c|c|c|c|c|c|c|c}
\hline
\hline
ID & $\alpha$ & $\delta$ & Offset & \R & \I & \z & S/N & Photo-$z$ & $\chi^2$ & Spec-$z$ & $P_{\textrm{Ia}}$ & Post-$z$ & $\chi^2$ & Type & Adopted-$z$ \\
(1) & (2) & (3) & (4) & (5) & (6) & (7) & (8) & (9) & (10) & (11) & (12) & (13) & (14) & (15) & (16) \\
\hline

SNSDF0503.21 & 24:50.36 & 45:16.52 & 0.26(14) & $>27.28$ & 26.07(15) & 25.34(19) & 12 & 1.70 & 0.95 & ... & 0.73 & 1.62 & 0.31 & Ia & 1.62 \\

SNSDF0702.28 & 24:47.92 & 44:36.92 & 0.64(10) & $>28.09$ & 27.24(27) & 26.42(26) & 5 & 2.05 & 1.65 & ... & 1.00 & 1.99 & 4.64 & Ia & 1.99 \\

SNSDF0705.25 & 25:30.61 & 12:59.39 & 0.58(11) & $>26.98$ & $>27.33$ & 25.77(25) & 4 & 1.55 & 1.44 & ... & 0.95 & 1.54 & 5.50 & Ia & 1.55 \\

SNSDF0705.29 & 25:01.80 & 18:38.87 & 0.24(12) & $>26.98$ & $>27.33$ & 26.29(32) & 3 & 1.61 & 3.57 & ... & 0.93 & 1.51 & 3.47 & Ia & 1.61 \\

SNSDF0806.31 & 24:19.53 & 29:59.53 & 0.10(11) & $>27.19$ & 26.91(24) & 25.70(16) & 11 & 1.83 & 4.46 & ... & 1.00 & 1.83 & 0.05 & Ia & 1.83 \\

SNSDF0806.32 & 25:20.44 & 43:08.62 & 0.36(12) & $>27.19$ & 25.89(10) & 25.72(17) & 10 & 1.92 & 7.33 & ... & 1.00 & 1.66 & 6.54 & Ia & 1.66 \\

SNSDF0806.38 & 23:33.39 & 14:20.86 & 0.56(11) & $>27.19$ & $>27.80$ & 25.86(19) & 3 & 1.71 & 10.20 & ... & 0.83 & 1.83 & 2.79 & Ia & 1.71 \\

SNSDF0806.46 & 24:29.97 & 14:08.90 & 0.23(11) & $>27.19$ & 27.12(27) & 26.25(24) & 6 & 1.56 & 3.98 & ... & 0.97 & 1.53 & 0.57 & Ia & 1.56 \\

SNSDF0806.50 & 23:46.04 & 39:00.42 & 0.86(13) & $>27.19$ & 27.00(25) & 26.26(25) & 6 & 1.66 & 5.45 & ... & 0.99 & 1.66 & 0.92 & Ia & 1.66 \\

SNSDF0806.57 & 25:33.63 & 28:03.32 & 0.46(13) & $>27.19$ & $>27.80$ & 26.63(30) & 4 & 1.55 & 2.53 & ... & 0.90 & 1.54 & 3.46 & Ia & 1.55 \\

\hline
\multicolumn{16}{l}{(1) -- SN identification.} \\
\multicolumn{16}{l}{(2)--(3) -- Right ascensions (J2000; starting at 13$^{\rm h}$) and declinations (J2000; starting at +27\textdegree).} \\
\multicolumn{16}{l}{(4) -- SN offset from host galaxy, in arcseconds. Uncertainties appear in parenthesis, and have been multiplied by 100.} \\
\multicolumn{16}{l}{(5)--(7) -- SN photometry in the \R, \I, and \z\ bands, in magnitudes. Uncertainties appear in parenthesis, and have been multiplied by 100.} \\
\multicolumn{16}{l}{(8) -- Signal-to-noise ratio of the SN, as measured in the \z-band image.} \\
\multicolumn{16}{l}{(9)--(10) -- Photometric redshift of SN host galaxy, with reduced $\chi^2$, as derived with ZEBRA.} \\
\multicolumn{16}{l}{(11) -- Spectroscopic redshift of SN host galaxy, where available.} \\
\multicolumn{16}{l}{(12)--(14) -- Probability of a SN being a Type Ia, or CC~SN, as derived with the SNABC, together with its posterior redshift and reduced $\chi^2$.} \\
\multicolumn{16}{l}{(15)--(16) -- Final adopted SN type and redshift.} \\

\multicolumn{16}{l}{} \\
\multicolumn{16}{l}{} \\
\multicolumn{16}{l}{}

\end{tabular}
\label{table:SNe}
\end{table}
\end{landscape}

\newpage
\begin{landscape}
\begin{table}
\caption{$1.5<z<2.0$ SN host galaxies. The full table, including the entire sample, is available in the electronic version of the paper.}
\begin{tabular}{l|c|c|c|c|c|c|c|c|c|c|c|c|c}
\hline
\hline
ID & $\alpha$ & $\delta$ & $FUV$ & $NUV$ & \B & \V & \R & \I & \z & $NB$816 & $NB$921 & \J & \K \\
(1) & (2) & (3) & (4) & (5) & (6) & (7) & (8) & (9) & (10) & (11) & (12) & (13) & (14) \\
\hline

hSDF0503.21 & 24:50.38 & 45:16.55 & $-1$ & 0 & 26.81(13) & 26.97(28) & 26.45(19) & 26.60(25) & $>26.62$ & $>26.63$ & 26.27(32) & ... & ... \\

hSDF0702.28 & 24:47.96 & 44:37.39 & $-1$ & 0 & 24.71(03) & 24.55(05) & 24.49(05) & 24.41(06) & 24.48(09) & 24.28(08) & 24.57(11) & ... & 23.32(15) \\

hSDF0705.25 & 25:30.64 & 12:59.84 & $-1$ & 0 & 24.49(03) & 24.46(05) & 24.17(04) & 24.05(05) & 23.86(06) & 23.94(06) & 24.37(10) & 23.19(18) & 23.16(14) \\

hSDF0705.29 & 25:01.78 & 18:38.96 & $-1$ & 0 & 24.17(02) & 23.95(04) & 23.59(03) & 23.12(03) & 22.71(02) & 22.75(02) & 22.74(02) & ... & 20.81(03) \\

hSDF0806.31 & 24:19.54 & 29:59.51 & $-1$ & 0 & 26.17(09) & 25.85(14) & 25.42(09) & 24.69(07) & 24.01(06) & 24.56(10) & 24.11(08) & ... & 20.80(03) \\

hSDF0806.32 & 25:20.46 & 43:08.35 & $-1$ & 0 & 25.74(06) & 25.21(09) & 25.43(10) & 25.46(12) & 25.17(15) & 26.62(34) & $>26.54$ & ... & ... \\

hSDF0806.38 & 23:33.35 & 14:20.69 & 0 & 0 & 24.93(03) & 24.29(02) & 24.23(02) & 23.93(02) & 23.82(04) & 23.56(03) & 23.65(04) & 22.81(15) & 22.30(08) \\

hSDF0806.46 & 24:29.98 & 14:09.13 & $-1$ & 0 & 27.23(17) & 25.87(14) & 25.75(12) & 24.58(07) & 23.83(06) & 24.10(07) & 23.78(06) & ... & 21.23(04) \\

hSDF0806.50 & 23:46.02 & 38:59.59 & $-1$ & 0 & 25.36(05) & 24.56(05) & 24.28(04) & 23.95(04) & 23.15(03) & 23.91(06) & 23.23(04) & 21.68(08) & 20.61(03) \\

hSDF0806.57 & 25:33.67 & 28:03.27 & $-1$ & 0 & 25.66(06) & 25.65(12) & 25.16(08) & 24.88(08) & 24.47(09) & 24.78(12) & 24.77(13) & ... & 22.51(08) \\

\hline

\multicolumn{14}{l}{Note - magnitude limits are $3\sigma$.} \\
\multicolumn{14}{l}{(1) -- SN identification.} \\
\multicolumn{14}{l}{(2)--(3) -- Right ascensions (J2000; starting at 13$^{\rm h}$) and declinations (J2000; starting at +27\textdegree).} \\
\multicolumn{14}{l}{(4)--(5) -- {\it GALEX FUV} and {\it NUV} photometry. $-1$ means no UV signal observed in this band; 1 means a clear UV signal associated with the target galaxy;} \\
\multicolumn{14}{l}{\qquad\quad\quad\, and 0 means the UV signal could not be unequivocally matched to the target galaxy.} \\
\multicolumn{14}{l}{(6)--(12) -- Subaru optical photometry, in magnitudes. Uncertainties appear in parenthesis, and have been multiplied by 100.} \\
\multicolumn{14}{l}{(13)--(14) -- UKIRT \J\ and \K\ photometry, in magnitudes. Uncertainties appear in parenthesis, and have been multiplied by 100.} \\

\end{tabular}
\label{table:SNe_hosts}
\end{table}
\end{landscape}

\end{document}